\documentclass[article,longbibliography,nofootinbib]{revtex4-2}
\usepackage{graphicx} 
\usepackage{xcolor}
\usepackage{tikz}
\usepackage{amsmath}
\usepackage{subcaption}
\usepackage[version=4]{mhchem}
\usepackage{placeins}
\usepackage{caption}
\usepackage[colorlinks=true, urlcolor=blue, linkcolor=black, citecolor=black]{hyperref}
\usetikzlibrary{arrows.meta, positioning, shapes.geometric,
                 shapes.arrows, calc, backgrounds, fit}
                 \usepackage[utf8]{inputenc}
\usepackage{graphicx}
\usepackage{xcolor}
\usepackage{tikz}
\usepackage[margin=0.5in]{geometry}

\definecolor{prlblue}{RGB}{31,78,121}
\definecolor{prlred}{RGB}{153,31,35}
\definecolor{prlgray}{RGB}{68,68,68}
\definecolor{prllg}{RGB}{120,120,120}
\definecolor{prlbluetint}{RGB}{234,240,247}
\definecolor{prlredtint}{RGB}{248,234,233}

\definecolor{corefill}{RGB}{233,222,198}
\definecolor{shellfill}{RGB}{201,216,236}
\definecolor{outline}{RGB}{55,60,66}
\definecolor{muted}{RGB}{102,108,118}
\definecolor{accent}{RGB}{69,116,170}
\definecolor{hydrox}{RGB}{74,127,182}
\definecolor{calcined}{RGB}{182,112,83}
\definecolor{treated}{RGB}{126,106,171}
\definecolor{cardbg}{RGB}{248,249,251}
\definecolor{marker}{RGB}{45,150,120}
\definecolor{isotope}{RGB}{190,92,140}

\newcommand{\reqid}[1]{\hypertarget{req:#1}{\textbf{#1}}}

\newcommand{\reqrefs}[1]{\hyperlink{req:#1}{#1}}

\definecolor{lcaccent}{RGB}{70,107,178}
\definecolor{lcaccentlight}{RGB}{232,238,249}
\definecolor{lcsoftgray}{RGB}{243,245,248}
\definecolor{lctextgray}{RGB}{58,63,71}
\definecolor{lcparticlegray}{RGB}{188,193,200}

\tikzset{
  lcphasecircle/.style={
    circle,
    draw=lcaccent!92,
    line width=0.95pt,
    fill=white,
    minimum size=1.15cm,
    inner sep=0pt
  },
  lcphaselabel/.style={
    font=\sffamily\footnotesize,
    text=lctextgray,
    align=center
  },
  lcicon/.style={
    draw=lcaccent!92,
    line width=0.9pt,
    line cap=round,
    line join=round
  },
  lciconfill/.style={
    draw=lcaccent!92,
    fill=lcaccentlight,
    line width=0.85pt,
    line cap=round,
    line join=round
  }
}

\usepackage{xcolor}

\usepackage{textcomp}

\usepackage{placeins}

\usepackage[version=4]{mhchem}
\usepackage[margin=0.5in]{geometry}

\usepackage{eso-pic}

\usepackage{transparent}

\definecolor{prlblue}{RGB}{31,78,121}
\begin{document}

\title{Composite sub-micron solid particles engineered to enable \\ 
safe, controllable, efficient, and practical SAI}

\author{Amyad Spector$^*$$^\dagger$, Tzemah Kislev$^*$$^\ddagger$, Yair Segev, Yanai Yedvab, Eli Waxman$^\S$}
\thanks{These authors contributed equally to this work.\\ $^\dagger$\href{mailto:a.spector@stardust-initiative.com}{a.spector@stardust-initiative.com}; $^\ddagger$\href{mailto:t.kislev@stardust-initiative.com}{t.kislev@stardust-initiative.com}; $^\S$\href{mailto:e.waxman@stardust-initiative.com}{e.waxman@stardust-initiative.com} }
\affiliation{Stardust Labs, Ness Ziona, Israel; \url{www.stardust-initiative.com}}

\begin{abstract}

The properties of candidate particles for solar radiation management (SRM) through stratospheric aerosol injection (SAI) should comply with safety, controllability, and functionality requirements. Following the proposal in \cite{2026SafetyWP} of safety and controllability requirements, we define here a set of functionality requirements on the particles' properties - optical properties, stratospheric residence time, scalable manufacturing compatibility, and aerial dispersion compatibility- that, if met, ensure the feasibility of practical implementation, providing reflection of $\sim1$\% of the solar flux. We then present design principles and fabrication methods for sub-micron solid particles that may enable satisfying the combined requirements. 

These requirements do not identify a unique solution, but favor sub-micron particles with tightly controlled size distributions and stable properties over their stratospheric lifetime, and motivate a composite design where 
the bulk core composition is selected primarily for optimal radiative properties 
and the outer shell surface is engineered to control atmospheric chemistry and aging, 
and enhance aerial dispersion compatibility. 
We present two specific particle designs that are viable for meeting the coupled requirements: 
amorphous silica spheres and calcium-carbonate cores surrounded by spherical silica shells (both with appropriate surface treatment). The former is at an advanced stage of experimental verification  (described in detail in a companion paper) of meeting all requirements, provides a practical platform for surface engineering, and will enable reaching a substantial fraction of $\sim\!1$\% solar flux reflection. The latter is under development and will enable reaching $>1$\% reflection. 

\end{abstract}

\maketitle

\section{Introduction}

As climate risks intensify and emissions reductions may not proceed fast enough to limit near-term warming \citep{unep2025emissions,WMO2025StateClimate,lee2023ipcc,IPCC_WGIII}, solar radiation management (SRM) is receiving growing attention as a possible supplementary intervention method \citep{blackstock2009climate,2021NASEM.RS,2023OSTP.SRM,2023UNEP.SRMReport,2024EC.SRM,2025RoyalSoc.SRM}. The most widely discussed SRM approach is stratospheric aerosol injection (SAI) \citep{teller1997global,crutzen2006albedo,rasch2008overview}, in which submicron aerosols or aerosol precursors are injected into the stratosphere to increase solar flux reflection. If it can be developed and implemented safely, SAI may help reduce near-term warming and some of its associated impacts while buying time for emissions reductions and other long-term climate responses.

Any serious assessment of SAI development and implementation requires explicit requirements. In a companion paper~\cite{2026SafetyWP}, we described an initial proposal for a complete set of safety and controllability requirements that SAI systems should satisfy. Those requirements are intended to address risks of adverse or unintended effects within three categories: on humans, ecosystems, and the environment; on atmospheric chemistry and composition; and on the climate system (the latter are addressed through requirements on the predictability, controllability, and monitoring ability of the induced radiative forcing). Although safety and controllability requirements should be met by, and are therefore defined at the level of, the SAI system, they translate directly into constraints on the properties of the dispersed particles, particularly their composition (including impurities), morphology, size distribution, surface chemistry, aging behavior, and traceability.

The focus of the present paper is complementary and distinct. We define a set of functionality requirements on the particles’ properties - optical properties, stratospheric residence time, scalable manufacturing compatibility, and aerial dispersion compatibility- that, if met, ensures the feasibility of practical implementation, providing a reflection of $\sim1$\% of the solar flux ($\sim$2.5~W/m$^2$, comparable to the radiative forcing due to accumulating atmospheric greenhouse gases over the past 50 years). The requirements we set for particle properties are intended to be sufficient to enable the construction of a complete SAI system, including particle production, lofting, and dispersion, as well as efficient monitoring, based on technology that is currently available or could be developed over a few-year period. They are thus tightly related to the planned implementation route, which we consider to be through aerial dispersion by a fleet of airplanes flying in the lower stratosphere. Our assumptions regarding fleet capabilities that can be reasonably achieved within the coming few years are described in \S~\ref{sub-sec:functionality}.

Quantitative requirements on particle properties derived from functionality requirements depend on the quantitative functional properties of the SAI system, including the manufacturing, lofting, dispersion, and monitoring sub-systems. Our goal in the current document is thus not to define exact functionality-driven quantitative requirements, but rather to identify the particle properties that need to be constrained and set ballpark quantitative requirements that will enable practical implementation. It is, furthermore, important to note that the constraints we set on particles' properties are sufficient but not necessarily necessary for a practical implementation, in the sense that there may be alternative implementation routes that differ from the one considered here.

Following the definition of particle functionality requirements, we present design principles and fabrication methods for sub-micron solid particles that may enable satisfying the combined safety, controllability, and functionality requirements. Our approach differs from that of many earlier analyses, which examine the safety aspects of particles selected for their attractive radiative performance. We follow an alternative path, examining whether materials selected for their attractive safety potential can be used to construct engineered particles that also meet the functionality requirements.

While the combined requirements do not identify a unique solution, they do favor sub-micron particles with tightly controlled size distributions and stable properties over their stratospheric lifetime, and motivate a composite design where the bulk core composition is selected primarily for optimal radiative properties, and the outer shell surface is engineered to control atmospheric chemistry and aging, and enhance aerial dispersion compatibility.

We present two specific particle designs that are viable for meeting the coupled
requirements. "GEN1" particles, see Fig.~\ref{fig:powder_dispersion_workflow}, are near-spherical sub-micron amorphous silica particles produced by a bottom-up St\"ober process followed by surface treatment including calcination. It is a practical realization of the requirements-driven approach, allowing, in particular, tight control over size and morphology, as well as surface engineering to regulate cohesion, wetting, adhesion, and heterogeneous interactions. GEN1 particles are at an advanced stage of experimental verification of meeting all requirements. The main results of this experimental work, which is described in detail in a companion paper \citep{sai_fabrication_heterogeneous_chemistry}, are described in \S~\ref{sub-sec:GEN1}. GEN1 particles will enable reaching a substantial fraction of 1\% solar flux reflection, limited by the safety requirement limit on stratospheric heating \cite{2026SafetyWP}. GEN1 particles also provide a practical platform for surface engineering and for the development of the other components of the SAI system. GEN2 particles, which are under development, are calcium-carbonate cores surrounded by spherical amorphous silica shells with reduced infrared absorption that may enable $>1\%$ solar flux reflection while meeting the safety limit on stratospheric heating.

Several companion papers \cite{sai_atmospheric_transport,sai_climatic_response,sai_dispersion_subsystem,sai_manufacturing_processes,sai_monitoring_subsystem,sai_particle_architectures,sai_tagging_encoding_technology} describe in detail our R\&D results for developing all components of a full SAI system, which will enable a practical implementation using GEN1 and GEN2 particles, including the dispersion subsystem, manufacturing processes and their scale-up compatibility, 
and atmospheric transport and radiative transfer modeling. 
Additional studies, including monitoring subsystem design, estimation of climatic response as a function of controlled radiative forcing, further GEN1 and GEN2 particle studies, 
and exploration of additional particle architectures, are ongoing and will be described in future publications. 

Earlier works explored solid-particle alternatives to sulfate aerosols, emphasizing their potential to reduce stratospheric heating, improve optical performance, and mitigate some heterogeneous-chemistry risks \citep{teller1997global,dykema2016improved,pope2012stratospheric,Vattioni2023,Vattioni2024,weisenstein2015solar}. These alternatives address some fundamental limitations of sulfate-based approaches, in particular the significant infrared radiation absorption and significant reactive surface area for heterogeneous chemistry of sulfate-based aerosols. Here, we go beyond these earlier works by providing a complete analysis of the particle properties required to ensure compatibility with all system-level requirements (including safety, controllability, lofting and dispersing system requirements), and by setting quantitative requirements. 
It is important to note that the sulfate-based approach SAI is subject to additional limitations. It is poorly suited to small-scale field testing, due to the large natural stratospheric sulfate aerosol background \citep{sparc2006assessment,sparc2017data,Kovilakam2020GloSSAC,Liu2012}, which hinders the possibility of carrying out small-scale experiments in which limited aerosol quantities are dispersed with small radiative forcing and atmospheric chemistry impacts (especially when deployment relies on the relatively slow atmospheric conversion of \ce{SO2} to aerosol). Furthermore, SAI based on the injection of liquid aerosols (or aerosol precursor gases) is, in general, subject to large uncertainties in predicting radiative forcing and atmospheric chemical impacts. This is largely due to the long term evolution of particle size, composition, and mixing during their residence in the stratosphere.

The remainder of this paper is organized as follows. The requirements on particle properties derived from safety and controllability requirements and from functionality requirements are described in \S~\ref{sec:requirements}. Design principles and manufacturing methods that may enable meeting the requirements are described in \S~\ref{sec:design}. The GEN1 and GEN2 particles are presented in \S~\ref{sec:composite}. \S~\ref{sec:composite} presents both the fabrication process and the main results of the experimental program for validating that the particle properties meet both safety and functionality requirements. The main conclusions are briefly summarized in \S~\ref{sub-sec:GEN1}.

\section{SAI Requirements}
\label{sec:requirements}

In \S~\ref{sub-sec:safety} we discuss the requirements on particle properties that are derived from the safety and controllability requirements that SRM systems should meet, following \cite{2026SafetyWP}. The presentation is highly concise, and the reader is referred to \cite{2026SafetyWP} for a detailed discussion and explanation of the safety and controllability requirements. In \S~\ref{sub-sec:functionality} we present a set of requirements on particle properties that, if met, will enable practical implementation of SAI on a meaningful scale, i.e., providing $\sim1$\% solar flux reflection. 

The requirements discussed in \S~\ref{sub-sec:safety} are aimed at ensuring that the particles are admissible for SAI use, while those in \S~\ref{sub-sec:functionality} are intended to ensure the feasibility of a practical implementation (note that increased efficiency that reduces the total particle burden contributes to meeting safety requirements by reducing the particles' concentration and their atmospheric/environmental impacts). The combined set of particle-related requirements is 
summarized in Appendix~\ref{app:requirements}.

\subsection{Safety and Controllability} 
\label{sub-sec:safety}

The constraints on particle properties that are driven by safety and controllability requirements  are summarized in Appendix~\ref{app:requirements}. They are divided into three categories: Requirements intended to address risks of adverse or unintended effects on humans, ecosystems, and the environment (\reqrefs{A1}--\reqrefs{A7}); on atmospheric chemistry and composition  (\reqrefs{B1a}--\reqrefs{B4}); and on the climate system. The latter are addressed through requirements on the 
ability to measure, predict, control, attribute, and phase down the induced radiative forcing (\reqrefs{C1}--\reqrefs{C4d}). 
Constraints on composition, morphology, surface properties, environmental fate, and reproducibility under large-scale manufacturing should be derived from these constraints rather than chosen solely for radiative properties.
\begin{itemize}
    \item [A.] Composition and impurity content must satisfy material-screening, impurity-control, and hazardous-material exclusion criteria given in \reqrefs{A1}, \reqrefs{A1a}, \reqrefs{A1b}; The size distribution must avoid a non-negligible sub-0.1~$\mu$m fraction according to \reqrefs{A1c}; Ground-level and environmental exposure must remain compatible with toxicological and ecological thresholds according to \reqrefs{A2}--\reqrefs{A4}; Candidate materials should exclude mutagenic or carcinogenic classes according to \reqrefs{A5}; After deposition to soils or other environmental reservoirs, the particles should enter viable elimination pathways without persistence or bioaccumulation, according to \reqrefs{A6} and \reqrefs{A7}. 
    \item [B.] Particle composition and surface properties should enable meeting limits on the modification of the locally averaged (over 1~month-1000~km scale) column densities of stratospheric gases due to direct interactions (i.e., the release of particle components or reaction products to the stratosphere and the incorporation of stratospheric constituents into the particles) according to \reqrefs{B1a} and \reqrefs{B1b}, on ozone modification in both near-global and high-latitude regimes according to \reqrefs{B2} and \reqrefs{B2'}, on excess sulfate surface area and Polar Stratospheric Cloud (PSC) amplification according to \reqrefs{B2a} and \reqrefs{B2b}, and on upper-troposphere cloud radiative forcing modification according to \reqrefs{B3}. 
    \item [C.] The particle properties should be compatible with enabling an accurate determination of the induced radiative forcing during gradual ramp-up, where the quantity of dispersed particles and their radiative forcing is small ($\ll$0.1\% solar), according to \reqrefs{C1} and \reqrefs{C2}; with enabling accurately predictable and controllable radiative forcing on 1~month-1000~km-temporal and spatial scales, including controlled and rapid phase down, according to \reqrefs{C3a}, \reqrefs{C3b}, \reqrefs{C4a} and \reqrefs{C4b};  
    and with limits on lower-stratospheric heating according to \reqrefs{C4c}.
\end{itemize}

Aging of the particles during stratospheric residence (due to prolonged exposure to UV radiation, trace gases, and aerosols) should not compromise meeting the above constraints. This sets limits on aging changes in surface chemistry, morphology, optical properties and aggregation, in accordance with \reqrefs{A7} and \reqrefs{B4}. 
As discussed in \cite{2026SafetyWP}, meeting the safety and controllability requirements requires an effective monitoring system, which, in particular, allows a measurement of the (locally, 1~month-1000~km, averaged)
stratospheric particle concentrations and size distributions, that in turn enables determining the resulting (locally averaged) radiative forcing (based on known optical particle properties) to $\sim10$\% accuracy for a low mass of dispersed particles corresponding to $\ll 0.1$\% (0.3~W$/$m$^2$) modification of
the radiative flux at the top of the atmosphere. The measurement of the particles' distribution and their properties may be achieved by carrying out in situ measurements or by recovering samples for ground measurements. In both cases, particle tagging that enables attribution, i.e. determining the origin (time and location) of a particle observed at a given (space and time) point, is highly advantageous (\reqrefs{C4d}) and will significantly reduce the minimum mass of dispersed particles that enables particle distribution measurements. A suite of monitoring capabilities based on available technologies, including remote sensing and particle measurements, is described in a companion article currently in preparation~\citep{sai_monitoring_subsystem}. Beyond its value for physical measurements, particle tagging may also be an additional technological enabler for governance, compliance verification, and post-mixing attribution, as discussed in a companion article~\citep{sai_technological_capabilities_for_governance}.

\subsection{Functionality}
\label{sub-sec:functionality}

We describe and explain below the constraints on particle properties that are intended to ensure the feasibility of a practical SAI system. We consider aerial dispersion by a fleet of airplanes flying at the lower stratosphere. The lofting system must be capable of reaching the required dispersion altitude with adequate payload per sortie, and sortie rates must be sufficient to deliver the annual mass budget across the intended latitude bands. In order to derive the constraints on particle properties, we first provide a crude estimate of the expected capabilities of a reasonable aircraft fleet.

To ensure a stratospheric residence time of $\sim1$~yr and enable the required flexibility in dispersing at different latitude bands (see appendix \ref{app:requirements}), the lofting airplanes should be capable of reaching an altitude which is sufficiently above the tropical tropopause altitude of 16--17 km, implying a capability of reaching $>18.5$~km ($>60$ kft) altitude. 
We consider aircraft carrying $\sim 5$~t or $\sim 20$~t per sortie, representing a smaller initial platform and a larger or more specialized second phase platform. The required aircraft number $N$ may then be written as $$N \sim 10^3 \left(\dot{M}/(10\,\mathrm{Mt\,yr^{-1}})\right)\left(T_{\mathrm{flight}}/(3000\,\mathrm{h\,yr^{-1}}\right))^{-1}\left(f_{\mathrm{disp}}/0.5\right)^{-1}\left(m_{\mathrm{sortie}}/10\,\mathrm{t}\right)^{-1},$$ where $\dot{M}$ is the target deployment rate, $T_{\mathrm{flight}}$ is the annual flight time per aircraft, $f_{\mathrm{disp}}$ is the fraction of flight time spent actively dispersing, $m_{\mathrm{sortie}}$ is the dispersed mass per sortie, and $t_{\mathrm{disp}}$ is the effective dispersion time per sortie. Adopting plausible fiducial values of $T_{\mathrm{flight}}\sim 3000$ h\,yr$^{-1}$, $f_{\mathrm{disp}}\sim 0.5$, $m_{\mathrm{sortie}}\sim 5$--20 t, and $t_{\mathrm{disp}}\sim 1$--2~h (with dispersion hardware mass ideally limited to $\lesssim 0.2\,m_{\mathrm{sortie}}$), a fleet of $\simeq500$ airplanes is required for deployment of 10~Mt per year. A practical target of an all-in delivered cost is at most a few dollars per kilogram dispersed, inclusive of fuel, maintenance, flight operations and crew, and lease-equivalent capital costs.

The constraints on particle properties are derived using the fiducial fleet estimate above: for a 1~year stratospheric residence time of the particles a radiative forcing of $\sim 0.25{\rm W/m^2/Mton}$ is required to produce $2.5{\rm W/m^2}$ forcing, and the dispersion system should enable dispersion at a rate of few tons per hour from the aircraft (to enable several sorties per day). The derived requirements are summarized in tables~\ref{tab:req_eff_or} -- \ref{tab:req_eff_df} of Appendix~\ref{app:requirements}.

\begin{enumerate}

 \item \textbf{Optical properties} (requirements \reqrefs{Oa}-\reqrefs{Oc} in table~\ref{tab:req_eff_or})\textbf{.} Efficient sunlight scattering requires particle sizes comparable to visible wavelengths, with a dielectric contrast sufficiently above unity at visible wavelengths. This implies particle diameters of 0.2--0.8~$\mu$m and refractive indices well above 1 at $\lambda \sim 550$~nm, yielding forcing efficiencies of few $10^{-1}\;\mathrm{W\,m^{-2}\,Tg^{-1}}$, i.e. few $\;\mathrm{W\,m^{-2}\, (10\,Mton)^{-1}}$. Shortwave absorption at visible wavelengths should be negligible so that incoming solar radiation is predominantly scattered rather than absorbed, which would otherwise reduce the net cooling effect and heat the stratosphere. Likewise, infrared absorption, particularly within the 8--13~$\mu$m atmospheric window, should be low in order not to decrease substantially the radiative forcing efficiency and to avoid significant stratospheric heating \citep{dykema2016improved,weisenstein2015solar,pope2012stratospheric}. These optical properties must remain stable over the particles' stratospheric lifetime. A quantitative analysis of the dependence of the radiative forcing and stratospheric heating per unit mass on the particles' size distribution and optical properties is given in a companion paper dedicated to transport and radiative-transfer modeling \citep{sai_atmospheric_transport}.

  \item \textbf{Stratospheric residence time} (requirements \reqrefs{Ra}-\reqrefs{Rc} in table~\ref{tab:req_eff_or})\textbf{.}   Allowing residence times of order one year in the lower stratosphere requires the particles' size to be small enough to yield a sedimentation velocity that is below the tropical Brewer--Dobson upwelling velocity of approximately 0.1~mm/s \citep{weisenstein2015solar}. Since smaller particles remain aloft longer but scatter less efficiently, the particle's size distribution should be optimized considering jointly residence time and optical performance. This generally favors diameters somewhat smaller than those obtained by considering optical properties alone. Stratospheric aging, including possible coagulation and agglomeration, should not significantly alter the particles' properties (to an extent that affects their ability to meet these requirements). A quantitative analysis of the dependence of particles' transport in the stratosphere on their size distribution and on the spatial and temporal dispersion patterns is given in  the companion paper 
  \citep{sai_atmospheric_transport}.

  \item \textbf{Manufacturability} (requirements \reqrefs{Ma}-\reqrefs{Me} in table~\ref{tab:req_eff_m})\textbf{.} Production must be feasible at the relevant scale of 10 million tons per year at a fabrication cost comparable to the lofting cost of roughly a few dollars per kilogram. This requires a manufacturing process that enables meeting strict quality assurance requirements (particularly with a small variance of constrained particle properties across batches), access to feedstock and manufacturing equipment at the necessary scale, 
  realistic timelines and capital requirements for scale-up, and practical storage and transport logistics from factory to airfield. Taken together, these considerations favor well-established chemical processes supported by a credible scale-up plan. Manufacturing scale-up is discussed in detail in a dedicated companion paper \citep{sai_manufacturing_processes}.

  \item \textbf{Dispersion compatibility} (requirements \reqrefs{Da}-\reqrefs{De} in table~\ref{tab:req_eff_df})\textbf{.} The dispersion system should enable the release of particles at a rate of a few tons per hour from the lofting airplane, with a stratospheric particle size distribution meeting safety and functionality requirements determined above. The particle properties should enable overcoming the challenges of transport and agglomeration of sub-micron particles (at the particle containers and dispersion system as well as in the expanding wake), so as to enable the construction of a dispersion system compatible with the power and airflow that may be provided by the aircraft. The particles should also be designed to ensure that engine exhaust and plume do not degrade their properties. Quantitative experimental results and a model describing solid-aerosol de-agglomeration as a function of turbulent forces and particle microscopic properties are presented in a companion paper dedicated to the dispersion system \citep{sai_dispersion_subsystem}.

\end{enumerate}

\section{Design considerations and fabrication methods}
\label{sec:design}

We present in this section considerations for the choice of the particles' materials (\S~\ref{sub-sec:materials}), fabrication methods (\S\ref{sub-sec:fabrication}), and surface treatment methods (\S~\ref{sub-sec:surface}), that follow from, and may enable meeting, the safety and functionality requirements. These considerations take into account challenges to meeting requirements across the 

full life-cycle of the particles, from fabrication and transportation through deployment, atmospheric transport and residence, and eventual settling to the surface.

\subsection{Choosing biomineralized materials for the bulk composition }
\label{sub-sec:materials}

To increase the feasibility of meeting safety requirements for humans, biosphere, and the environment, we choose naturally abundant, bio-mineralized materials \citep{mann2001biomineralization,lowenstam1989biomineralization} that already participate in natural biogeochemical cycles, with low absorption and sufficiently high refractive indices in the optical bands. Selecting materials that are already widely ingested, implanted, and environmentally cycled allows us to build on an extensive body of human, biotic, and environmental safety validation. 

Key examples include amorphous silica (\ce{SiO2}), found in diatom frustules, radiolarian skeletons, sponge spicules, testate amoebae, and plant phytoliths; calcium carbonate (\ce{CaCO3}), appearing as calcite in coccoliths, foraminifera, and eggshells, aragonite in coral skeletons and many mollusk shells, vaterite in certain gastropods and ascidians, magnesian calcite in echinoderms and coralline algae, and amorphous calcium carbonate as a transient precursor in many organisms; and hydroxyapatite (\ce{Ca10(PO4)6(OH)2}), the primary mineral component of vertebrate bones, teeth, and fish scales \citep{mann2001biomineralization}.  
These materials differ in their infrared absorption properties, particularly within the atmospheric infrared window of 8–13$\mu$m. Calcium carbonate (pure or with partial magnesium substitution) exhibits lower infrared absorption than silica and hydroxyapatite, which is advantageous because infrared absorption can both increase radiative forcing (partially reducing the impact of negative forcing from visible light reflection) and cause stratospheric heating. 

Amorphous silica, calcium carbonate, and hydroxyapatite are produced in a range of established forms and are already part of everyday human life. Both calcium carbonate and silica are classified as Generally Recognized as Safe (GRAS) by the FDA \citep{fda_calcium_carbonate_21cfr1841191,fda_silicon_dioxide_21cfr172480} and are approved food additives in the EU under designations E170 and E551 \citep{codex_gsfa_silicon_dioxide_551,codex_gsfa_calcium_carbonate_170i}, related magnesium carbonates are also approved as E504 \citep{codex_gsfa_magnesium_carbonate_504i,codex_gsfa_magnesium_hydroxide_carbonate_504ii}. Calcium carbonate is widely used as a dietary calcium supplement, an over-the-counter antacid, a food fortifier in plant-based milks and cereals, and a white colorant in confectionery. Amorphous silica is widely used in everyday products as an anti-caking agent in powdered foods and spices (e.g., salt, dried egg, instant coffee), as an anti-foaming agent in food processing (e.g., sugar refining, brewing, juice production), as a carrier for liquid active ingredients such as vitamins and enzymes in human and animal nutrition, and as a mild abrasive in toothpaste. Silica is also naturally present in drinking water and many plant-based foods, where it occurs as biogenic (phytolith) silica. Hydroxyapatite, being chemically identical to the mineral phase of human bone, is routinely used in bone grafts, dental implants \citep{habibah2022hydroxyapatite}, and increasingly in toothpaste formulations as a remineralizing agent \citep{amaechi2019hydroxyapatite_toothpaste}; Related calcium-phosphate materials are also present in the food supply as E341(iii), i.e. tricalcium phosphate \citep{codex_gsfa_tricalcium_phosphate_341iii}. 

For synthetic amorphous silica and calcium-carbonate materials, reported repeated-inhalation no-observed-adverse-effect concentrations  \citep{oecd2004silica,atsdr2019silica,echa_calcium_carbonate} are far above the global-mean exposure levels estimated below for particle injection rates enabling $\sim1\%$ solar flux reflection. For Mg-substituted calcite and hydroxyapatite, the available evidence is better viewed as supporting biosafety precedent rather than as a material-specific inhalation NOAEC; candidate-specific inhalation testing would therefore still be required for any final particle. Amorphous silica has also been used as an airborne particulate tracer in dense urban transit environments, including DHS airflow studies conducted in the New York City subway system \citep{dhs_utr_2024,dhs2021tracer}.

As estimated in Appendix~\ref{app:groundconc}, the global-mean near-surface
concentration of descended SAI particles is given approximately by $\bar{C}_g \simeq 0.03\;\mu\mathrm{g\,m^{-3}} (\dot{M}/(1\;\mathrm{Mt\,yr^{-1}}))$, corresponding to a fiducial value of $\bar{C}_g \simeq 0.3\;\mu\mathrm{g\,m^{-3}}$ for a large illustrative injection rate of $10\;\mathrm{Mt\,yr^{-1}}$. Although regional variability in transport, mixing, and scavenging is not explicitly resolved, this global-mean estimate is roughly two orders of magnitude below present global population-weighted annual-mean ambient concentrations of particles with diameters smaller than $2.5\mu$m (PM$_{2.5}$) and more than three orders of magnitude below the sub-chronic inhalation no-observed-adverse-effect concentration (NOAEC) reported for synthetic amorphous silica \citep{atsdr2019silica,oecd2004silica,ecetoc2014silica}. These indicative
margins are consistent with the exposure-related requirements \reqrefs{A2}--\reqrefs{A4}.

For silica- or calcium carbonate- based particles, the post-deposition environmental fate is well studied, compatible with the elimination pathway and aging related requirements summarized in Appendix~\ref{app:requirements}, particularly \reqrefs{A6} and \reqrefs{A7}, and complements the low estimated near-surface exposure discussed above (\reqrefs{A2}--\reqrefs{A4}). As estimated in Appendix~C, amorphous silica spheres are expected to dissolve on environmentally relevant timescales into dissolved silicic acid, feeding into the pre-existing marine silica cycle rather than persisting indefinitely as an insoluble particulate phase \citep{ecetoc2014silica,demaster1981supply,spitzmuller2023dissolution,diedrich2012dissolution}. The same applies to calcium carbonate: both are the standard biomineral building blocks of marine organisms \citep{morse2007calcium_carbonate,lowenstam1989biomineralization} (diatoms use silica; coccolithophores, foraminifera, and corals use CaCO$_3$), so added particles are incorporated into the existing biological flux rather than accumulating as a novel phase.

\usetikzlibrary{positioning, calc}

\definecolor{darkcard}{HTML}{2D3748}
\definecolor{slidebg}{HTML}{EDF2F7}
\definecolor{labeloverlay}{HTML}{2D3748}
\definecolor{dividercolor}{HTML}{4A5568}

\graphicspath{{images/}}


\definecolor{darkcard}{HTML}{2D3748}
\definecolor{slidebg}{HTML}{EDF2F7}
\definecolor{labeloverlay}{HTML}{2D3748}
\definecolor{dividercolor}{HTML}{4A5568}

\begin{figure}[t]
\centering
\resizebox{\textwidth}{!}{%
\begin{tikzpicture}

\def\cw{3.4}          
\def\ch{2.3}          
\def\gx{0.28}         
\def\capskip{0.30}    
\def\capHeight{0.45}  
\def\rowGap{0.45}     
\def\sectionGap{0.55} 
\def\dividerGap{0.50} 
\def\titleHeight{0.55}
\def\bgPad{0.40}      
\pgfmathsetmacro{\colStep}{\cw + \gx}
\pgfmathsetmacro{\figW}{\colStep*4 + \cw}
\pgfmathsetmacro{\figCx}{\figW/2}
\pgfmathsetmacro{\bgLeft}{-\bgPad}
\pgfmathsetmacro{\bgRight}{\figW + \bgPad}

\newcommand{\organism}[5]{%
    \pgfmathsetmacro{\xL}{#1 * \colStep}%
    \pgfmathsetmacro{\xR}{\xL + \cw}%
    \pgfmathsetmacro{\yT}{#2}%
    \pgfmathsetmacro{\yB}{\yT - \ch}%
    \fill[darkcard, rounded corners=3pt]
        (\xL, \yT) rectangle (\xR, \yB);
    \begin{scope}
        \clip[rounded corners=3pt] (\xL, \yT) rectangle (\xR, \yB);
        \node[anchor=center, inner sep=0pt] at ({(\xL+\xR)/2}, {(\yT+\yB)/2})
            {\includegraphics[width=\cw cm, height=\ch cm]{#3}};
    \end{scope}
    \node[anchor=north west,
          fill=labeloverlay, fill opacity=0.78, text opacity=1,
          text=white, font=\sffamily\fontsize{8}{9.5}\selectfont\bfseries,
          rounded corners=2pt, inner xsep=4pt, inner ysep=2.8pt]
        at ({\xL+0.10}, {\yT-0.10}) {#4};
    \node[anchor=north, font=\sffamily\fontsize{9}{11}\selectfont,
          text=darkcard]
        at ({(\xL+\xR)/2}, {\yB - \capskip}) {#5};
}


\pgfmathsetmacro{\secOneTitleY}{0}
\node[font=\sffamily\bfseries\fontsize{14}{17}\selectfont, text=darkcard]
    at (\figCx, {\secOneTitleY - \titleHeight/2})
    {Amorphous Silica (SiO\textsubscript{2})};

\pgfmathsetmacro{\rowOneY}{\secOneTitleY - \titleHeight - \sectionGap}
\organism{0}{\rowOneY}{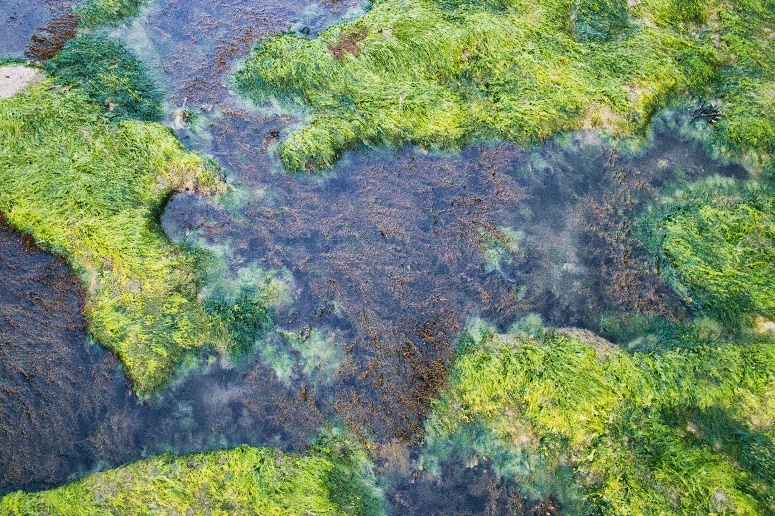}{Amorphous Silica}{Diatom}
\organism{1}{\rowOneY}{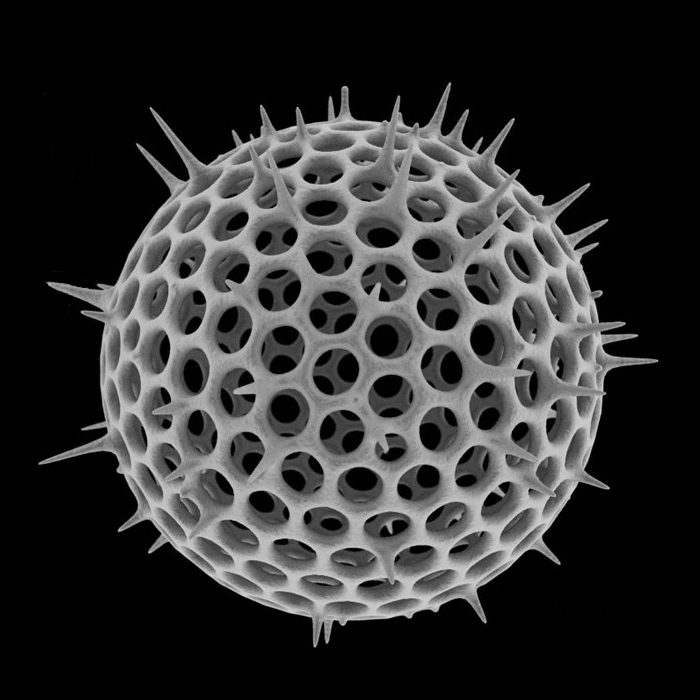}{Amorphous Silica}{Radiolarian}
\organism{2}{\rowOneY}{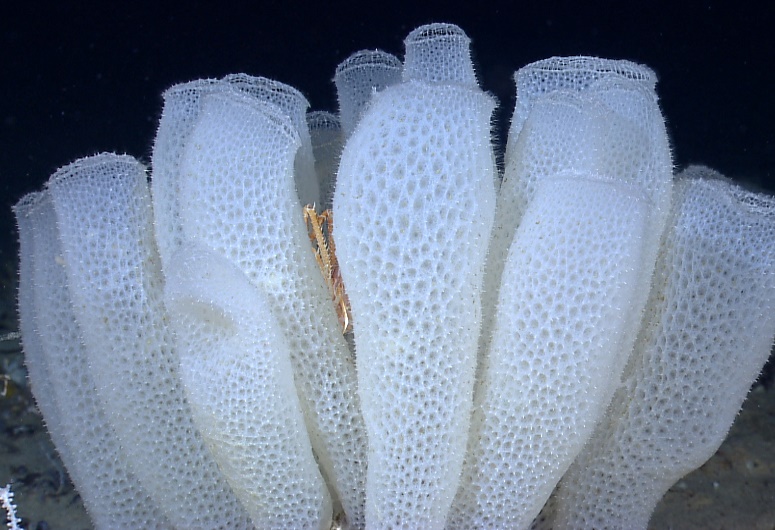}{Amorphous Silica}{Sponge Spicule}
\organism{3}{\rowOneY}{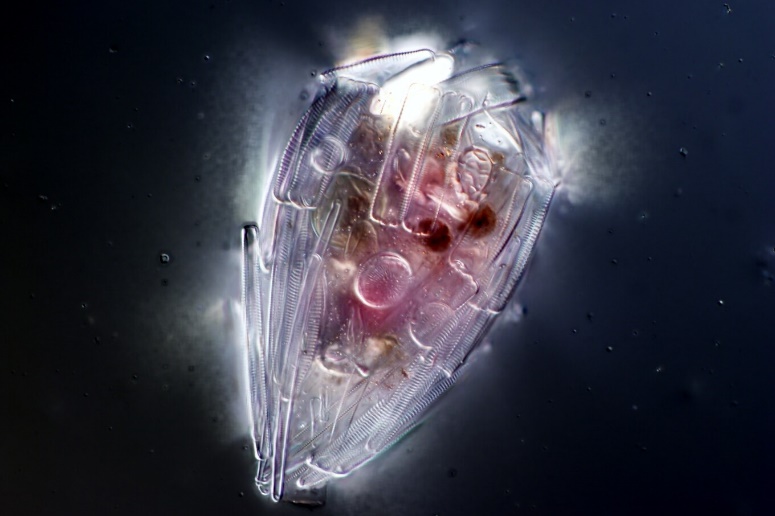}{Amorphous Silica}{Testate Amoeba}
\organism{4}{\rowOneY}{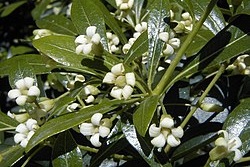}{Amorphous Silica}{Plant Phytolith}

\pgfmathsetmacro{\rowOneCapBot}{\rowOneY - \ch - \capskip - \capHeight}
\pgfmathsetmacro{\divY}{\rowOneCapBot - \dividerGap}
\draw[dividercolor, line width=0.8pt]
    (0.3, \divY) -- ({\figW - 0.3}, \divY);

\pgfmathsetmacro{\secTwoTitleY}{\divY - \sectionGap}
\node[font=\sffamily\bfseries\fontsize{14}{17}\selectfont, text=darkcard]
    at (\figCx, {\secTwoTitleY - \titleHeight/2})
    {Calcium Carbonate (CaCO\textsubscript{3})};

\pgfmathsetmacro{\rowTwoY}{\secTwoTitleY - \titleHeight - \sectionGap}
\organism{0}{\rowTwoY}{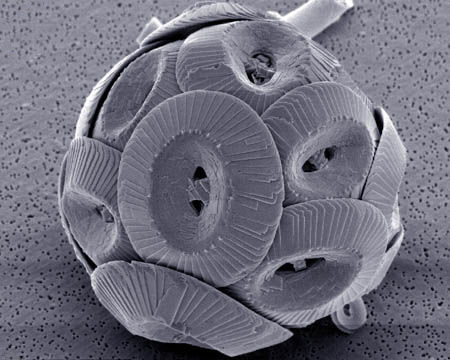}{Calcite}{Coccolithophore}
\organism{1}{\rowTwoY}{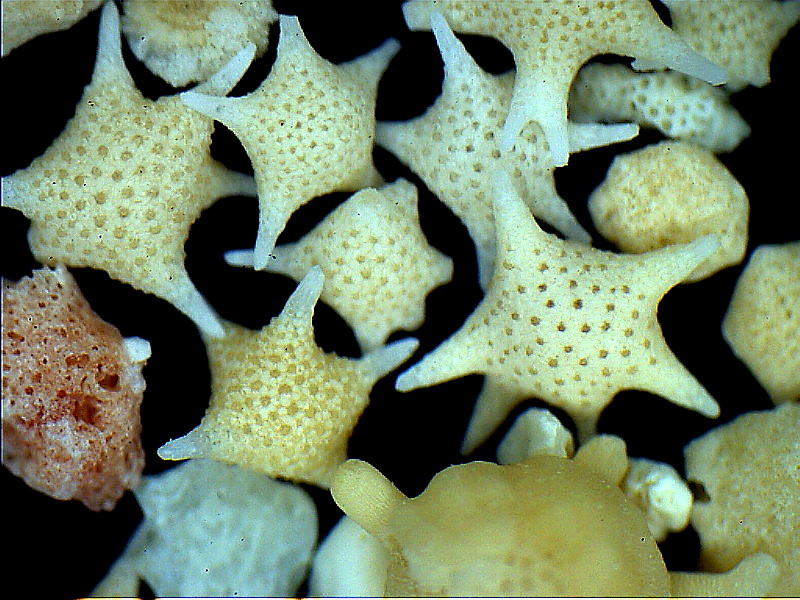}{Calcite}{Foraminifer}
\organism{2}{\rowTwoY}{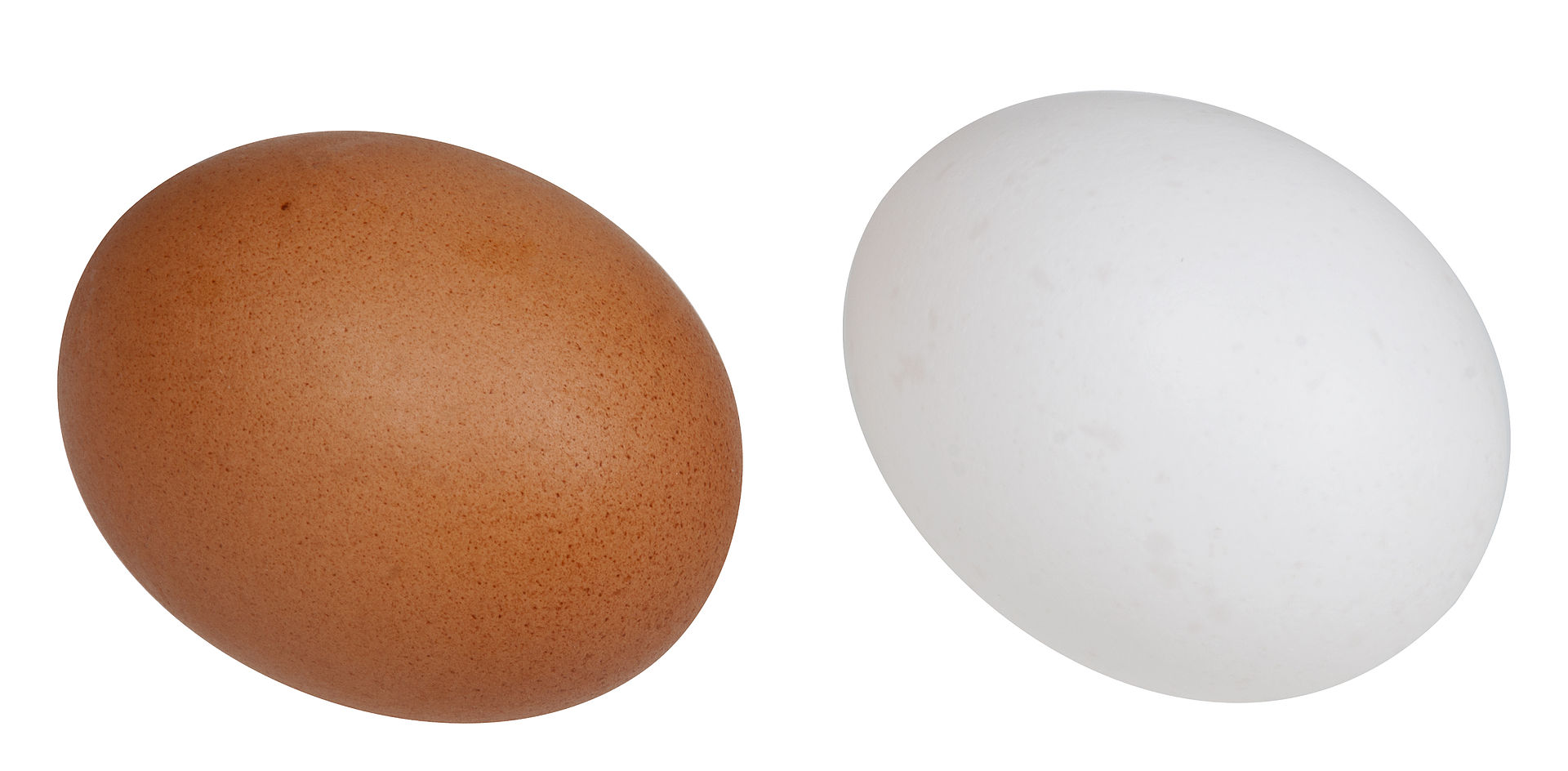}{Calcite}{Eggshell}
\organism{3}{\rowTwoY}{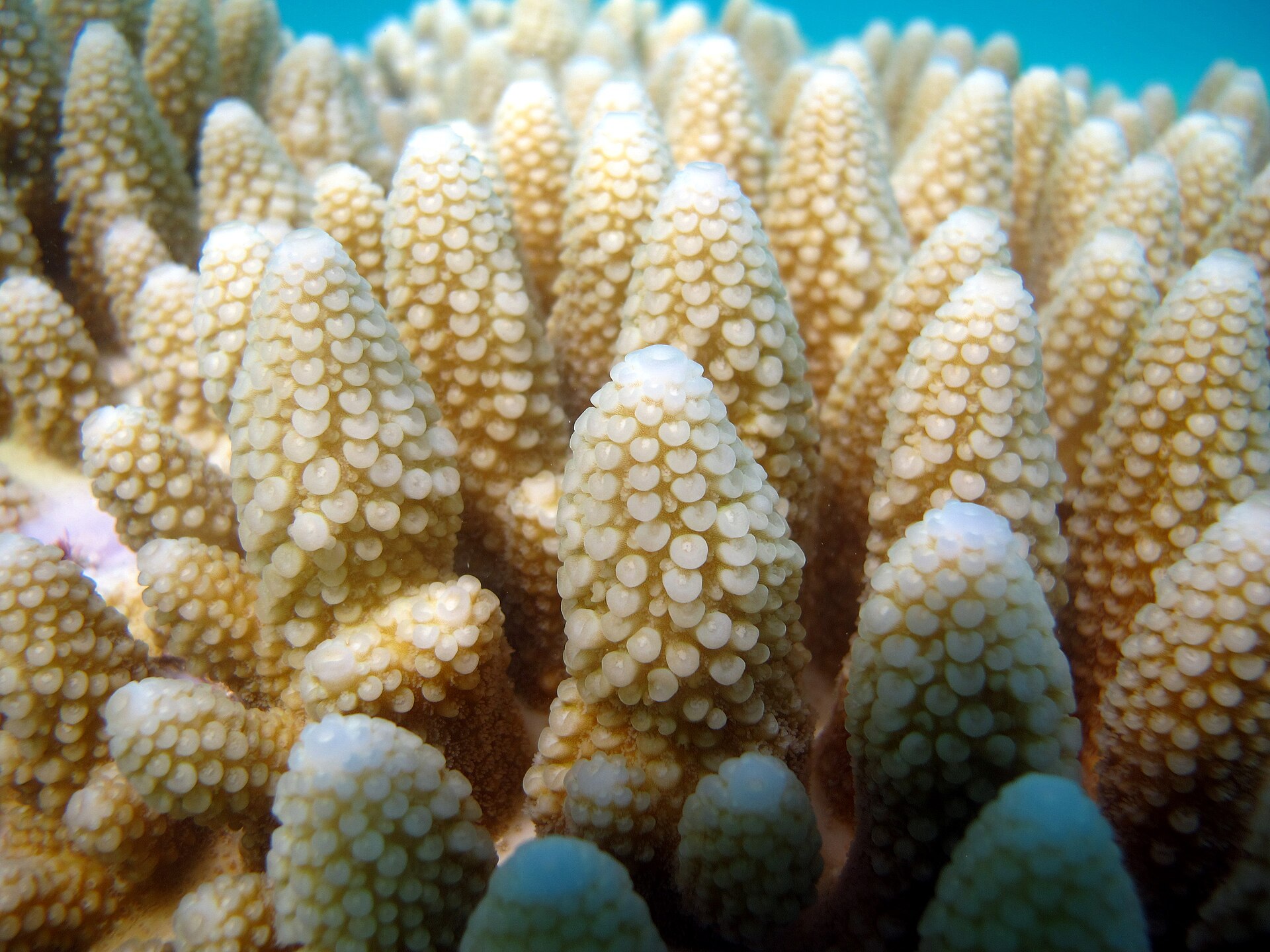}{Aragonite}{Coral}
\organism{4}{\rowTwoY}{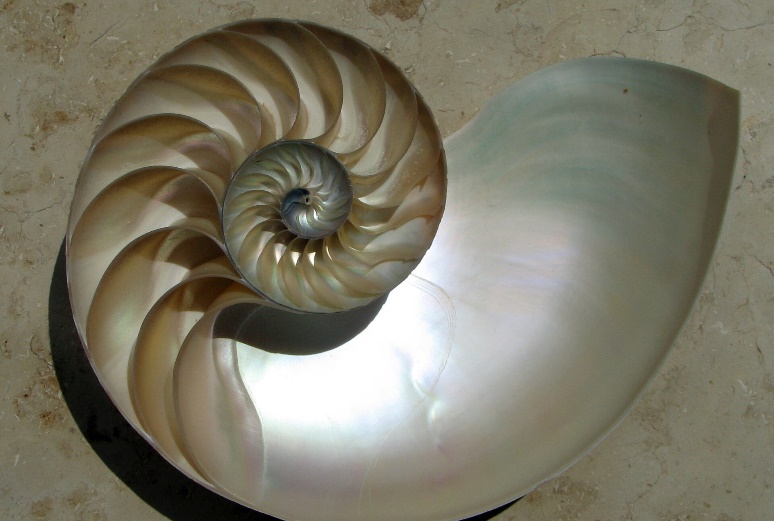}{Aragonite}{Nacreous Shell}

\pgfmathsetmacro{\rowTwoCapBot}{\rowTwoY - \ch - \capskip - \capHeight}
\pgfmathsetmacro{\rowThreeY}{\rowTwoCapBot - \rowGap}
\organism{0}{\rowThreeY}{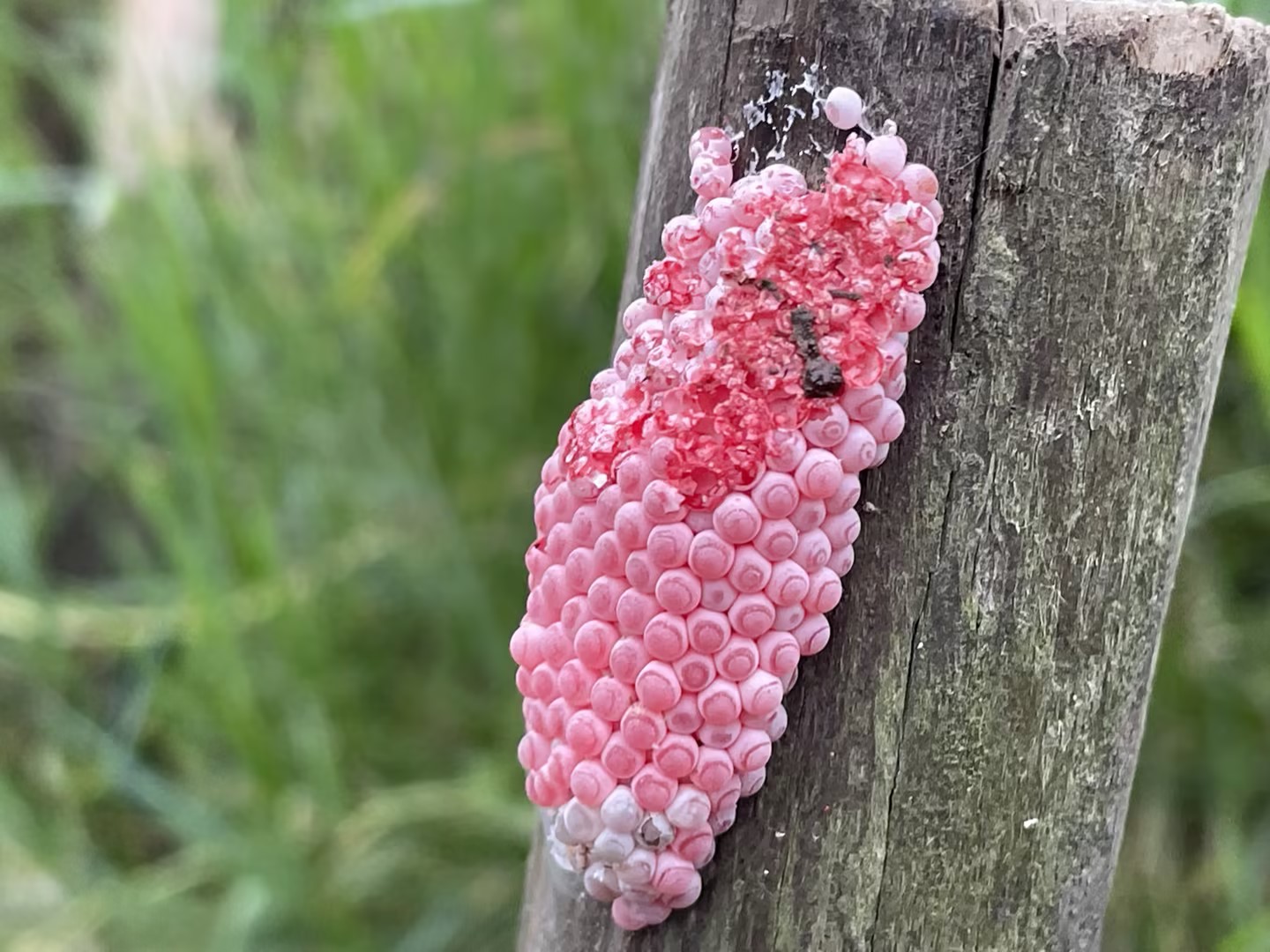}{Vaterite}{Gastropod Eggs}
\organism{1}{\rowThreeY}{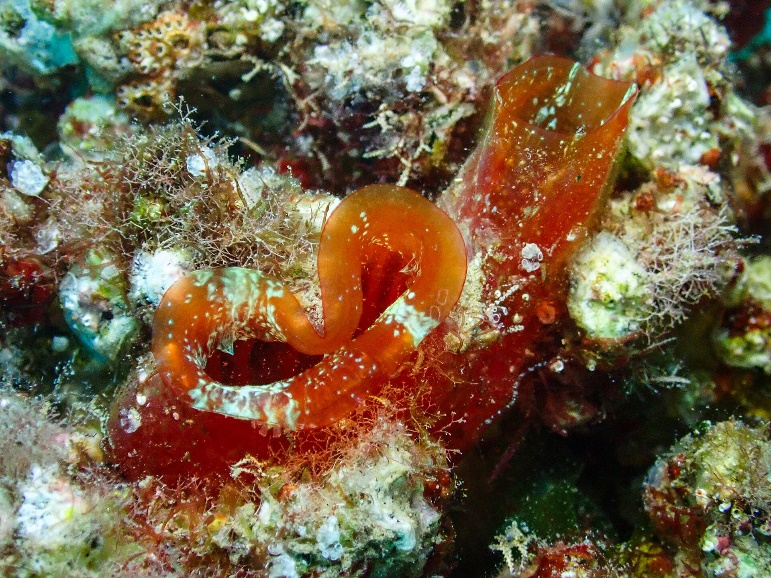}{Vaterite}{Ascidian (Herdmania)}
\organism{2}{\rowThreeY}{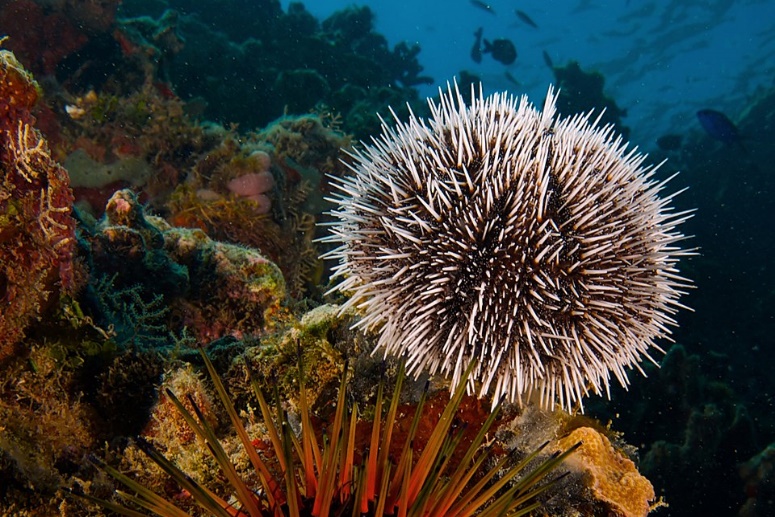}{Mg-calcite}{Sea Urchin}
\organism{3}{\rowThreeY}{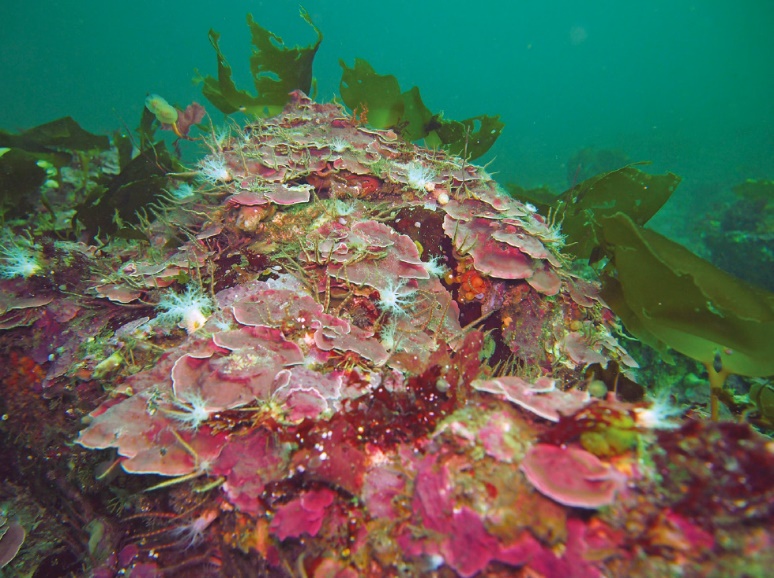}{Mg-calcite}{Coralline Alga}
\organism{4}{\rowThreeY}{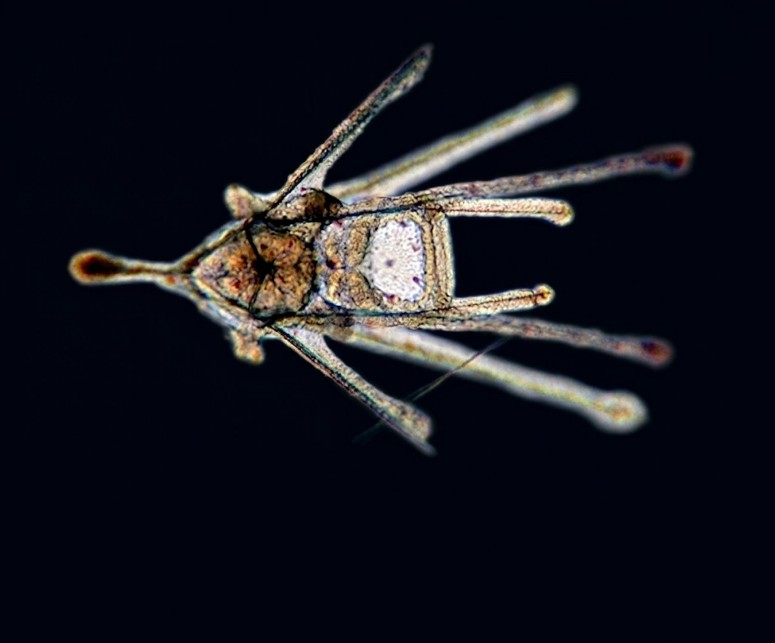}{Amorphous CaCO$_3$}{Sea Urchin Larva}

\pgfmathsetmacro{\bgTop}{\secOneTitleY + \bgPad}
\pgfmathsetmacro{\rowThreeCapBot}{\rowThreeY - \ch - \capskip - \capHeight}
\pgfmathsetmacro{\bgBot}{\rowThreeCapBot - \bgPad}
\begin{scope}[on background layer]
\fill[slidebg, rounded corners=5pt]
    (\bgLeft, \bgTop) rectangle (\bgRight, \bgBot);
\end{scope}

\useasboundingbox (\bgLeft, \bgTop) rectangle (\bgRight, \bgBot);
\node[anchor=south east, font=\sffamily\fontsize{5}{6}\selectfont, text=darkcard!50] 
    at (\bgRight-0.1, \bgBot+0.1) {Image credits: Wikimedia Commons};
\end{tikzpicture}%
}
\caption{Representative examples of biomineralized mineral forms relevant to candidate SAI particle bulk composition. The examples illustrate that amorphous silica and calcium carbonate already occur in natural biogeochemical cycles and in familiar human-use contexts. This supports their consideration as candidate particle materials enabling meeting safety requirements, 
while recognizing that SAI suitability ultimately depends not on bulk composition alone but on additional particle properties, including size, morphology, purity, surface state, and atmospheric behavior.}
\label{fig:stardust_shells}

\end{figure}

\newpage
\subsection{Scalable fabrication methods enabling control of geometry and morphology}
\label{sub-sec:fabrication}

The functionality and safety requirements translate into a challenging set of requirements on the particles' size distribution, geometry, and morphology. 
\begin{enumerate}
    \item The particle size distribution should be narrow and concentrated around the visible light wavelength of $\sim0.5\mu$m. This is set by the requirements for shortwave scattering and stratosphere, and by the safety requirement to minimize the fraction of sub-100~nm particles (\reqrefs{A1c}).
    \item Safety-related constraints (\reqrefs{A1}--\reqrefs{A7}) favor particles with no hard, sharp edges. 
    \item These particles' properties should be maintained across their lifetime. In particular, ultrafine fragment generation during dispersion should be avoided (see \reqrefs{Db}).
\end{enumerate}

The manufacturing process should enable production at the scale of millions of tons per year, consistently and accurately meeting the above requirements (\reqrefs{Mb}), as well as the bulk composition and morphology requirements implied by the optical properties' requirements (\reqrefs{Oa}-\reqrefs{Oc}).

These constraints favor "bottom-up" fabrication methods, where particles are grown from their molecular constituents through controlled nucleation and growth, over "top-down" methods such as grinding, where large particles are broken down into smaller particles. The latter are disfavored because they offer poor control over particle size and shape, making it difficult to achieve the target size distribution (\reqrefs{Da}) and inevitably producing sharp-edged fragments. Bottom-up methods include wet-chemical synthesis methods like sol-gel synthesis (e.g., the St\"ober process for amorphous spherical silica) and controlled precipitation of inorganic carbonates (e.g., spherical calcite or Mg-calcite). Under carefully controlled conditions, they yield smooth, spherical particles with tunable and reproducible size distributions compatible with \reqrefs{Oa}, \reqrefs{Rb}, and \reqrefs{Mb}.

Wet-chemical synthesis is already compatible with industrial scale-up: precipitated silica is produced commercially through aqueous silicate--acid precipitation processes, and the existing market is already at multi-million-ton-per-year scale. Recent market estimates place global precipitated-silica volumes above \(3~\mathrm{Mt\,yr^{-1}}\), with bulk prices of order \(\sim 1~\mathrm{\$\,kg^{-1}}\), depending on grade, region, and application~\cite{ecetoc2006synthetic_amorphous_silica,mordor2026precipitated_silica,procurementresource2025precipitated_silica_price}. To preserve potential manufacturability, synthesis routes should be prioritized when they already operate at meaningful industrial scale, well beyond pilot level, and when they provide a credible path toward the throughput, cost, supply-chain, and quality-control constraints summarized in \reqrefs{M}--\reqrefs{Me}. St\"ober sol-gel silica is one such example \citep{stober1968controlled_growth}, already produced at hundreds of tons per year. Colloidal silica sol goes further still, with precipitated and colloidal silica representing the current industrial leaders in liquid-phase specialty silica manufacturing. Both are grounded in the same bottom-up, controlled-nucleation chemistry, demonstrating that this class of synthesis approach is technically mature and already operating at industrially relevant scales.

\begin{figure}[htbp]
    \centering
    \includegraphics[width=\textwidth]{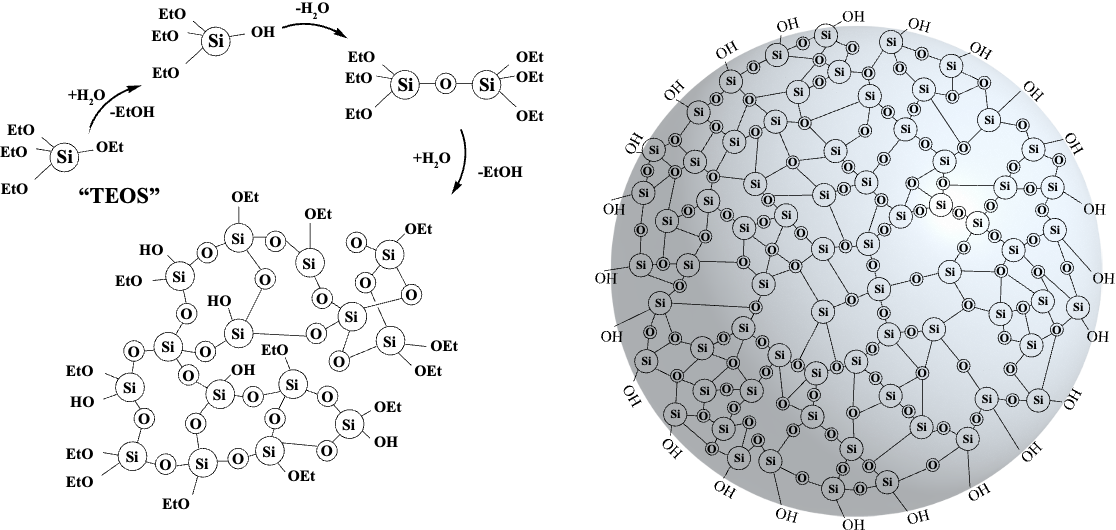}
    \caption{ A schematic illustration of silica nanoparticle formation via the St\"ober process. Tetraethyl orthosilicate (TEOS, Si(OEt)\textsubscript{4}), where Et denotes the ethyl group (--C\textsubscript{2}H\textsubscript{5}), undergoes base-catalyzed hydrolysis and condensation in an ethanol--water--ammonia mixture to produce monodisperse silica nanoparticles (left). The magnified view (right) shows the amorphous Si--O network that constitutes the internal structure of a single particle.}
    \label{fig:stober}
\end{figure}

The bottom-up synthesis route is illustrated, for the St\"ober case, in Fig.~\ref{fig:stober}. In this process, TEOS undergoes hydrolysis and condensation to form monodisperse silica spheres. The magnified inset emphasizes that the resulting particles consist of an amorphous Si--O network rather than a crystalline quartz-like structure. This distinction is important because the same synthesis strategy that provides geometric control also preserves the non-crystalline morphology preferred for safety while maintaining the intended bulk properties relevant to optical performance (\reqrefs{Oa}--\reqrefs{Oc}).

\subsection{Engineering surface properties}
\label{sub-sec:surface}

The surface properties of particles strongly affect their atmospheric chemistry, aging, and the ability to disperse at the required rates while avoiding agglomeration. Proper engineering of the surface properties is therefore essential for meeting both safety and functionality requirements. 
\begin{enumerate}
    \item The surface should be sufficiently inert to enable meeting the limits on modification of trace gases' column densities  (\reqrefs{B1a}, \reqrefs{B1b}).
    \item It should enable avoiding modifications of the ozone column density via catalytic reactions (\reqrefs{B1a}, \reqrefs{B1b}, \reqrefs{B2}) or increasing the stratospheric sulfate surface area.
    \item It should support the ability to meet requirements on impact on Polar Stratospheric Clouds (PSC) and cloud formation in the upper troposphere (\reqrefs{B2b}, \reqrefs{B3}). 
    \item It should enable avoiding aging modifications of the particle properties due to  UV radiation (particularly near the stratospheric UV window at approximately 210~nm) and reactions with ozone or trace gases (\reqrefs{A7}, \reqrefs{B4}), or due to agglomeration and coagulation/wetting (\reqrefs{B2a}, \reqrefs{B2b}).
    \item It should support meeting the dispersion rate requirements by reducing the agglomeration tendency of the sub-micron particles (\reqrefs{D}, \reqrefs{Da}--\reqrefs{Dc}).
\end{enumerate}

The atmospheric chemistry and dispersion requirements both suggest that a non-porous, non-polar, hydrophobic surface is highly advantageous. Such a surface will not only suppress heterogeneous reactions, adhesion, and wetting but also reduce inter-particle cohesion forces, thereby enhancing the ability to overcome agglomeration in the dispersion system. We note that large-scale deviations from sphericity that do not significantly increase surface area may further reduce cohesion forces while not undermining other required particle properties.

\begin{figure*}
    \centering
    \includegraphics[width=0.9\linewidth]{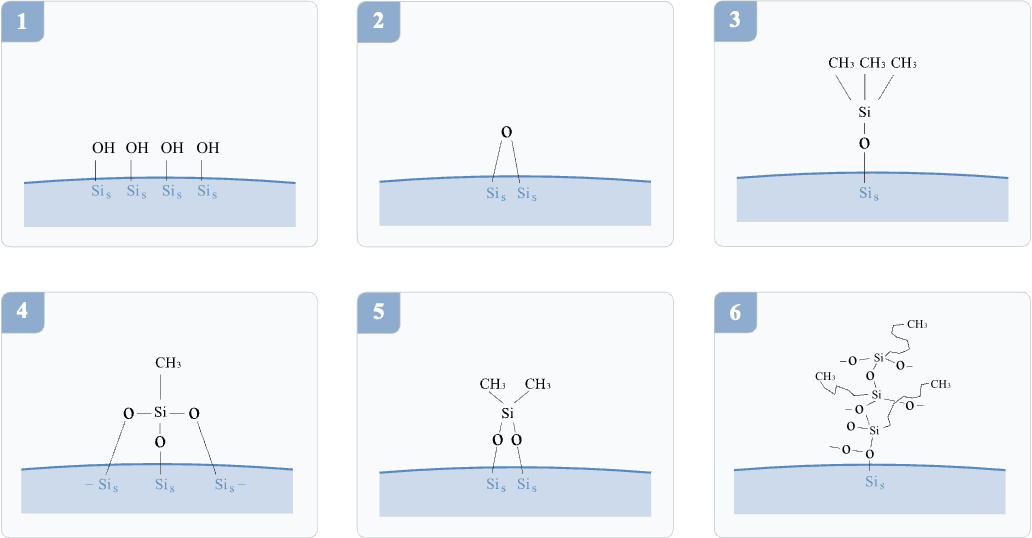}
    \caption{Representative surface motifs on amorphous silica relevant to silanization. Shown schematically are (1) a hydroxylated silanol-rich surface, (2) calcinated surface, (3) a monodentate trimethylsilyl species, (4) a tridentate methylsilyl species, (5) a bidentate dimethylsilyl bridge, and (6) a crosslinked or polymerized methylsiloxane overlayer. The schematic is illustrative of possible attachment geometries and network formation pathways.}
    \label{fig:silane_configurations}
\end{figure*}

Let us consider first meeting the above requirements using amorphous silica particles. While silica is relatively inert in bulk form, surface modification may still be required to meet the requirements. Organosilane coatings on silica (see Fig.~\ref{fig:silane_configurations}) have relevant regulatory precedents in human oral/pharmaceutical and animal-feed applications, including hydrophobic colloidal silica/silica dimethyl silylate as an oral non-medicinal ingredient and pharmaceutical excipient, and silane-treated hydrophobic silica as an FDA-affirmed GRAS anticaking/free-flowing agent in feed preparations~\cite{healthcanada_silica_dimethyl_silylate,evonik_aerosil_r972_pharma,fda1996_hydrophobic_silica_feed,ecfr_21cfr584700_hydrophobic_silicas}, may be useful in this context since they can reduce surface polarity and hydroxyl-mediated reactivity, increase hydrophobicity, and provide steric shielding that suppresses adsorption of polar species and water. Related silanized synthetic amorphous silica materials also have direct inhalation-toxicity precedent, with reported repeated-inhalation no-observed-adverse-effect concentrations (NOAECs) or closely related no-effect/effect-threshold concentrations in the $\mathrm{mg\,m^{-3}}$ range \citep{atsdr2019silica,oecd2004silica,ecetoc2014silica}.

Short-chain silane-derived surface terminations that do not contain C–C bonds are particularly useful, as shorter silane-derived surface terminations better preserve the hydrophobic properties and apparent surface integrity under irradiation, and avoiding C–C bonds enhances UV durability (C–C bonds are significantly degraded under strong ultraviolet radiation). Surface treatments should minimize hydroxyl groups, as these reactive sites can facilitate unwanted interactions with atmospheric species. Sterically hindering groups should be incorporated where useful limit sulfate coating and associated reactivity (e.g., reducing sulfuric-acid wetting of untreated silica~\citep{mcgrory2022sulfuric_silica}), and restrict reactive surface area, thereby reducing chemical evolution during stratospheric residence (and making the resulting radiative forcing more predictable over time). More generally, Fig.~3 highlights that silanization and calcination do not define a single surface state but rather allow a broad design space, with multiple degrees of freedom in attachment geometry, terminal group structure, steric bulk, lateral crosslinking, and network formation, allowing the silica interface to be tuned for different combinations of dissolution rate \citep{spitzmuller2023dissolution}, stability, reactivity, and dispersion compatibility. The right-hand side of Fig.~\ref{fig:core_shell_compare_triple_surface_twocol} presents the three surface structures discussed in this work: hydroxylated, calcined, and silanized/treated silica.

Calcium carbonate and Mg-calcite are, by contrast, chemically reactive \citep{dai2020experimental,Vattioni2024}, motivating a targeted surface treatment, or a core--shell structure in which a reactive \ce{CaCO3} or Mg-calcite core is encapsulated within a thin and less reactive amorphous silica shell. This can regulate surface reactivity while largely preserving the favorable infrared absorption properties of the core, since a thin silica shell contributes negligibly to infrared absorption and allows the bulk optical advantages of the core to be retained (\reqrefs{Ob}, \reqrefs{C4c}).

\subsection{Particle tagging}
\label{sub-sec:tagging}

A "bottom-up" particle production process naturally permits the incorporation of tracers in the form of atomic or stable isotopic impurities (see Fig.~\ref{fig:core_shell_compare_triple_surface_twocol}), which may enable attribution (\reqrefs{C4d}). This is highly advantageous, as discussed at the end of \S~\ref{sub-sec:safety}. Varying the tracers' identities, concentrations, or isotopic ratios will enable attaching unique "fingerprints" to particles produced/dispersed at different times and locations. Identification of tracers in airborne-collected particles may be achievable, for example, using laser-based mass spectrometry instruments operated both on the ground and on aircraft\footnote{Doping with fluorescent materials may provide another means for attribution, provided they satisfy the same stringent safety conditions outlined above.}. Such measurements will support not only attribution but also monitoring of particle evolution during atmospheric aging, providing a means to detect unexpected physicochemical transformations. 

To maintain compatibility with human, biotic, and environmental safety requirements, as well as with atmospheric-chemistry and aging requirements, tracers must satisfy several criteria: They must be limited in both composition and concentration (\reqrefs{A1}, \reqrefs{A1a}, \reqrefs{A1b}), remain stable under synthesis and deployment conditions (\reqrefs{A7}, \reqrefs{B4}), be detectable using established analytical techniques (\reqrefs{C4d}, \reqrefs{Fe}), and have minimal impact on the physicochemical properties of the host particle, including optical behavior, atmospheric evolution, and dispersion compatibility (\reqrefs{Oa}--\reqrefs{Oc}, \reqrefs{Da}--\reqrefs{De}). Together, these constraints restrict the feasible implantation and detection methods. 

Suitable tracer candidates include elemental dopants that, in their oxidized state and at sufficiently low concentration, are considered compatible with human and environmental exposure. Representative atomic tracer examples include Mg, Zr, Ca, and Fe, which are either naturally abundant or toxicologically well characterized. Stable-isotope tracers may include 
$^{18}\mathrm{O}$, $^{29}\mathrm{Si}$, or $^{44}\mathrm{Ca}$. 


\section{Stardust's composite particles}
\label{sec:composite}

Based on the considerations discussed in \S~\ref{sec:design}, we have chosen to design near-spherical sub-micron particles produced by controllable bottom-up synthesis, with bulk material chosen mainly to meet optical, thermal, and manufacturability requirements, and an outer surface tailored to meet atmospheric chemistry, aging, cohesion, and dispersion requirements. Our approach enables flexibility by controlling optical properties through the choice of bulk composition, surface chemistry through the selection of the outer-shell material and surface treatment, and the particles' "fingerprints" through the choice of tracers embedded during synthesis.

Our work focused on two closely related types of sub-micron particles (see Figure~\ref{fig:core_shell_compare_triple_surface_twocol}). The first is a bare amorphous \ce{SiO2} sphere produced by a bottom-up St\"ober sol-gel process. The second is a core--shell architecture in which a \ce{CaCO3} or Mg-calcite core is enclosed within a thin amorphous silica shell. Both types are compatible with controlled bottom-up synthesis, post-synthetic surface modification, and tracer incorporation. 

The amorphous silica GEN1 particles enabled a well-controlled basis for studying how particle size, morphology, hydroxyl density, and terminal surface groups can be modified to enable meeting requirements. In particular, calcination and silane-based surface treatments were used to reduce silanol density and introduce lower-polarity, sterically hindering surface terminations. These modifications are intended to regulate atmospheric interactions, improve dispersion compatibility by reducing interparticle cohesion, and preserve the advantages of near-spherical amorphous silica particles. GEN1 particles are at an advanced stage of experimental verification, described in detail in a companion paper~\cite{sai_fabrication_heterogeneous_chemistry}, demonstrating compatibility with particle properties requirements. 

The validation program is briefly summarized in table~\ref{tab:gen1_validation_map}, which lists the processes addressed in the GEN1 particles’ experimental validation program (including a description of the potential impacts of these processes, the mitigation strategies, and the experimental validation
methods), and in figure~\ref{fig:gen1_validation_scales}, describing the time periods within the particles' life cycle over which the examined processes operate. The current key results of this experimental study are summarized in \S~\ref{sub-sec:GEN1}. Using GEN1 particles is expected to enable reaching a substantial fraction of 1\% solar flux
reflection, limited by the safety requirement limit on stratospheric heating \cite{2026SafetyWP}. GEN1 particles provide a practical platform for both surface engineering and for the development of the other components of the SAI system.

The GEN2 core--shell particles are calcium-carbonate core-silica shell spheres, where the calcium-carbonate core enables reduced infrared absorption, 
while the silica shell enables meeting surface properties requirements.  They are described in some more detail in \S~\ref{sub-sec:GEN2}. 
The reduced infrared absorption GEN2 particles, which are under development, may enable
reaching $>$1\% solar flux reflection while meeting the safety limit on stratospheric heating. The validation program for GEN2 particles follows that of the GEN1 program, and many of the experimental results obtained for GEN1 particles are applicable to GEN2 particles with similar surface structures and treatments.

\definecolor{outline}{HTML}{2B2B2B}
\definecolor{cardbg}{HTML}{E8EDF2}
\definecolor{shellfill}{HTML}{B0C4DE}
\definecolor{corefill}{HTML}{F5DEB3}
\definecolor{marker}{HTML}{C0392B}
\definecolor{isotope}{HTML}{2980B9}
\definecolor{treated}{HTML}{27AE60}
\definecolor{accent}{HTML}{5B7FA5}
\definecolor{muted}{HTML}{7F8C8D}
\definecolor{hydrox}{HTML}{3498DB}
\definecolor{calcined}{HTML}{E67E22}
 
\begin{figure*}[t]
\resizebox{0.98\textwidth}{!}{%
\begin{tikzpicture}[x=1cm,y=1cm, line cap=round, line join=round, >=Latex, font=\footnotesize]
 
    \def\Rshell{1.22}
    \def\rcore{1.05}
    \def\dotR{0.055}
 
    \coordinate (Msw) at (0.00,1.30);
    \coordinate (Mne) at (4.90,7.20);
 
    \fill[white] (Msw) rectangle (Mne);
    \draw[outline, rounded corners=1.4mm, line width=0.72pt] (Msw) rectangle (Mne);
 
    \coordinate (C1) at (7.80,6.30);
    \coordinate (C2) at (7.80,2.30);
 
    \draw[accent!70, dashed, line width=0.55pt]
        ($(C1)+(180:\Rshell)$) -- (4.90,5.85);
    \draw[accent!70, dashed, line width=0.55pt]
        ($(C2)+(180:\Rshell)$) -- (4.90,2.35);
 
    \fill[cardbg] (0.05,6.58) rectangle (4.85,7.15);
    \node[anchor=west, text=outline, font=\bfseries\scriptsize] at (0.22,6.86)
        {(c) optional traceability markers};
 
    \node[anchor=west, text=outline, font=\bfseries\scriptsize] at (0.25,6.20)
        {elemental dopant};
 
    \fill[shellfill!92] (1.40,5.35) circle (0.52);
    \draw[outline, line width=0.55pt] (1.40,5.35) circle (0.52);
    \fill[marker] (1.18,5.52) circle (\dotR);
    \fill[marker] (1.55,5.56) circle (\dotR);
    \fill[marker] (1.38,5.18) circle (\dotR);
    \fill[marker] (1.62,5.38) circle (\dotR);
    \fill[marker] (1.24,5.34) circle (\dotR);
    \fill[marker] (1.48,5.45) circle (\dotR);
    \fill[marker] (1.32,5.50) circle (\dotR);
    \fill[marker] (1.56,5.22) circle (\dotR);
 
    \node[anchor=west, text=marker!80!black, font=\scriptsize] at (2.10,5.35)
        {e.g.\ $\mathrm{Zr,\ Mg,\ Ca,\ Fe}$};
 
    \node[anchor=west, text=outline, font=\bfseries\scriptsize] at (0.25,4.52)
        {stable-isotope label};
 
    \fill[shellfill!92] (1.40,3.68) circle (0.52);
    \draw[outline, line width=0.55pt] (1.40,3.68) circle (0.52);
    \draw[isotope, line width=0.55pt] (1.18,3.85) circle (0.065);
    \draw[isotope, line width=0.55pt] (1.55,3.89) circle (0.065);
    \draw[isotope, line width=0.55pt] (1.38,3.51) circle (0.065);
    \draw[isotope, line width=0.55pt] (1.62,3.71) circle (0.065);
    \draw[isotope, line width=0.55pt] (1.24,3.67) circle (0.065);
    \draw[isotope, line width=0.55pt] (1.48,3.78) circle (0.065);
    \draw[isotope, line width=0.55pt] (1.32,3.83) circle (0.065);
    \draw[isotope, line width=0.55pt] (1.56,3.55) circle (0.065);
 
    \node[anchor=west, text=isotope!85!black, font=\scriptsize] at (2.10,3.88)
        {e.g.\ ${}^{18}\mathrm{O},\ {}^{29}\mathrm{Si},$};
    \node[anchor=west, text=isotope!85!black, font=\scriptsize] at (2.10,3.55)
        {${}^{44}\mathrm{Ca}$};
 
    \node[anchor=west, text=outline, font=\bfseries\scriptsize] at (0.25,2.85)
        {fluorescent / optical tag};
 
    \fill[shellfill!92] (1.40,2.00) circle (0.52);
    \draw[outline, line width=0.55pt] (1.40,2.00) circle (0.52);
    \foreach \x/\y in {1.18/2.17, 1.56/2.20, 1.36/1.82, 1.62/2.02, 1.25/1.98} {
        \node[star, star points=5, star point ratio=2.1,
              minimum size=2.8mm, inner sep=0pt,
              fill=treated!88, draw=treated!55!black, line width=0.22pt] at (\x,\y) {};
    }
 
    \node[anchor=west, text=treated!85!black, font=\scriptsize] at (2.10,2.00)
        {e.g. dye};
 
    \coordinate (Bsw) at (10.80,0.45);
    \coordinate (Bne) at (19.60,8.55);
    \fill[white] (Bsw) rectangle (Bne);
    \draw[outline, rounded corners=1.8mm, line width=0.82pt] (Bsw) rectangle (Bne);
    \fill[cardbg] (10.85,7.78) rectangle (19.55,8.50);
    \fill[hydrox!5]   (10.95,5.30) rectangle (19.40,7.55);
    \fill[calcined!6]  (10.95,2.88) rectangle (19.40,5.10);
    \fill[treated!6]   (10.95,0.55) rectangle (19.40,2.68);
 
 
    \node[anchor=east, text=outline, font=\bfseries\footnotesize] at (6.36,6.30) {(a)};
 
    \fill[black!8, opacity=0.22] ($(C1)+(0.07,-0.07)$) circle (\Rshell);
    \fill[shellfill] (C1) circle (\Rshell);
    \fill[corefill]  (C1) circle (\rcore);
    \draw[outline, line width=0.82pt] (C1) circle (\Rshell);
    \draw[outline, line width=0.68pt] (C1) circle (\rcore);
    \draw[white, line width=0.90pt, opacity=0.35]
        ($(C1)+(145:1.02)$) arc[start angle=145,end angle=218,radius=1.02];
 
    \draw[accent!95, line width=1.10pt]
        ($(C1)+(18:\Rshell)$) arc[start angle=18,end angle=52,radius=\Rshell];
    \draw[accent!35, line width=3.8pt, opacity=0.12]
        ($(C1)+(18:\Rshell)$) arc[start angle=18,end angle=52,radius=\Rshell];
 
    \node[anchor=south, text=outline, font=\bfseries\footnotesize] at ($(C1)+(0,1.55)$)
        {$\mathrm{carbonate@SiO_2}$ core--shell};
 
    \draw[outline, line width=0.58pt, -{Latex[length=1.8mm]}]
        (8.90,7.50) -- ($(C1)+(50:1.12)$);
    \node[anchor=south west, text=outline, font=\bfseries\scriptsize] at (8.92,7.53)
        {$\mathrm{SiO_2}$ shell};
 
    \draw[outline, line width=0.58pt, -{Latex[length=1.8mm]}]
        (7.90,5.10) -- ($(C1)+(310:0.56)$);
    \node[anchor=north west, text=outline, font=\bfseries\scriptsize] at (7.92,5.07)
        {carbonate core};
    \node[anchor=north west, text=muted, font=\scriptsize] at (7.92,4.77)
        {\ce{CaCO3} or Mg-calcite};
 
    \draw[outline!75, line width=0.48pt, <->]
        ($(C1)+(128:\rcore)$) -- ($(C1)+(128:\Rshell)$)
        node[midway, fill=white, inner sep=0.8pt, text=muted, font=\scriptsize] {$t_{\mathrm{shell}}$};
 
    \node[anchor=east, text=outline, font=\bfseries\footnotesize] at (6.36,2.30) {(b)};
 
    \fill[black!8, opacity=0.22] ($(C2)+(0.07,-0.07)$) circle (\Rshell);
    \fill[shellfill] (C2) circle (\Rshell);
    \draw[outline, line width=0.82pt] (C2) circle (\Rshell);
    \draw[white, line width=0.90pt, opacity=0.35]
        ($(C2)+(145:1.02)$) arc[start angle=145,end angle=218,radius=1.02];
 
    \draw[accent!95, line width=1.10pt]
        ($(C2)+(18:\Rshell)$) arc[start angle=18,end angle=52,radius=\Rshell];
    \draw[accent!35, line width=3.8pt, opacity=0.12]
        ($(C2)+(18:\Rshell)$) arc[start angle=18,end angle=52,radius=\Rshell];
 
    \node[anchor=north, text=outline, font=\bfseries\footnotesize] at ($(C2)+(0,-1.50)$)
        {bare amorphous $\mathrm{SiO_2}$};
 
    \draw[outline, line width=0.58pt, -{Latex[length=1.8mm]}]
        (8.90,1.35) -- ($(C2)+(320:0.56)$);
    \node[anchor=north west, text=outline, font=\bfseries\scriptsize] at (8.92,1.32)
        {$\mathrm{SiO_2}$ matrix};
 
 
    \draw[accent!75, dashed, line width=0.55pt]
        ($(C1)+(52:\Rshell)$) -- (10.80,7.05);
    \draw[accent!75, dashed, line width=0.55pt]
        ($(C1)+(18:\Rshell)$) -- (10.80,6.15);
    \draw[accent!75, dashed, line width=0.55pt]
        ($(C2)+(52:\Rshell)$) -- (10.80,2.92);
    \draw[accent!75, dashed, line width=0.55pt]
        ($(C2)+(18:\Rshell)$) -- (10.80,2.05);
 
    \draw[outline, rounded corners=1.8mm, line width=0.82pt] (Bsw) rectangle (Bne);
 
    \node[anchor=west, text=outline, font=\bfseries\footnotesize] at (11.05,8.14)
        {Representative silica surface states};
    \node[anchor=east, text=muted, font=\scriptsize] at (19.35,8.13)
        {schematic};
 
    \draw[outline!18, rounded corners=1.0mm, line width=0.35pt] (10.95,5.30) rectangle (19.40,7.55);
    \draw[outline!18, rounded corners=1.0mm, line width=0.35pt] (10.95,2.88) rectangle (19.40,5.10);
    \draw[outline!18, rounded corners=1.0mm, line width=0.35pt] (10.95,0.55) rectangle (19.40,2.68);
 
    \fill[hydrox] (11.19,7.18) circle (0.09);
    \node[anchor=west, text=outline, font=\bfseries\footnotesize] at (11.40,7.18)
        {(i) hydroxylated surface};
    \node[anchor=west, text=muted, font=\scriptsize] at (11.40,6.88)
        {high silanol density};
 
    \fill[shellfill!97]
        (13.90,5.45) -- (18.80,5.45) -- (18.80,6.00)
        .. controls (17.50,6.18) and (15.60,6.20) .. (13.90,6.05) -- cycle;
    \draw[hydrox!92, line width=0.82pt]
        (13.90,6.05) .. controls (15.60,6.20) and (17.50,6.18) .. (18.80,6.00);
 
    \foreach \x/\y in {14.60/5.68, 15.70/5.62, 16.80/5.70, 17.90/5.64} {
        \fill[accent!60] (\x,\y) circle (0.055);
    }
    \draw[accent!48, line width=0.35pt] (14.60,5.68)--(15.70,5.62)--(16.80,5.70)--(17.90,5.64);
    \draw[accent!38, line width=0.32pt] (14.60,5.68)--(14.65,5.96);
    \draw[accent!38, line width=0.32pt] (15.70,5.62)--(15.72,6.02);
    \draw[accent!38, line width=0.32pt] (16.80,5.70)--(16.75,6.00);
    \draw[accent!38, line width=0.32pt] (17.90,5.64)--(17.85,5.92);
 
    \draw[outline, line width=0.48pt] (14.65,6.08)--++(100:0.42) coordinate (OH1);
    \node[font=\scriptsize, text=outline, inner sep=0.2pt, anchor=south] at (OH1) {$\mathrm{OH}$};
    \draw[outline, line width=0.48pt] (15.72,6.18)--++(88:0.42) coordinate (OH2);
    \node[font=\scriptsize, text=outline, inner sep=0.2pt, anchor=south] at (OH2) {$\mathrm{OH}$};
    \draw[outline, line width=0.48pt] (16.75,6.12)--++(82:0.42) coordinate (OH3);
    \node[font=\scriptsize, text=outline, inner sep=0.2pt, anchor=south] at (OH3) {$\mathrm{OH}$};
    \draw[outline, line width=0.48pt] (17.85,6.02)--++(94:0.42) coordinate (OH4);
    \node[font=\scriptsize, text=outline, inner sep=0.2pt, anchor=south] at (OH4) {$\mathrm{OH}$};
 
    \node[anchor=west, text=hydrox!85!black, font=\scriptsize] at (13.92,5.35)
        {$\equiv\!\mathrm{Si{-}OH}$ dominant termination};
 
    \fill[calcined] (11.19,4.73) circle (0.09);
    \node[anchor=west, text=outline, font=\bfseries\footnotesize] at (11.40,4.73)
        {(ii) calcined surface};
    \node[anchor=west, text=muted, font=\scriptsize] at (11.40,4.43)
        {reduced silanol density};
 
    \fill[shellfill!97]
        (13.90,3.08) -- (18.80,3.08) -- (18.80,3.64)
        .. controls (17.50,3.82) and (15.60,3.84) .. (13.90,3.68) -- cycle;
    \draw[calcined!90, line width=0.82pt]
        (13.90,3.68) .. controls (15.60,3.84) and (17.50,3.82) .. (18.80,3.64);
 
    \foreach \x/\y in {14.60/3.32, 15.70/3.26, 16.80/3.34, 17.90/3.28} {
        \fill[accent!60] (\x,\y) circle (0.055);
    }
    \draw[accent!48, line width=0.35pt] (14.60,3.32)--(15.70,3.26)--(16.80,3.34)--(17.90,3.28);
    \draw[accent!38, line width=0.32pt] (14.60,3.32)--(14.65,3.60);
    \draw[accent!38, line width=0.32pt] (15.70,3.26)--(15.72,3.66);
    \draw[accent!38, line width=0.32pt] (16.80,3.34)--(16.75,3.64);
    \draw[accent!38, line width=0.32pt] (17.90,3.28)--(17.85,3.56);
 
    \draw[outline, line width=0.46pt] (14.65,3.72)--++(100:0.38) coordinate (COH1);
    \node[font=\scriptsize, text=outline, inner sep=0.2pt, anchor=south] at (COH1) {$\mathrm{OH}$};
    \draw[outline, line width=0.46pt] (17.85,3.66)--++(88:0.36) coordinate (COH2);
    \node[font=\scriptsize, text=outline, inner sep=0.2pt, anchor=south] at (COH2) {$\mathrm{OH}$};
 
    \draw[outline, line width=0.50pt] (15.72,3.82)
        .. controls +(90:0.38) and +(90:0.38) .. (16.75,3.76);
    \node[font=\scriptsize, text=calcined!90!black, inner sep=0.2pt, anchor=south] at (16.24,4.18) {$\mathrm{O}$};
 
    \node[anchor=west, text=calcined!85!black, font=\scriptsize] at (13.92,2.96)
        {$\equiv\!\mathrm{Si{-}O{-}Si}\equiv$ condensed motifs};
 
    \fill[treated] (11.19,2.31) circle (0.09);
    \node[anchor=west, text=outline, font=\bfseries\footnotesize] at (11.40,2.31)
        {(iii) silanized / treated surface};
    \node[anchor=west, text=muted, font=\scriptsize] at (11.40,2.01)
        {terminal $\equiv\!\mathrm{Si{-}O{-}R}$ groups};
 
    \fill[shellfill!97]
        (13.90,0.72) -- (18.80,0.72) -- (18.80,1.28)
        .. controls (17.50,1.46) and (15.60,1.48) .. (13.90,1.32) -- cycle;
    \draw[treated!90, line width=0.82pt]
        (13.90,1.32) .. controls (15.60,1.48) and (17.50,1.46) .. (18.80,1.28);
 
    \foreach \x/\y in {14.60/0.96, 15.70/0.90, 16.80/0.98, 17.90/0.92} {
        \fill[accent!60] (\x,\y) circle (0.055);
    }
    \draw[accent!48, line width=0.35pt] (14.60,0.96)--(15.70,0.90)--(16.80,0.98)--(17.90,0.92);
    \draw[accent!38, line width=0.32pt] (14.60,0.96)--(14.65,1.24);
    \draw[accent!38, line width=0.32pt] (15.70,0.90)--(15.72,1.30);
    \draw[accent!38, line width=0.32pt] (16.80,0.98)--(16.75,1.28);
    \draw[accent!38, line width=0.32pt] (17.90,0.92)--(17.85,1.20);
 
    \draw[outline, line width=0.46pt] (14.65,1.36)--++(92:0.26) coordinate (EO1);
    \draw[outline, line width=0.46pt] (EO1)--++(25:0.24) coordinate (ER1);
    \node[font=\scriptsize, text=treated!92!black, inner sep=0.2pt, anchor=west] at (ER1) {$\mathrm{R_1}$};
 
    \draw[outline, line width=0.46pt] (15.72,1.46)--++(88:0.26) coordinate (EO2);
    \draw[outline, line width=0.46pt] (EO2)--++(0:0.24) coordinate (ER2);
    \node[font=\scriptsize, text=treated!92!black, inner sep=0.2pt, anchor=west] at (ER2) {$\mathrm{R_2}$};
 
    \draw[outline, line width=0.46pt] (16.75,1.40)--++(84:0.26) coordinate (EO3);
    \draw[outline, line width=0.46pt] (EO3)--++(-15:0.24) coordinate (ER3);
    \node[font=\scriptsize, text=treated!92!black, inner sep=0.2pt, anchor=west] at (ER3) {$\mathrm{R_3}$};
 
    \draw[outline, line width=0.46pt] (17.85,1.30)--++(94:0.26) coordinate (EO4);
    \draw[outline, line width=0.46pt] (EO4)--++(10:0.22) coordinate (ER4);
    \node[font=\scriptsize, text=treated!92!black, inner sep=0.2pt, anchor=west] at (ER4) {$\mathrm{R_n}$};
 
    \node[anchor=west, text=treated!88!black, font=\scriptsize] at (13.92,0.60)
        {$\mathrm{R}$: generic terminal group};
 
\end{tikzpicture}%
}
\caption{A schematic comparison of (a) a carbonate@\ce{SiO2} core--shell particle, where the core may be \ce{CaCO3} or Mg-calcite, and (b) a bare amorphous \ce{SiO2} particle. Panel~(c) illustrates optional traceability markers incorporated during synthesis: elemental dopants (filled symbols), stable-isotope labels (open symbols), and fluorescent or optically distinct tags (star symbols) such as biosafe UV-resistant fluorescent dyes. Markers may be embedded in the core, shell, or silica matrix of either particle type. The right panel summarizes three representative silica surface structure considered in this work: (i) hydroxylated ($\equiv\!\mathrm{Si{-}OH}$), (ii) calcined with condensed siloxane linkages ($\equiv\!\mathrm{Si{-}O{-}Si}\equiv$), and (iii) silanized with terminal $\equiv\!\mathrm{Si{-}O{-}R}$ groups. Representative terminal moieties may include methyl termination, while other silane end groups may also be used. Cross-sections are schematic and not to scale; throughout, ``spherical'' should be understood to include near-spherical particles.}
\label{fig:core_shell_compare_triple_surface_twocol}
\end{figure*}
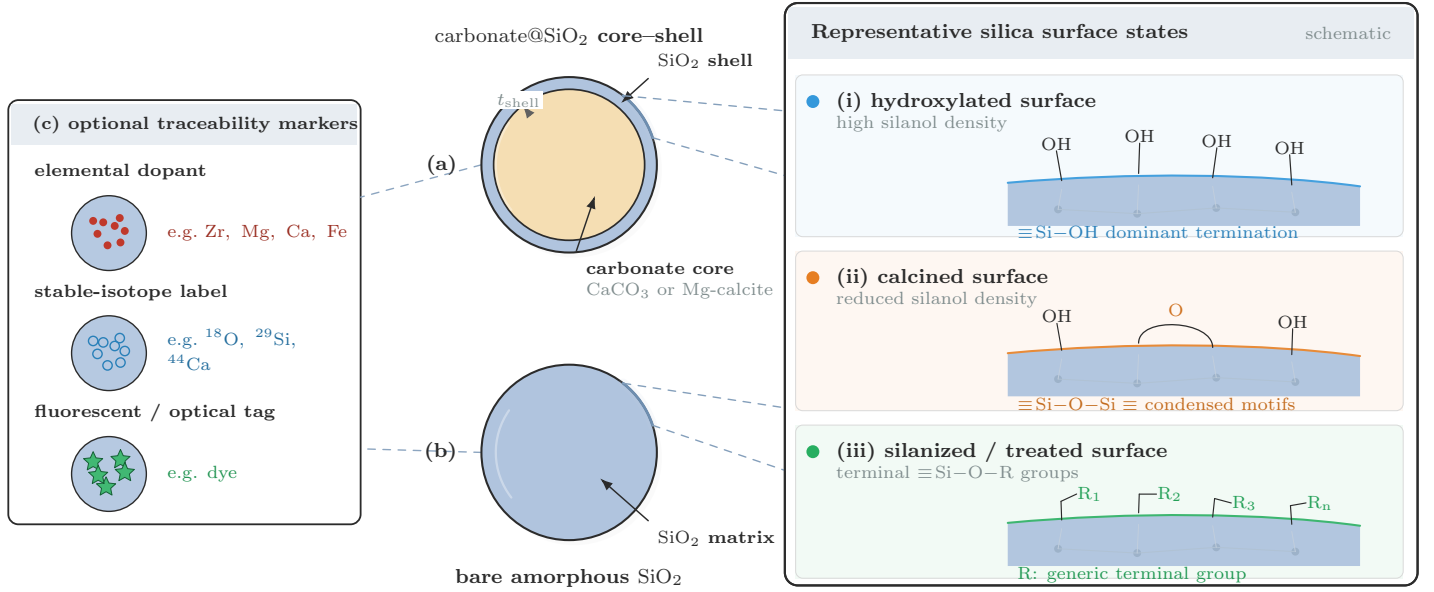
\newpage

\subsection{GEN1: Amorphous silica spheres}
\label{sub-sec:GEN1}

In \citep{sai_fabrication_heterogeneous_chemistry}, we provide a detailed description of the synthesis process and of the characterization of the physicochemical properties of GEN1 particles. It includes a description of the surface treatments, tracer incorporation, and their single-particle detectability; it demonstrates the compatibility of particle properties with the safety and functionality requirements identified in this paper regarding particle size distribution and morphology, surface structure and environmental stability, optical properties, and dispersion compatibility; it presents initial results regarding compatibility with atmospheric chemistry safety requirements. 
A brief summary of the main results reported in \citep{sai_fabrication_heterogeneous_chemistry} is given below.

\begin{enumerate}
    \item {\it Size distribution and surface morphology}. A successful synthesis is demonstrated of 
    amorphous silica spheres with tight control over size distribution and morphology. As shown in Fig.~\ref{fig:powder_dispersion_workflow}, the synthesized particles exhibit the near-spherical shape and narrow size distribution targeted by the requirements discussed above. SEM and BET measurements further indicate high surface smoothness, consistent with near spherical, low roughness morphology.  Detailed results and validation are provided in \citep{sai_fabrication_heterogeneous_chemistry}.

\definecolor{darkcard}{HTML}{2D3748}
\definecolor{slidebg}{HTML}{EDF2F7}
\definecolor{labeloverlay}{HTML}{2D3748}
\definecolor{dividercolor}{HTML}{4A5568}

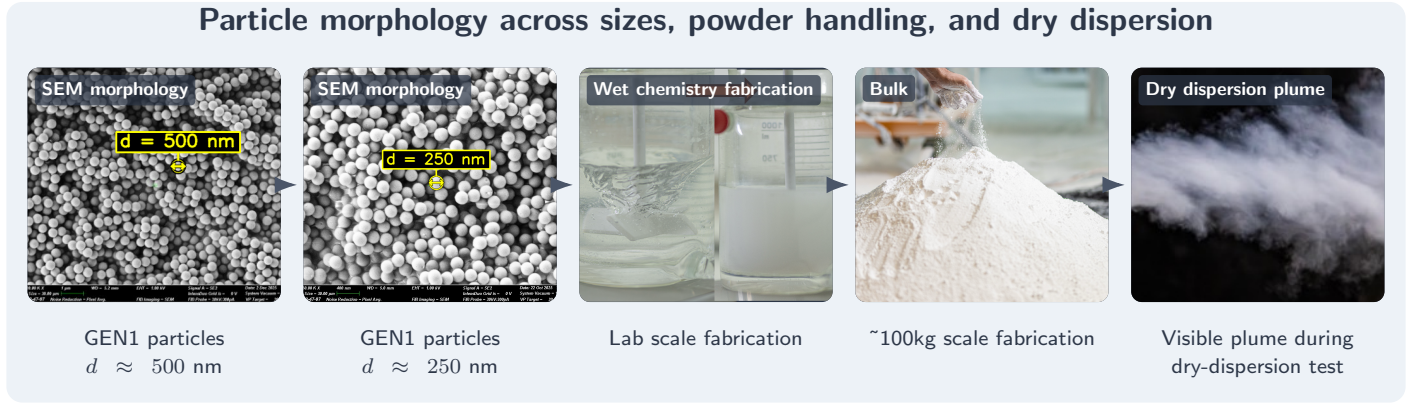
\begin{figure*}[t]
\centering
\begin{tikzpicture}
\def\cw{3.35}      
\def\ch{3.10}      
\def\gx{0.30}      
\def\capskip{0.26} 
\pgfmathsetmacro{\colStep}{\cw + \gx}
\pgfmathsetmacro{\totalW}{4*\colStep + \cw}   
\pgfmathsetmacro{\bgLeft}{-0.28}
\pgfmathsetmacro{\bgRight}{\totalW + 0.28}
\pgfmathsetmacro{\bgTop}{1.05}
\pgfmathsetmacro{\bgBot}{-\ch - 1.15}

\newcommand{\stagecard}[4]{%
    \pgfmathsetmacro{\xL}{#1 * \colStep}%
    \pgfmathsetmacro{\xR}{\xL + \cw}%
    \pgfmathsetmacro{\yT}{0.18}%
    \pgfmathsetmacro{\yB}{\yT - \ch}%
    \fill[darkcard, rounded corners=4pt]
        (\xL,\yT) rectangle (\xR,\yB);
    \begin{scope}
        \clip[rounded corners=4pt] (\xL,\yT) rectangle (\xR,\yB);
        \node[anchor=center, inner sep=0pt]
            at ({(\xL+\xR)/2}, {(\yT+\yB)/2})
            {\includegraphics[height=\ch cm]{#2}};
    \end{scope}
    \node[
        anchor=north west,
        fill=labeloverlay,
        fill opacity=0.72,
        text opacity=1,
        text=white,
        font=\sffamily\fontsize{7}{8.5}\selectfont\bfseries,
        rounded corners=1.6pt,
        inner sep=2.6pt
    ] at ({\xL+0.09}, {\yT-0.09}) {#3};
    \node[
        anchor=north,
        align=center,
        text width=\cw cm,
        text=darkcard,
        font=\sffamily\fontsize{8.5}{10.5}\selectfont
    ] at ({(\xL+\xR)/2}, {\yB-\capskip}) {#4};
}
\node[
    font=\sffamily\bfseries\fontsize{13}{16}\selectfont,
    text=darkcard
] at ({\totalW/2}, 0.78)
{Particle morphology across sizes, powder handling, and dry dispersion};
\stagecard{0}{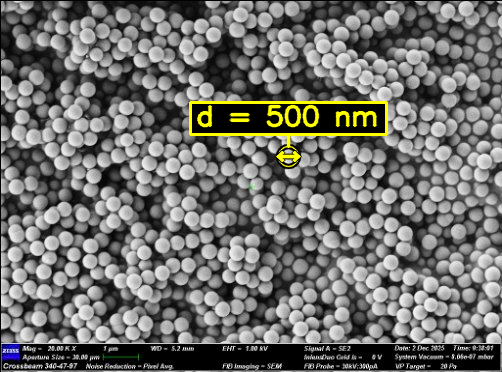}
    {SEM morphology}
    {GEN1 particles\\$d \,\approx\, 500$ nm}
\stagecard{1}{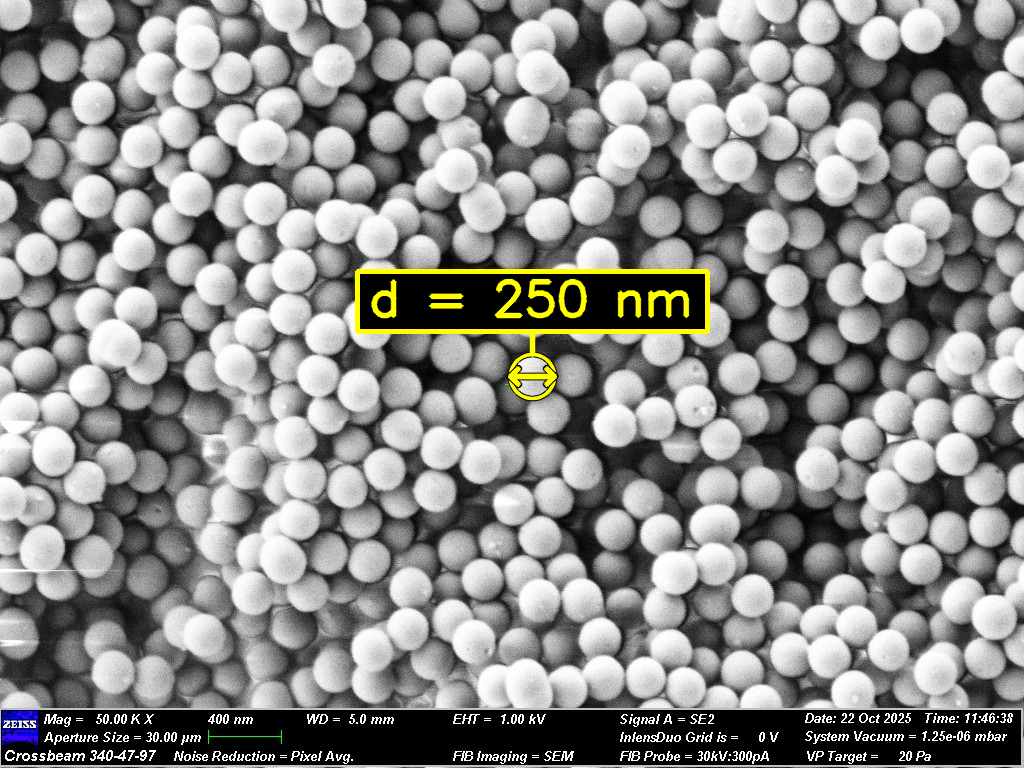}
    {SEM morphology}
    {GEN1 particles\\$d \,\approx\, 250$ nm}
\stagecard{2}{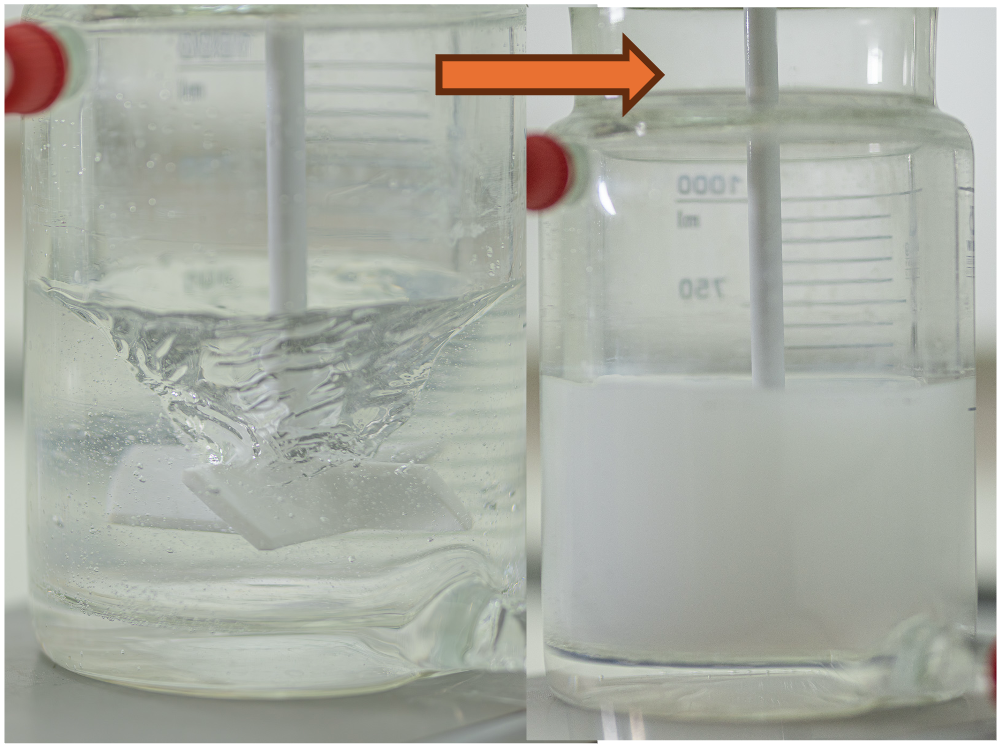}
    {Wet chemistry fabrication}  
    {Lab scale fabrication}
\stagecard{3}{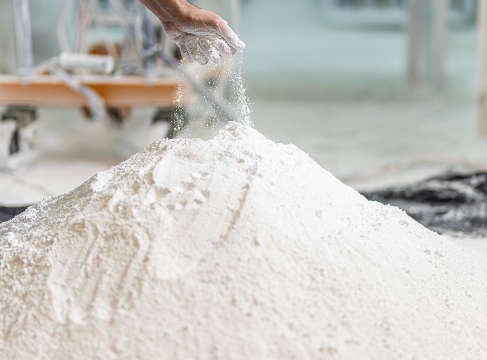}
    {Bulk}
    {\textasciitilde100kg scale fabrication}  
\stagecard{4}{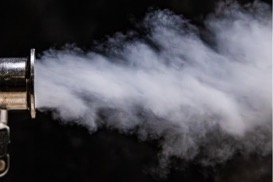}
    {Dry dispersion plume}
    {Visible plume during\\dry-dispersion test}
\draw[-{Latex[length=3mm]}, line width=1.05pt, draw=dividercolor]
    ({\cw+0.08}, {-0.5*\ch+0.18}) -- ({\colStep-0.08}, {-0.5*\ch+0.18});
\draw[-{Latex[length=3mm]}, line width=1.05pt, draw=dividercolor]
    ({\colStep+\cw+0.08}, {-0.5*\ch+0.18}) -- ({2*\colStep-0.08}, {-0.5*\ch+0.18});
\draw[-{Latex[length=3mm]}, line width=1.05pt, draw=dividercolor]
    ({2*\colStep+\cw+0.08}, {-0.5*\ch+0.18}) -- ({3*\colStep-0.08}, {-0.5*\ch+0.18});
\draw[-{Latex[length=3mm]}, line width=1.05pt, draw=dividercolor]   
    ({3*\colStep+\cw+0.08}, {-0.5*\ch+0.18}) -- ({4*\colStep-0.08}, {-0.5*\ch+0.18});
\begin{scope}[on background layer]
\fill[slidebg, rounded corners=6pt]
    (\bgLeft,\bgTop) rectangle (\bgRight,\bgBot);
\end{scope}
\useasboundingbox (\bgLeft,\bgTop) rectangle (\bgRight,\bgBot);
\end{tikzpicture}
\caption{
Representative stages in the dry-powder deployment workflow. Left two panels: SEM images of synthesized GEN1 silica microspheres with nominal diameters of approximately 500 nm and 250 nm, respectively, showing near-spherical morphology and narrow size distributions. Third panel: A wet-chemistry fabrication process demonstrating lab-scale synthesis. Fourth panel: bulk powder loaded into the feed reservoir prior to dispersion. Right panel: visible plume generated during dry-dispersion testing. 
}
\label{fig:powder_dispersion_workflow}
\end{figure*}

\newpage
    \item \textit{Surface treatment and atmospheric chemistry safety requirements}. An extensive set of hydrophobic coatings and surface treatments (e.g., calcination with different parameters) was designed and tested to reduce wetting by polar species, provide steric hindrance, and suppress undesired heterogeneous interactions. From these candidates, we selected a short-chain nonpolar monolayer coating that is compatible with the silica surface, contains only Si--C bonds, and is expected and validated to be more resistant to UV degradation than longer-chain silanes with C--C backbones, while also being regarded as bio-safe (see Table~\ref{tab:gen1_validation_map} for a list of experiments designed for testing the effectiveness of the surface treatment in addressing key compatibility requirements). 
    \begin{enumerate}
        \item The treated particles exhibit very large contact angles, of order $130^\circ$ (see Fig. \ref{fig:contact}), for both water and concentrated sulfuric acid. This strongly hydrophobic character suggests a low propensity for wetting, for nucleation of water- or sulfate-rich surface layers under stratospheric conditions, and for PSC nucleation. Detailed results and validation are provided in\citep{sai_fabrication_heterogeneous_chemistry}.
        
    \item Knudsen-cell and flow-through reactor measurements in packed-bed configuration, using stratospherically relevant trace gases, constrain the corresponding uptake coefficients \citep{sai_fabrication_heterogeneous_chemistry,sai_uptake_coefficients}. These constraints, in turn, support estimates of the associated ozone-depletion potential and provide experimental evidence that ozone-depletion risk can be substantially reduced through particle design, without relying on highly reactive particle surfaces. 
    
        \item Environmental aging measurements support the stability of the treated surface. Contact-angle, XPS, and NMR measurements indicate that the short silane coating remains preserved under UV irradiation exposure corresponding to approximately 54 weeks in the lower stratosphere, and under ozone exposure corresponding to roughly a year under stratospheric conditions (e.g. fumes of hydrochloric acid (HCl), nitric acid (HNO$_3$), and sulfuric acid (H$_2$SO$_4$) under controlled conditions as accelerated chemical-aging tests of the hydrophobic coating). Representative images of these environmental-exposure tests are shown in Figure~\ref{fig:environmental_stability_tests}. After exposure, the samples were subjected to contact angle analysis to determine whether the treatment retained its non-wetting properties. These results support the expectation that the intended GEN1 surface properties remain sufficiently stable 
        over the targeted stratospheric residence time. Detailed results are provided in \citep{sai_fabrication_heterogeneous_chemistry}. 
    \end{enumerate}

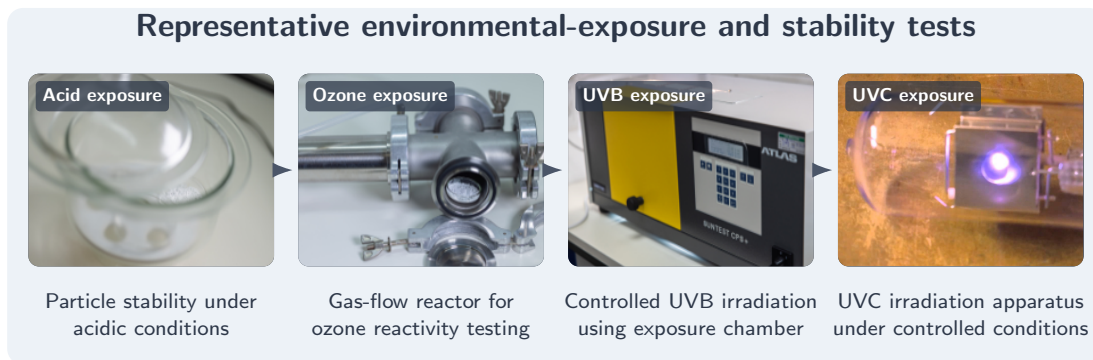
\begin{figure*}[t]
\centering
\begin{tikzpicture}
\def\cw{3.25}      
\def\ch{2.55}      
\def\gx{0.32}      
\def\capskip{0.24} 

\pgfmathsetmacro{\colStep}{\cw + \gx}
\pgfmathsetmacro{\totalW}{3*\colStep + \cw}
\pgfmathsetmacro{\bgLeft}{-0.28}
\pgfmathsetmacro{\bgRight}{\totalW + 0.28}
\pgfmathsetmacro{\bgTop}{1.05}
\pgfmathsetmacro{\bgBot}{-\ch - 1.12}

\newcommand{\stagecard}[5]{%
    \pgfmathsetmacro{\xL}{#1 * \colStep}%
    \pgfmathsetmacro{\xR}{\xL + \cw}%
    \pgfmathsetmacro{\yT}{0.18}%
    \pgfmathsetmacro{\yB}{\yT - \ch}%
    \fill[darkcard, rounded corners=4pt]
        (\xL,\yT) rectangle (\xR,\yB);
    \begin{scope}
        \clip[rounded corners=4pt] (\xL,\yT) rectangle (\xR,\yB);
        \node[anchor=center, inner sep=0pt]
            at ({(\xL+\xR)/2}, {(\yT+\yB)/2})
            {\includegraphics[width=\cw cm,height=\ch cm,#2]{#3}};
    \end{scope}
    \node[
        anchor=north west,
        fill=labeloverlay,
        fill opacity=0.72,
        text opacity=1,
        text=white,
        font=\sffamily\fontsize{7}{8.5}\selectfont\bfseries,
        rounded corners=1.6pt,
        inner sep=2.6pt
    ] at ({\xL+0.09}, {\yT-0.09}) {#4};
    \node[
        anchor=north,
        align=center,
        text=darkcard,
        font=\sffamily\fontsize{8.4}{10}\selectfont
    ] at ({(\xL+\xR)/2}, {\yB-\capskip}) {#5};
}

\node[
    font=\sffamily\bfseries\fontsize{13}{16}\selectfont,
    text=darkcard
] at ({\totalW/2}, 0.78)
{Representative environmental-exposure and stability tests};

\stagecard{0}{trim=0 0 0 0,clip}{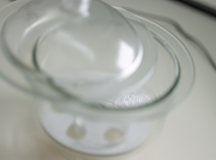}
    {Acid exposure}
    {Particle stability under\\acidic conditions}

\stagecard{1}{trim=0 0 0 0,clip}{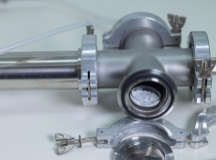}
    {Ozone exposure}
    {Gas-flow reactor for\\ozone reactivity testing}

\stagecard{2}{trim=0 0 0 0,clip}{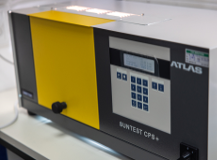}
    {UVB exposure}
    {Controlled UVB irradiation\\using exposure chamber}

\stagecard{3}{trim=0 0 0 0,clip}{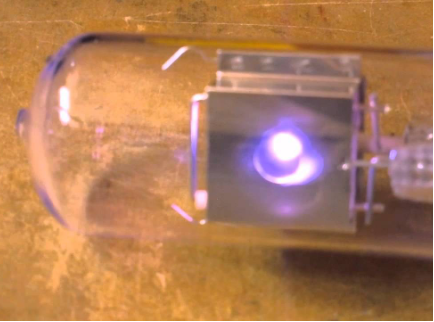}
    {UVC exposure}
    {UVC irradiation apparatus\\under controlled conditions}

\draw[-{Latex[length=2.7mm]}, line width=0.95pt, draw=dividercolor]
    ({\cw+0.07}, {-0.5*\ch+0.18}) -- ({\colStep-0.07}, {-0.5*\ch+0.18});

\draw[-{Latex[length=2.7mm]}, line width=0.95pt, draw=dividercolor]
    ({\colStep+\cw+0.07}, {-0.5*\ch+0.18}) -- ({2*\colStep-0.07}, {-0.5*\ch+0.18});

\draw[-{Latex[length=2.7mm]}, line width=0.95pt, draw=dividercolor]
    ({2*\colStep+\cw+0.07}, {-0.5*\ch+0.18}) -- ({3*\colStep-0.07}, {-0.5*\ch+0.18});

\begin{scope}[on background layer]
\fill[slidebg, rounded corners=6pt]
    (\bgLeft,\bgTop) rectangle (\bgRight,\bgBot);
\end{scope}

\useasboundingbox (\bgLeft,\bgTop) rectangle (\bgRight,\bgBot);
\end{tikzpicture}

\caption{
Representative environmental-exposure and stability-test images. Left to right: acid-exposure test of particles under concentrated acidic-fume conditions; ozone-exposure setup using a dedicated chamber; UVB irradiation using a controlled exposure chamber; and UVC irradiation apparatus. Together, these tests provide a visual overview of the experimental stress conditions used to assess particle stability and chemical robustness.
}
\label{fig:environmental_stability_tests}
\end{figure*}


\definecolor{panelbg}{RGB}{230,233,238}
\definecolor{titlecol}{RGB}{46,57,74}
\definecolor{labelbg}{RGB}{47,55,71}

\begin{figure*}[b]
\centering
\begin{tikzpicture}

\def\cw{4.05}      
\def\ch{3.10}      
\def\gx{0.38}      
\def\capskip{0.24} 

\pgfmathsetmacro{\colStep}{\cw + \gx}
\pgfmathsetmacro{\totalW}{2*\colStep + \cw}
\pgfmathsetmacro{\bgLeft}{-0.28}
\pgfmathsetmacro{\bgRight}{\totalW + 0.28}
\pgfmathsetmacro{\bgTop}{1.05}
\pgfmathsetmacro{\bgBot}{-\ch - 1.02}

\newcommand{\stagecard}[5]{%
    \pgfmathsetmacro{\xL}{#1 * \colStep}%
    \pgfmathsetmacro{\xR}{\xL + \cw}%
    \pgfmathsetmacro{\yT}{0.18}%
    \pgfmathsetmacro{\yB}{\yT - \ch}%
    \fill[darkcard, rounded corners=4pt]
        (\xL,\yT) rectangle (\xR,\yB);
    \begin{scope}
        \clip[rounded corners=4pt] (\xL,\yT) rectangle (\xR,\yB);
        \node[anchor=center, inner sep=0pt]
            at ({(\xL+\xR)/2}, {(\yT+\yB)/2})
            {\includegraphics[width=\cw cm,height=\ch cm,#2]{#3}};
    \end{scope}
    \node[
        anchor=north west,
        fill=labeloverlay,
        fill opacity=0.72,
        text opacity=1,
        text=white,
        font=\sffamily\fontsize{7}{8.5}\selectfont\bfseries,
        rounded corners=1.6pt,
        inner sep=2.6pt
    ] at ({\xL+0.09}, {\yT-0.09}) {#4};
    \node[
        anchor=north,
        align=center,
        text=darkcard,
        font=\sffamily\fontsize{9}{11}\selectfont
    ] at ({(\xL+\xR)/2}, {\yB-\capskip}) {#5};
}

\node[
    font=\sffamily\bfseries\fontsize{13}{16}\selectfont,
    text=darkcard
] at ({\totalW/2}, 0.78)
{Macroscopic wetting and contact-angle characterization};

\stagecard{0}{trim=0 0 0 0,clip}{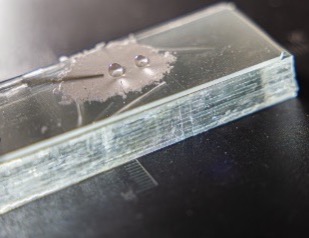}
    {Powder-coated slide}
    {Droplets deposited on\\coated glass slide}

\stagecard{1}{trim=0 0 0 0,clip}{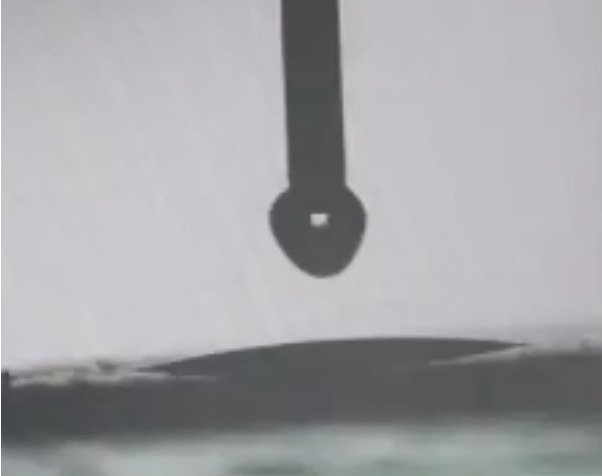}
    {Untreated silica}
    {Hydrophilic surface}

\stagecard{2}{trim=0 0 0 0,clip}{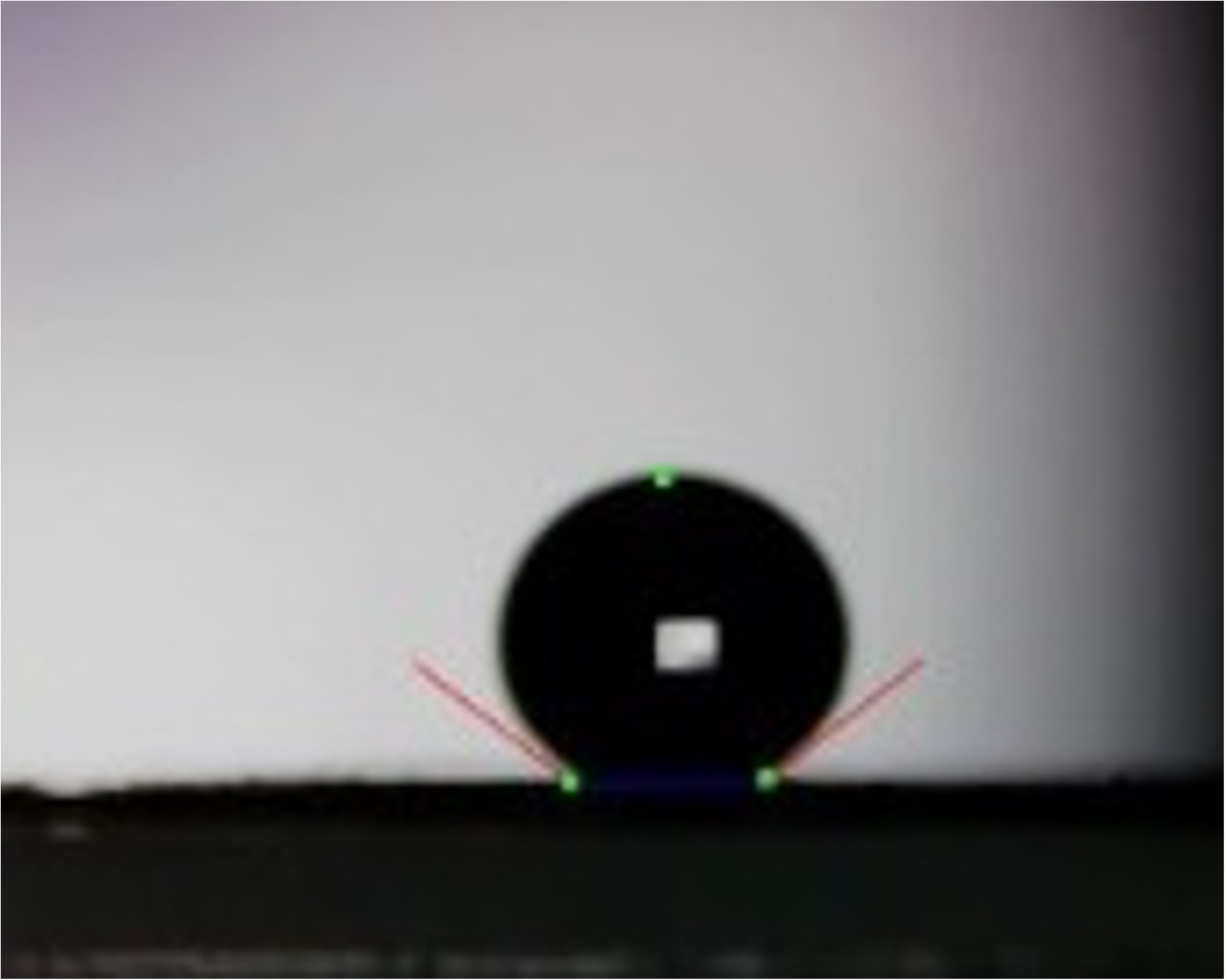}
    {Silane-treated silica}
    {Higher contact angle\\increased hydrophobicity}

\draw[-{Latex[length=3mm]}, line width=1.05pt, draw=dividercolor]
    ({\cw+0.08}, {-0.5*\ch+0.18}) -- ({\colStep-0.08}, {-0.5*\ch+0.18});

\draw[-{Latex[length=3mm]}, line width=1.05pt, draw=dividercolor]
    ({\colStep+\cw+0.08}, {-0.5*\ch+0.18}) -- ({2*\colStep-0.08}, {-0.5*\ch+0.18});

\begin{scope}[on background layer]
\fill[slidebg, rounded corners=6pt]
    (\bgLeft,\bgTop) rectangle (\bgRight,\bgBot);
\end{scope}

\useasboundingbox (\bgLeft,\bgTop) rectangle (\bgRight,\bgBot);

\end{tikzpicture}
\caption{
Representative wetting-characterization images. Left: macroscopic view of droplets deposited on a powder-coated glass slide. Center: side-view contact-angle image for untreated silica, showing complete wetting. Right: side-view contact-angle image after Silane treatment, showing reduced wetting and a higher contact angle, consistent with increased hydrophobicity.
}
\label{fig:contact}
\end{figure*}
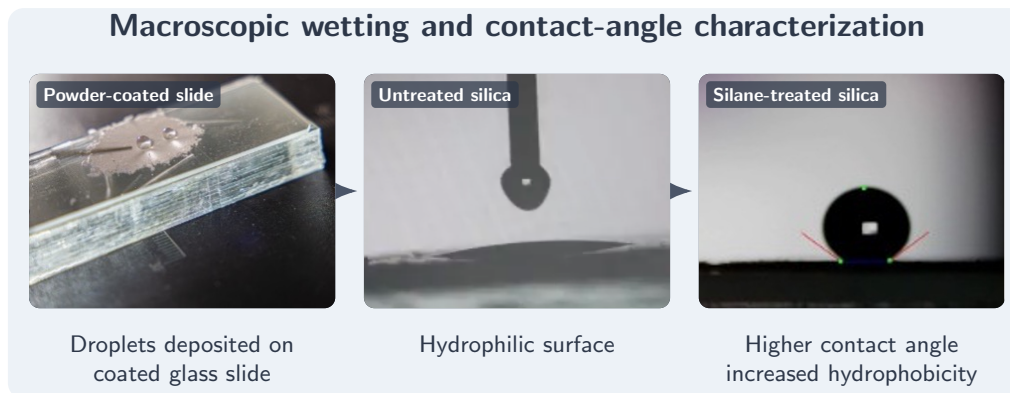

    \item {\it Scalable manufacturability}. We have carried out \citep{sai_manufacturing_processes} a detailed manufacturing analysis outlining the overall unit economics, CAPEX, and fabrication timeline, while assessing raw-material availability to ensure that no major supply bottlenecks are expected 
    and showing that a viable scale-up plan consistent with the above target price can be achieved.
    
    \item {\it Optical properties}. The optical properties of amorphous silica are well known, and the visible refractive index was measured and experimentally verified for representative GEN1 samples; infrared absorption coefficients were separately validated. The chemical-inertness results above suggest that the particle size distribution, which affects light scattering properties, is unlikely to evolve significantly during stratospheric residence due to wetting, nucleation, or sulfate-mediated coagulation. Shortwave and longwave optical properties' measurements are given in 
    \citep{sai_fabrication_heterogeneous_chemistry}.


    \item {\it Dispersion compatibility}. A dedicated spray tunnel was built to examine de-agglomeration under realistic pressurized-air conditions (Figure~\ref{fig:spray_tunnel}). These measurements, together with the associated physical analysis of the controlling mechanisms, indicate that effective de-agglomeration can be achieved under conditions compatible with feasible aerial deployment platforms. Details are provided in \citep{sai_dispersion_subsystem}.

    \begin{figure}[t]
    \centering
    \includegraphics[width=0.9\linewidth]{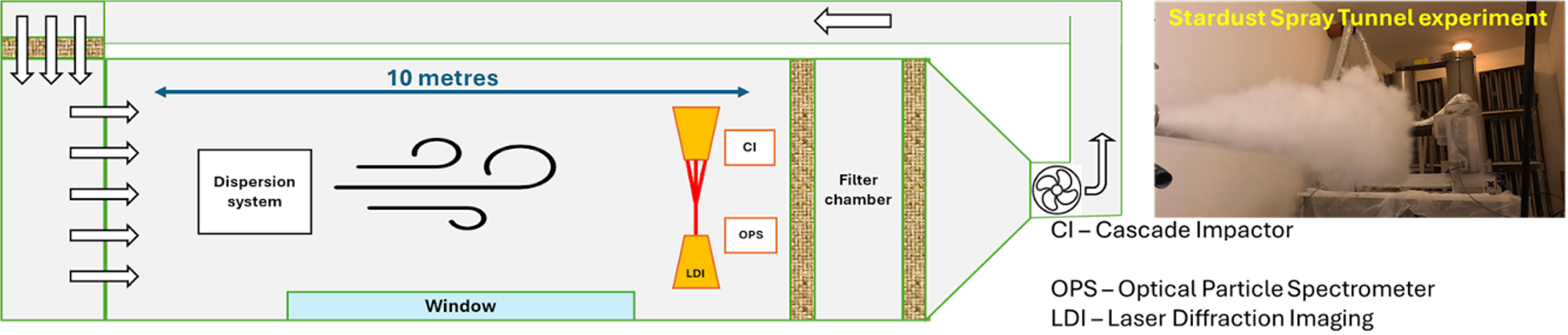}
    \caption{Spray tunnel experimental setup for testing particle de-agglomeration under pressurized-air conditions representative of aerial deployment conditions. The apparatus allows controlled injection and dispersal testing of candidate SAI particles to validate de-agglomeration performance under realistic flow conditions.}
    \label{fig:spray_tunnel}
    \end{figure}

    \item {\it Stratospheric residence time and implied radiative forcing efficiency}. The sedimentation velocity of the $\sim 0.5\,\mu$m particles is below the Brewer--Dobson tropical upwelling velocity in the lower stratosphere, supporting residence times of order one year. Using a stratospheric transport model calibrated against tracer-evolution measurements 
    \citep{sai_atmospheric_transport}, we further estimate that, given the measured and inferred GEN1 particle optical properties, a net negative radiative forcing of about $2.5\;\mathrm{W\,m^{-2}}$ could be achieved with a lower-stratospheric dispersion rate of approximately $10$ million tonnes per year. Details are provided in \citep{sai_atmospheric_transport}. 

    \item {\it Tagging}. Initial tagging results \citep{sai_tagging_encoding_technology} demonstrate the ability to identify the origin of individual particles by introducing and measuring atomic impurities. This provides preliminary support for the traceability concept developed above. Ongoing work is extending this approach toward fuller characterization of detection limits, encoding robustness, and the possible use of isotopic tagging.

    \item {\it Human, biota and environmental safety}. An overall safety assessment, including degradation experiments in water, was carried out, based on the broader literature on synthetic amorphous silica, including both hydrophilic and hydrophobic forms, to evaluate consistency with the biological-safety constraints discussed in \S~\ref{sub-sec:safety}. The broader natural silicon-cycle context and the laboratory dissolution-testing setup are shown in Figure~\ref{fig:silica_cycle_dissolution}. Details are given in a companion paper dedicated to bio-safety \citep{sai_biosafety} and in a companion paper under preparation dedicated to dissolution, degradation, and optimization~\citep{sai_degradability_dissolution}.
    
\end{enumerate}

\begin{figure*}
\centering
\begin{tikzpicture}

\definecolor{slidebg}{RGB}{235,240,248}
\definecolor{darkcard}{RGB}{35,45,60}
\definecolor{labeloverlay}{RGB}{25,32,45}
\definecolor{dividercolor}{RGB}{95,110,130}

\def\cwA{8.55}      
\def\cwB{4.05}      
\def\ch{4.20}       
\def\gx{0.45}       
\def\capskip{0.24}  

\pgfmathsetmacro{\xA}{0}
\pgfmathsetmacro{\xB}{\cwA + \gx}
\pgfmathsetmacro{\totalW}{\cwA + \gx + \cwB}
\pgfmathsetmacro{\bgLeft}{-0.28}
\pgfmathsetmacro{\bgRight}{\totalW + 0.28}
\pgfmathsetmacro{\bgTop}{1.05}
\pgfmathsetmacro{\bgBot}{-\ch - 1.02}

\newcommand{\silicacard}[6]{%
    \pgfmathsetmacro{\xL}{#1}%
    \pgfmathsetmacro{\xR}{\xL + #2}%
    \pgfmathsetmacro{\yT}{0.18}%
    \pgfmathsetmacro{\yB}{\yT - \ch}%
    \fill[darkcard, rounded corners=4pt]
        (\xL,\yT) rectangle (\xR,\yB);
    \begin{scope}
        \clip[rounded corners=4pt] (\xL,\yT) rectangle (\xR,\yB);
        \node[anchor=center, inner sep=0pt]
            at ({(\xL+\xR)/2}, {(\yT+\yB)/2})
            {\includegraphics[width=#2 cm,height=\ch cm,#3]{#4}};
    \end{scope}
    \node[
        anchor=north west,
        fill=labeloverlay,
        fill opacity=0.72,
        text opacity=1,
        text=white,
        font=\sffamily\fontsize{7}{8.5}\selectfont\bfseries,
        rounded corners=1.6pt,
        inner sep=2.6pt
    ] at ({\xL+0.09}, {\yT-0.09}) {#5};
    \node[
        anchor=north,
        align=center,
        text=darkcard,
        font=\sffamily\fontsize{9}{11}\selectfont
    ] at ({(\xL+\xR)/2}, {\yB-\capskip}) {#6};
}

\node[
    font=\sffamily\bfseries\fontsize{13}{16}\selectfont,
    text=darkcard
] at ({\totalW/2}, 0.78)
{Biogenic silicon cycling and laboratory dissolution testing};

\silicacard{\xA}{\cwA}{trim=0 0 0 0,clip}{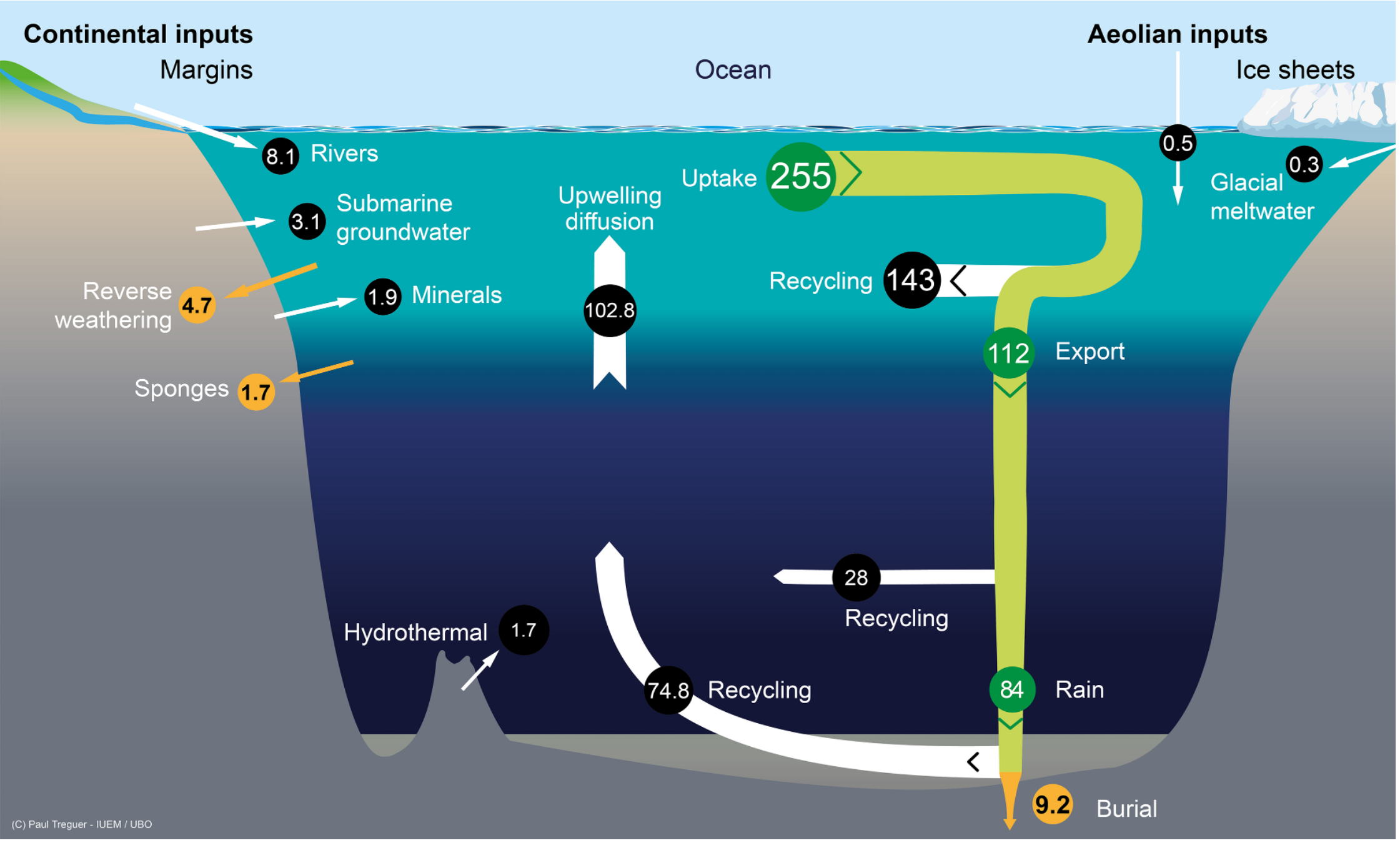}
    {Biogenic ocean silicon cycle}
    {Biogenic natural silicon inputs, uptake, recycling,\\export, rain, and burial fluxes}

\silicacard{\xB}{\cwB}{trim=0 0 0 0,clip}{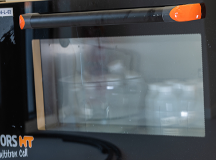}
    {Dissolution experiment}
    {Laboratory setup for controlled\\aqueous exposure testing}

\draw[-{Latex[length=3mm]}, line width=1.05pt, draw=dividercolor]
    ({\cwA+0.10}, {-0.5*\ch+0.18}) --
    ({\cwA+\gx-0.10}, {-0.5*\ch+0.18});

\begin{scope}[on background layer]
\fill[slidebg, rounded corners=6pt]
    (\bgLeft,\bgTop) rectangle (\bgRight,\bgBot);
\end{scope}

\useasboundingbox (\bgLeft,\bgTop) rectangle (\bgRight,\bgBot);

\end{tikzpicture}

\caption{
Natural silicon cycling and experimental evaluation of silica-relevant materials.
Left: schematic representation of the biogenic silicon cycle and balance, illustrating major natural inputs, biological uptake, recycling, export, rain, and burial fluxes. Reproduced from Tréguer et al.~\citep{treguer2021biogeochemical}, distributed under the Creative Commons Attribution 4.0 License.
Right: photograph of the laboratory experimental setup used to evaluate particle behavior under controlled aqueous exposure conditions.
}
\label{fig:silica_cycle_dissolution}
\end{figure*}
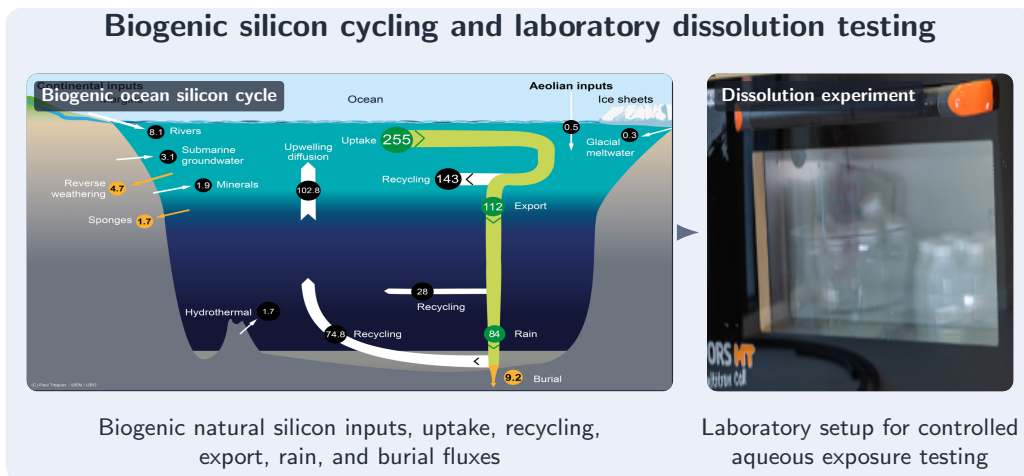

\newpage
\subsection{GEN2 core--shell particle development.}
\label{sub-sec:GEN2}

\begin{figure*}[htbp]
\centering
\begin{tikzpicture}
\def\cw{3.05}      
\def\ch{2.65}      
\def\gx{0.30}      
\def\capskip{0.24} 

\pgfmathsetmacro{\colStep}{\cw + \gx}
\pgfmathsetmacro{\totalW}{3*\colStep + \cw}
\pgfmathsetmacro{\bgLeft}{-0.28}
\pgfmathsetmacro{\bgRight}{\totalW + 0.28}
\pgfmathsetmacro{\bgTop}{1.05}
\pgfmathsetmacro{\bgBot}{-\ch - 1.08}

\newcommand{\genTwostagecard}[5]{%
    \pgfmathsetmacro{\xL}{#1 * \colStep}%
    \pgfmathsetmacro{\xR}{\xL + \cw}%
    \pgfmathsetmacro{\yT}{0.18}%
    \pgfmathsetmacro{\yB}{\yT - \ch}%
    \fill[darkcard, rounded corners=4pt]
        (\xL,\yT) rectangle (\xR,\yB);
    \begin{scope}
        \clip[rounded corners=4pt] (\xL,\yT) rectangle (\xR,\yB);
        \node[anchor=center, inner sep=0pt]
            at ({(\xL+\xR)/2}, {(\yT+\yB)/2})
            {\includegraphics[width=\cw cm,height=\ch cm,#2]{#3}};
    \end{scope}
    \node[
        anchor=north west,
        fill=labeloverlay,
        fill opacity=0.72,
        text opacity=1,
        text=white,
        font=\sffamily\fontsize{7}{8.5}\selectfont\bfseries,
        rounded corners=1.6pt,
        inner sep=2.6pt
    ] at ({\xL+0.09}, {\yT-0.09}) {#4};
    \node[
        anchor=north,
        align=center,
        text=darkcard,
        font=\sffamily\fontsize{8.2}{9.8}\selectfont
    ] at ({(\xL+\xR)/2}, {\yB-\capskip}) {#5};
}

\node[
    font=\sffamily\bfseries\large,
    text=darkcard
] at ({\totalW/2}, 0.78)
{GEN2 carbonate-core and core--shell particle development};

\genTwostagecard{0}{}{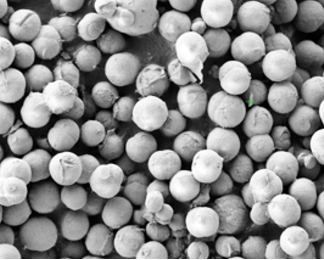}
    {Mg--calcite cores}
    {Spherical Mg--calcite\\core particles}

\genTwostagecard{1}{}{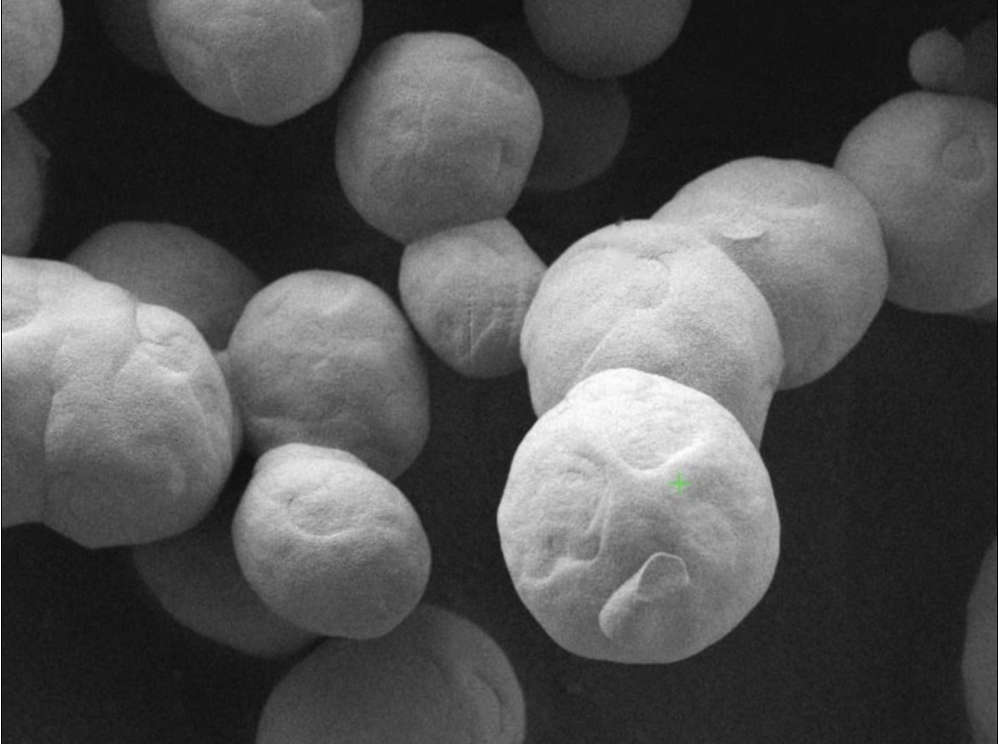}
    {CaCO\textsubscript{3} cores}
    {Spherical \ce{CaCO3}\\core particles}

\genTwostagecard{2}{}{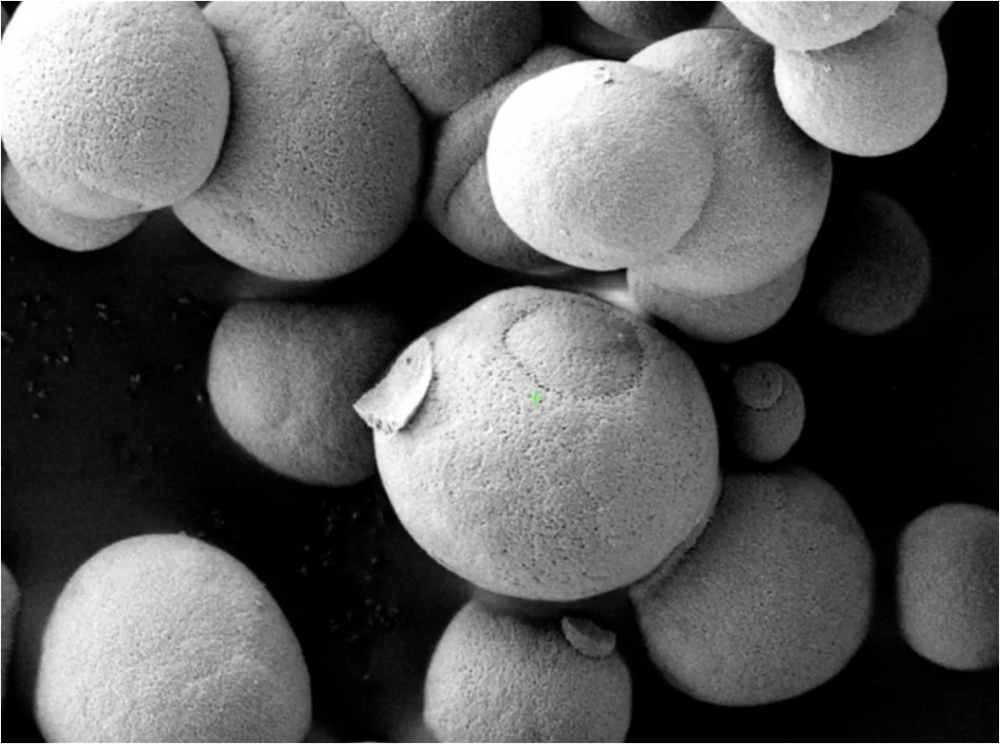}
    {Core--shell particles}
    {\ce{CaCO3}@\ce{SiO2}\\core--shell particles}
    
\genTwostagecard{3}{}{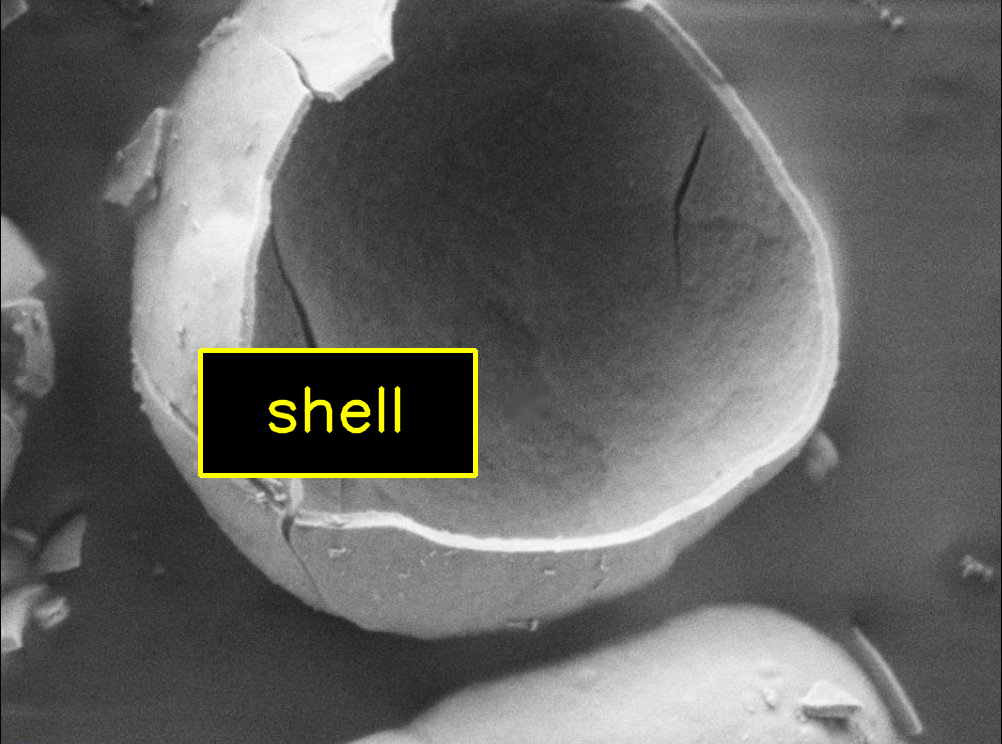}
    {Particle cross-section}
    {Intentionally fractured particle\\showing \ce{SiO2} shell}

\draw[-{Latex[length=2.7mm]}, line width=0.95pt, draw=dividercolor]
    ({\colStep+\cw+0.07}, {-0.5*\ch+0.18}) -- ({2*\colStep-0.07}, {-0.5*\ch+0.18});

\draw[-{Latex[length=2.7mm]}, line width=0.95pt, draw=dividercolor]
    ({2*\colStep+\cw+0.07}, {-0.5*\ch+0.18}) -- ({3*\colStep-0.07}, {-0.5*\ch+0.18});

\begin{scope}[on background layer]
\fill[slidebg, rounded corners=6pt]
    (\bgLeft,\bgTop) rectangle (\bgRight,\bgBot);
\end{scope}

\useasboundingbox (\bgLeft,\bgTop) rectangle (\bgRight,\bgBot);
\end{tikzpicture}

\caption{
GEN2 development results showing spherical Mg--calcite core particles as an additional carbonate-core architecture, micron-scale spherical \ce{CaCO3} particles developed and used as cores, the resulting \ce{CaCO3}@\ce{SiO2} core--shell particles after shell formation, and an intentionally fractured particle cross-section revealing the internal structure and the silica shell. The images demonstrate the feasibility of carbonate-core and core--shell routes, although optimization toward the final submicron size range and narrow monodispersity is still ongoing.
}
\label{fig:gen2_development}
\end{figure*}
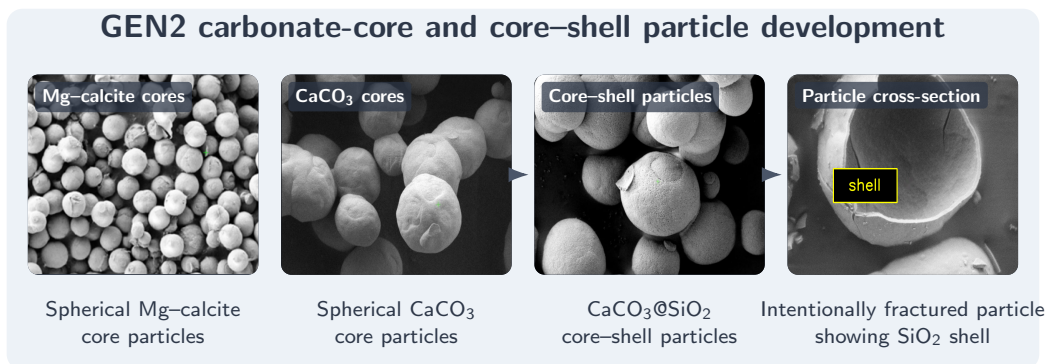

GEN2 particles are currently at an earlier development stage, where we are engineering core-shell structures to meet the same size distribution and morphology requirements established for GEN1. Initial synthesis trials have successfully produced near-spherical \ce{CaCO3} cores and demonstrated subsequent silica-shell encapsulation, supporting the feasibility of the core--shell approach. Figure~\ref{fig:gen2_development} shows representative spherical Mg--calcite and \ce{CaCO3} core samples, the corresponding \ce{CaCO3}$@$\ce{SiO2} particles after silica-shell formation, and a particle cross-section revealing the shell. Together, these images establish the feasibility of the GEN2 architecture. 
The size distribution has not yet been optimized to meet the target range 
but ongoing work on controlling nucleation and shell-growth conditions is directed toward that goal. Detailed results on  GEN2 core--shell particle development, fabrication, characterization, and validation of meeting requirements will be reported in a paper currently under preparation~\citep{sai_particle_architectures}.

\begin{table*}[t]
\caption{The main processes that may affect particle properties and atmospheric chemistry, addressed in the GEN1 particles' experimental validation program. The table lists the processes and describes their potential impacts, the triggers or conditions under which they operate, specific examples, the mitigation strategies, and the experimental validation methods.}
\label{tab:gen1_validation_map}
\scriptsize
\setlength{\tabcolsep}{3.5pt}
\renewcommand{\arraystretch}{1.04}
\begin{ruledtabular}
\begin{tabular}{p{0.115\textwidth} p{0.15\textwidth} p{0.16\textwidth} p{0.14\textwidth} p{0.10\textwidth} p{0.225\textwidth}}
\textbf{Process} & \textbf{Potential Impact} & \textbf{Trigger / condition} & \textbf{Examples} & \textbf{Mitigation} & \textbf{Experimental validation} \\
\hline

Jet-plume organic exposure &
Surface contamination and altered wettability or downstream aging behavior &
Exposure to exhaust vapors immediately after release &
Organic compounds, e.g.\ PAHs &
Low-energy inert surface &
Teflon chamber, PAM-OFR, or U-tube exposures, followed by comparison of fresh and exposed particles using contact-angle and surface-analysis measurements \\[2pt]

UV aging and oxidative exposure &
Loss or oxidation of surface groups, changed wettability, and modified heterogeneous reactivity &
Stratospheric UV radiation and oxidants during residence or accelerated laboratory exposure &
UVA/B; UVC (185--369~nm); \ce{O3}; photochemical truncation of trimethylsilyl groups &
Robust coating chemistry; calcined inert surface &
Stardust UV ovens, PAM-OFR, UVC sources, and ozone-chamber exposures, followed by XPS, NMR and contact-angle tests before and after treatment \\[2pt]

Uptake of stratospheric trace species &
Heterogeneous chemistry relevant to ozone, chlorine, and nitrogen budgets &
Low-temperature exposure to reactive trace gases, often enhanced after aging &
\ce{HCl}, \ce{HOCl}, \ce{ClONO2}, \ce{N2O5}, \ce{HNO3}; \ce{ClONO2 + HCl_{[s]} -> Cl2} &
Sterically hindered, low-affinity surface &
Knudsen cell, aerosol flow tube, and related heterogeneous-chemistry reactors; gas-depletion kinetics with CIMS and in-situ surface analysis; include fresh and pre-aged particles \\[2pt]

Sulfate deposition and wetting &
Thin-film formation, enhanced heterogeneous chemistry, and increased settling speed &
Coagulation with sulfate aerosol or sulfuric-acid deposition &
Sulfate aerosols; sulfuric-acid deposition; \ce{SO2 + H2O + oxidant} &
Hydrophobic coating &
Droplet-spraying and sulfate-generation experiments, combined with trapped-particle measurements, in-line holography, and in-situ Raman spectroscopy at low temperature and pressure \\[2pt]

Water, ice, and PSC nucleation &
CCN or ice activity, PSC chemistry, and possible microphysical changes during or after descent &
High RH or supersaturation at low temperature, in contrails, cirrus, PSCs, or tropospheric clouds &
Water, ice, NAT; PSC~I/II; \ce{ClONO2 + H2O_{[PSC]} -> Cl2} &
Hydrophobic coating &
CCN reactor, ice-nucleation setup, cloud-chamber tests, and contact-angle measurements, including low-temperature/low-pressure conditions and pre-aged particles \\[2pt]

\end{tabular}
\end{ruledtabular}
\end{table*}

\usetikzlibrary{shadings}

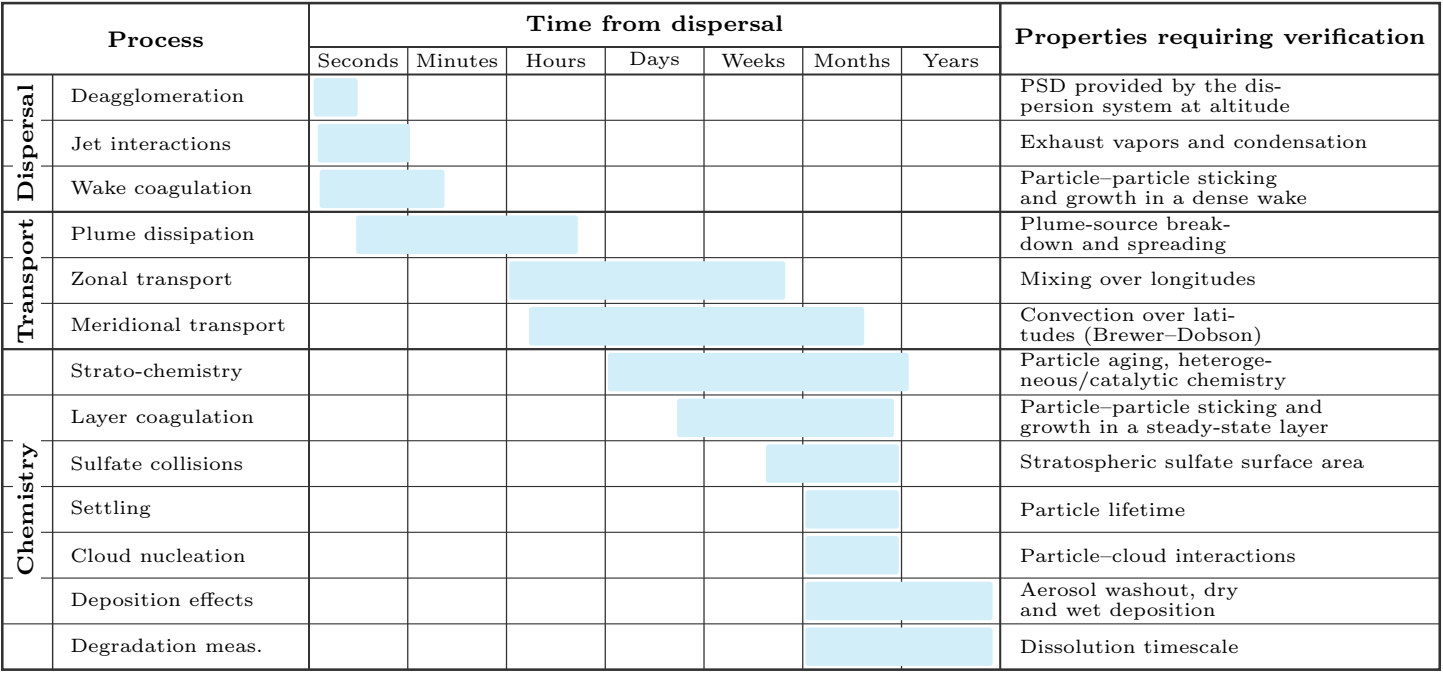
\begin{figure*}[!t]
\centering
\resizebox{\textwidth}{!}{%
\begin{tikzpicture}[x=1cm,y=-1cm]
\definecolor{barblue}{RGB}{31,170,225}
\definecolor{gridcol}{RGB}{50,50,50}

\def\wCat{0.55}
\def\wEff{2.80}
\def\wTime{7.55}
\def\wPar{4.80}
\def\hHeadA{0.48}
\def\hHeadB{0.30}
\def\hRow{0.50}
\def\nRows{13}

\pgfmathsetmacro{\W}{\wCat+\wEff+\wTime+\wPar}
\pgfmathsetmacro{\H}{\hHeadA+\hHeadB+\nRows*\hRow}
\pgfmathsetmacro{\yData}{\hHeadA+\hHeadB}
\pgfmathsetmacro{\dt}{\wTime/7.0}

\def\tbar#1#2#3{%
  \pgfmathsetmacro{\xb}{\wCat+\wEff + (#2)*\dt + 0.03}%
  \pgfmathsetmacro{\xe}{\wCat+\wEff + (#3)*\dt - 0.03}%
  \pgfmathsetmacro{\yb}{\yData + ((#1)-1)*\hRow + 0.04}%
  \pgfmathsetmacro{\ye}{\yb + \hRow - 0.08}%
  \shade[rounded corners=0.03cm,left color=barblue!20,middle color=barblue!70,right color=barblue!20]
  (\xb,\yb) rectangle (\xe,\ye);%
}

\draw[line width=0.8pt,color=gridcol] (0,0) rectangle (\W,\H);

\draw[line width=0.6pt,color=gridcol] (0,0) rectangle (\wCat+\wEff,\yData);
\draw[line width=0.6pt,color=gridcol] (\wCat+\wEff,0) rectangle (\wCat+\wEff+\wTime,\hHeadA);
\draw[line width=0.6pt,color=gridcol] (\wCat+\wEff,\hHeadA) rectangle (\wCat+\wEff+\wTime,\yData);
\draw[line width=0.6pt,color=gridcol] (\wCat+\wEff+\wTime,0) rectangle (\W,\yData);

\draw[line width=0.6pt,color=gridcol] (\wCat+\wEff,0) -- (\wCat+\wEff,\H);
\draw[line width=0.6pt,color=gridcol] (\wCat+\wEff+\wTime,0) -- (\wCat+\wEff+\wTime,\H);
\draw[line width=0.45pt,color=gridcol] (\wCat,\yData) -- (\wCat,\H);

\foreach \i in {0,...,7} {
  \draw[line width=0.4pt,color=gridcol]
  ({\wCat+\wEff+\i*\dt},\hHeadA) -- ({\wCat+\wEff+\i*\dt},\H);
}

\foreach \r in {0,...,13} {
  \draw[line width=0.35pt,color=gridcol] (0,{\yData+\r*\hRow}) -- (\W,{\yData+\r*\hRow});
}

\foreach \r in {3,6} {
  \draw[line width=0.7pt,color=gridcol] (0,{\yData+\r*\hRow}) -- (\W,{\yData+\r*\hRow});
}

\node[font=\bfseries\fontsize{6.6}{7.2}\selectfont] at ({0.5*(\wCat+\wEff)},{0.5*\yData}) {Process};
\node[font=\bfseries\fontsize{6.6}{7.2}\selectfont] at ({\wCat+\wEff+0.5*\wTime},{0.5*\hHeadA}) {Time from dispersal};
\node[font=\bfseries\fontsize{6.6}{7.2}\selectfont] at ({\wCat+\wEff+\wTime+0.5*\wPar},{0.5*\yData}) {Properties requiring verification};

\foreach \i/\lab in {
0/Seconds,
1/Minutes,
2/Hours,
3/Days,
4/Weeks,
5/Months,
6/Years} {
  \node[font=\fontsize{6.1}{6.4}\selectfont]
  at ({\wCat+\wEff+(\i+0.5)*\dt},{\hHeadA+0.5*\hHeadB}) {\lab};
}

\node[
  rotate=90,
  fill=white,
  inner sep=1.2pt,
  font=\bfseries\fontsize{6.8}{7.0}\selectfont
] at ({0.5*\wCat},{\yData+1.5*\hRow}) {Dispersal};

\node[
  rotate=90,
  fill=white,
  inner sep=1.2pt,
  font=\bfseries\fontsize{6.8}{7.0}\selectfont
] at ({0.5*\wCat},{\yData+4.5*\hRow}) {Transport};

\node[
  rotate=90,
  fill=white,
  inner sep=1.2pt,
  font=\bfseries\fontsize{6.8}{7.0}\selectfont
] at ({0.5*\wCat},{\yData+9.5*\hRow}) {Chemistry};

\foreach \i/\eff/\param in {
1/{Deagglomeration}/{PSD provided by the dispersion system at altitude},
2/{Jet interactions}/{Exhaust vapors and condensation},
3/{Wake coagulation}/{Particle--particle sticking and growth in a dense wake},
4/{Plume dissipation}/{Plume-source breakdown and spreading},
5/{Zonal transport}/{Mixing over longitudes},
6/{Meridional transport}/{Convection over latitudes (Brewer--Dobson)},
7/{Strato-chemistry}/{Particle aging, heterogeneous/catalytic chemistry},
8/{Layer coagulation}/{Particle--particle sticking and growth in a steady-state layer},
9/{Sulfate collisions}/{Stratospheric sulfate surface area},
10/{Settling}/{Particle lifetime},
11/{Cloud nucleation}/{Particle--cloud interactions},
12/{Deposition effects}/{Aerosol washout, dry and wet deposition},
13/{Degradation meas.}/{Dissolution timescale}
}{
  \pgfmathsetmacro{\yc}{\yData+(\i-0.5)*\hRow}
  \node[anchor=west,align=left,font=\fontsize{5.9}{6.2}\selectfont,text width=2.45cm]
  at ({\wCat+0.08},\yc) {\eff};
  \node[anchor=west,align=left,font=\fontsize{5.9}{6.2}\selectfont,text width=4.35cm]
  at ({\wCat+\wEff+\wTime+0.10},\yc) {\param};
}

\tbar{1}{0.02}{0.52}
\tbar{2}{0.06}{1.05}
\tbar{3}{0.08}{1.40}

\tbar{4}{0.45}{2.75}
\tbar{5}{2.00}{4.85}
\tbar{6}{2.20}{5.65}

\tbar{7}{3.00}{6.10}
\tbar{8}{3.70}{5.95}
\tbar{9}{4.60}{6.00}
\tbar{10}{5.00}{6.00}
\tbar{11}{5.00}{6.00}
\tbar{12}{5.00}{6.95}
\tbar{13}{5.00}{6.95}

\end{tikzpicture}%
}
\caption{A timescale view describing the time periods over which the processes examined within the GEN1 particles' validation program operate. Blue bands indicate approximate time intervals during which each process is expected to have its maximum impact.}
\label{fig:gen1_validation_scales}
\end{figure*}

\clearpage
\section{Conclusion}

Designing particles for stratospheric aerosol injection cannot be treated as a problem of radiative efficiency alone. When functionality, safety, and controllability constraints are considered jointly, the admissible design space narrows substantially, pointing toward sub-micron, near-spherical particles with a tightly controlled size distribution, stable physicochemical properties, and surface properties engineered to control atmospheric chemistry and aging and enhance compatibility with aerial dispersion. In this context, engineered solid particles may offer important advantages over sulfate-based aerosols. In sulfate-based SAI, the properties of the dispersed aerosols that determine the evolution of the particle population in the stratosphere and their shortwave radiative forcing also determine the reactive surface area available for heterogeneous chemistry and the aerosols' IR absorption. This does not allow controlling separately the wanted impacts, negative RF, and the unwanted impacts of heterogeneous chemistry and stratospheric heating. 
By contrast, the composite solid particles discussed here allow partial decoupling: bulk composition can be selected for radiative performance, while the outer surface can be engineered independently to limit reactivity, aging, cohesion, wetting, and adhesion. 

Composite solid particles are advantageous 
also for controllability and staged validation. Compared with sulfate-based aerosols, engineered solid particles with distinct composition and embedded traceability markers may be easier to detect, attribute, and study in small- and moderate-scale field experiments against the natural stratospheric background. The spherical amorphous silica particles discussed in this paper provide a demonstration of the feasibility of a practical SAI implementation meeting safety and controllability requirements while achieving significant negative radiative forcing, and provide a practical platform for surface engineering. The carbonate-core/silica-shell particles may enable larger negative forcing with reduced stratospheric heating. 

\bibliography{refs-bibfile}

@ARTICLE{2026SafetyWP,
       author = {{Waxman}, E. and {Spector}, A. and {Lederer}, Y. and {Segev}, Y. and {Kislev}, T. and {Yedvab}, Y. and {Kushnir}, D. and {Yahav}, R.},
        title = "{A proposal for the safety and controllability requirements that SRM systems should meet}",
      journal = {arXiv e-prints},
         year = 2026,
        month = apr,
          eid = {arXiv:2604.02283},
        pages = {arXiv:2604.02283},
archivePrefix = {arXiv},
       eprint = {2604.02283},
 primaryClass = {physics.ao-ph},
       adsurl = {https://ui.adsabs.harvard.edu/abs/2026arXiv260402283W}
}

@techreport{atsdr2019silica,
  author       = {{ATSDR}},
  title        = {Toxicological Profile for Silica},
  institution  = {Agency for Toxic Substances and Disease Registry},
  address      = {Atlanta, GA},
  year         = {2019},
  url          = {https://www.atsdr.cdc.gov/toxprofiles/tp211.pdf}
}

@techreport{oecd2004silica,
  author       = {{OECD}},
  title        = {{SIDS} Initial Assessment Report for Synthetic Amorphous Silica and Silicates},
  institution  = {OECD Screening Information Data Sets},
  year         = {2004},
  url          = {https://hpvchemicals.oecd.org/ui/handler.axd?id=f1d7c3f8-9db8-4ec5-b1e0-4a9b0ad75de1}
}

@techreport{ecetoc2014silica,
  author       = {{ECETOC}},
  title        = {Synthetic Amorphous Silica ({CAS} No.\ 7631-86-9)},
  institution  = {European Centre for Ecotoxicology and Toxicology of Chemicals},
  number       = {JACC Report No.\ 51},
  address      = {Brussels},
  year         = {2014},
  url          = {https://www.ecetoc.org/wp-content/uploads/2014/08/JACC-051.pdf}
}

@article{demaster1981supply,
  author       = {DeMaster, David J.},
  title        = {The supply and accumulation of silica in the marine environment},
  journal      = {Geochimica et Cosmochimica Acta},
  volume       = {45},
  number       = {10},
  pages        = {1715--1732},
  year         = {1981},
  doi          = {10.1016/0016-7037(81)90006-5}
}

@article{spitzmuller2023dissolution,
  author       = {Spitzm{\"u}ller, Laura and Nitschke, Fabian and Rudolph, Bastian and Berson, Jonathan and Schimmel, Thomas and Kohl, Thomas},
  title        = {Dissolution control and stability improvement of silica nanoparticles in aqueous media},
  journal      = {Journal of Nanoparticle Research},
  volume       = {25},
  number       = {3},
  pages        = {40},
  year         = {2023},
  doi          = {10.1007/s11051-023-05688-4}
}

@article{diedrich2012dissolution,
  author       = {Diedrich, Tamara and Dybowska, Agnieszka and Schott, Jacques and Valsami-Jones, Eugenia and Oelkers, Eric H.},
  title        = {The Dissolution Rates of {SiO$_2$} Nanoparticles As a Function of Particle Size},
  journal      = {Environmental Science \& Technology},
  volume       = {46},
  number       = {9},
  pages        = {4909--4915},
  year         = {2012},
  doi          = {10.1021/es2045053}
}

@book{mann2001biomineralization,
  author       = {Mann, Stephen},
  title        = {Biomineralization: Principles and Concepts in Bioinorganic Materials Chemistry},
  series       = {Oxford Chemistry Masters},
  publisher    = {Oxford University Press},
  address      = {Oxford},
  year         = {2001},
  isbn         = {978-0198508823}
}

@article{colarco2010online,
  author       = {Colarco, Peter and da Silva, Arlindo and Chin, Mian and Diehl, Thomas},
  title        = {Online simulations of global aerosol distributions in the {NASA} {GEOS-4} model and comparisons to satellite and ground-based aerosol optical depth},
  journal      = {Journal of Geophysical Research: Atmospheres},
  volume       = {115},
  number       = {D14},
  pages        = {D14207},
  year         = {2010},
  doi          = {10.1029/2009JD012820}
}

@misc{fda_calcium_carbonate_21cfr1841191,
  author       = {{U.S. Food and Drug Administration}},
  title        = {{21 CFR \S 184.1191 -- Calcium carbonate}},
  howpublished = {Electronic Code of Federal Regulations},
  year         = {2026},
  note         = {Direct food substance affirmed as generally recognized as safe}
}

@misc{fda_silicon_dioxide_21cfr172480,
  author       = {{U.S. Food and Drug Administration}},
  title        = {{21 CFR \S 172.480 -- Silicon dioxide}},
  howpublished = {Electronic Code of Federal Regulations},
  year         = {2026},
  note         = {Food additive permitted for use as an anticaking agent and beer stabilizer}
}

@misc{habibah2022hydroxyapatite,
  author       = {Habibah, Tutut Ummul and Amlani, Dharanshi V. and Brizuela, Melina},
  title        = {Hydroxyapatite Dental Material},
  howpublished = {StatPearls, NCBI Bookshelf},
  year         = {2022},
  note         = {Last update: September 12, 2022}
}

@misc{healthcanada_silica_dimethyl_silylate,
  author       = {{Health Canada}},
  title        = {Natural Health Products Ingredients Database: Silica Dimethyl Silylate},
  howpublished = {\url{https://webprod.hc-sc.gc.ca/nhpid-bdipsn/ingredReq?id=3550}},
  note         = {Lists silica dimethyl silylate / hydrophobic colloidal silica / silicon dioxide, silanated; oral route allowed as anticaking agent and glidant; accessed 29 April 2026}
}

@misc{evonik_aerosil_r972_pharma,
  author       = {{Evonik}},
  title        = {{AEROSIL R 972 Pharma}},
  howpublished = {\url{https://products.evonik.com/assets/29/19/AEROSIL_R_972_Pharma_EN_EN_132919.pdf}},
  note         = {Product information for high-purity hydrophobic colloidal silica for use as a pharmaceutical excipient; accessed 29 April 2026}
}

@article{toohey2025stratospheric_lifetime,
  author  = {Toohey, Matthew and Jia, Yue and Khanal, Sushant and Tegtmeier, Susann},
  title   = {Stratospheric residence time and the lifetime of volcanic stratospheric aerosols},
  journal = {Atmospheric Chemistry and Physics},
  volume  = {25},
  pages   = {3821--3839},
  year    = {2025},
  doi     = {10.5194/acp-25-3821-2025}
}

@article{mcgrory2022sulfuric_silica,
  author  = {McGrory, M. R. and Ward, A. D. and Jones, S. H. and King, M. D.},
  title   = {Mie scattering from optically levitated mixed sulfuric acid--silica core--shell aerosols: observation of core--shell morphology for atmospheric science},
  journal = {Physical Chemistry Chemical Physics},
  volume  = {24},
  pages   = {5813--5822},
  year    = {2022},
  doi     = {10.1039/D1CP04068E}
}

@misc{ihme2024gbd2021_air_pollution,
  author = {{Institute for Health Metrics and Evaluation}},
  title  = {Global Burden of Disease Study 2021 ({GBD} 2021) Air Pollution Exposure Estimates 1990--2021},
  year   = {2024},
  note   = {Seattle, WA: Institute for Health Metrics and Evaluation}
}

@article{feely1966fallout,
  author  = {Feely, H. W. and Seitz, H. and Lagomarsino, R. J. and Biscaye, P. E.},
  title   = {Transport and fallout of stratospheric radioactive debris},
  journal = {Tellus},
  volume  = {18},
  number  = {2--3},
  pages   = {316--328},
  year    = {1966},
  doi     = {10.3402/tellusa.v18i2-3.9302}
}

@misc{fda1996_hydrophobic_silica_feed,
  author       = {{U.S. Food and Drug Administration}},
  title        = {Food Substances Affirmed as Generally Recognized as Safe in Feed and Drinking Water of Animals; Hydrophobic Silica},
  howpublished = {Federal Register},
  volume       = {61},
  number       = {165},
  pages        = {43451--43454},
  year         = {1996},
  note         = {Affirms hydrophobic silica produced by hydrophobization of silicon dioxide with dichlorodimethylsilane as GRAS for use as an anticaking/free-flowing agent in vitamin preparations for animal feed}
}

@misc{ecfr_21cfr584700_hydrophobic_silicas,
  author       = {{U.S. Food and Drug Administration}},
  title        = {{21 CFR \S 584.700 -- Hydrophobic silicas}},
  howpublished = {\url{https://www.ecfr.gov/current/title-21/chapter-I/subchapter-E/part-584/subpart-B/section-584.700}},
  year         = {2026},
  note         = {Lists amorphous fumed hydrophobic silica and precipitated hydrophobic silica, silane dichlorodimethyl reaction products with silica, for use in vitamin preparations for animal feed; accessed 29 April 2026}
}

@article{stober1968controlled_growth,
  author  = {St{\"o}ber, Werner and Fink, Arthur and Bohn, Ernst},
  title   = {Controlled Growth of Monodisperse Silica Spheres in the Micron Size Range},
  journal = {Journal of Colloid and Interface Science},
  volume  = {26},
  number  = {1},
  pages   = {62--69},
  year    = {1968},
  doi     = {10.1016/0021-9797(68)90272-5}
}

@misc{mordor2026precipitated_silica,
  author       = {{Mordor Intelligence}},
  title        = {Precipitated Silica Market Size \& Share Analysis -- Growth Trends \& Forecasts 2026--2031},
  year         = {2026},
  howpublished = {\url{https://www.mordorintelligence.com/industry-reports/precipitated-silica-market}},
  note         = {Reports precipitated silica market volume of 3.25 million tons in 2025 and 3.38 million tons in 2026; accessed 29 April 2026}
}

@techreport{ecetoc2006synthetic_amorphous_silica,

  author      = {{European Centre for Ecotoxicology and Toxicology of Chemicals}},

  title       = {Synthetic Amorphous Silica ({CAS} No. 7631-86-9)},

  institution = {European Centre for Ecotoxicology and Toxicology of Chemicals},

  type        = {{JACC} Report},

  number      = {51},

  address     = {Brussels, Belgium},

  year        = {2006},

  url         = {https://www.ecetoc.org/publication/jacc-report-51-synthetic-amorphous-silica/},

  note        = {Joint Assessment of Commodity Chemicals}

}

@misc{procurementresource2025precipitated_silica_price,
  author       = {{Procurement Resource}},
  title        = {{SiO2 (Precipitated Silica) Price Trend, Index, News, Chart}},
  year         = {2025},
  howpublished = {\url{https://www.procurementresource.com/resource-center/precipitated-silica-price-trends}},
  note         = {Reports recent precipitated silica prices around 1300--1450 USD/MT in Europe; accessed 29 April 2026}
}

@misc{codex_gsfa_tricalcium_phosphate_341iii,
  author       = {{FAO/WHO Codex Alimentarius}},
  title        = {{GSFA Online Food Additive Details: Tricalcium phosphate (INS 341(iii))}},
  howpublished = {\url{https://www.fao.org/gsfaonline/additives/details.html?id=30}},
  note         = {General Standard for Food Additives Online; accessed 29 April 2026}
}

@book{lowenstam1989biomineralization,
  author    = {Lowenstam, Heinz A. and Weiner, Stephen},
  title     = {On Biomineralization},
  publisher = {Oxford University Press},
  year      = {1989}
}

@article{morse2007calcium_carbonate,
  author  = {Morse, John W. and Arvidson, Rolf S. and L{\"u}ttge, Andreas},
  title   = {Calcium Carbonate Formation and Dissolution},
  journal = {Chemical Reviews},
  volume  = {107},
  number  = {2},
  pages   = {342--381},
  year    = {2007},
  doi     = {10.1021/cr050358j}
}

@article{amaechi2019hydroxyapatite_toothpaste,
  author  = {Amaechi, Bennett T. and AbdulAzees, Parveez Ahamed and Alshareif, Dina Ossama and Shehata, Marina Adel and Lima, Patrícia Paula de Carvalho Sampaio and Abdollahi, Azadeh and Kalkhorani, Parisa Samadi and Evans, Veronica},
  title   = {Comparative efficacy of a hydroxyapatite and a fluoride toothpaste for prevention and remineralization of dental caries in children},
  journal = {BDJ Open},
  volume  = {5},
  pages   = {18},
  year    = {2019},
  doi     = {10.1038/s41405-019-0026-8}
}

@misc{codex_gsfa_silicon_dioxide_551,
  author       = {{FAO/WHO Codex Alimentarius}},
  title        = {{GSFA Online Food Additive Details: Silicon dioxide, amorphous (INS 551)}},
  howpublished = {\url{https://www.fao.org/gsfaonline/additives/details.html?id=284}},
  note         = {General Standard for Food Additives Online; accessed 29 April 2026}
}

@misc{codex_gsfa_calcium_carbonate_170i,
  author       = {{FAO/WHO Codex Alimentarius}},
  title        = {{GSFA Online Food Additive Details: Calcium carbonate (INS 170(i))}},
  howpublished = {\url{https://www.fao.org/gsfaonline/additives/details.html?id=185}},
  note         = {General Standard for Food Additives Online; accessed 29 April 2026}
}

@techreport{who2021air_quality_guidelines,
  author      = {{World Health Organization}},
  title       = {WHO Global Air Quality Guidelines: Particulate Matter (PM2.5 and PM10), Ozone, Nitrogen Dioxide, Sulfur Dioxide and Carbon Monoxide},
  institution = {World Health Organization},
  year        = {2021},
  isbn        = {9789240034228}
}

@book{2021NASEM.RS,

  author    = {{National Academies of Sciences, Engineering, and Medicine}},

  title     = {Reflecting Sunlight: Recommendations for Solar Geoengineering Research and Research Governance},

  year      = {2021},

  publisher = {The National Academies Press},

  address   = {Washington, DC},

  doi       = {10.17226/25762},

  url       = {https://doi.org/10.17226/25762}

}

@techreport{dhs2021tracer,
  author       = {{U.S. Department of Homeland Security, Science and Technology Directorate}},
  title        = {Final Environmental Assessment of Proposed Tracer Particle and Gas Releases for Chemical and Bio-Defense Testbed ({CBT}) Program and Urban Threat Dispersion ({UTD}) Program},
  institution  = {U.S. Department of Homeland Security},
  year         = {2021},
  url          = {https://www.dhs.gov/sites/default/files/publications/ea_utd_cbt_v7clean.pdf}
}

@techreport{dhs_utr_2024,
  title = {Environmental Assessment of Proposed {NYC} Subway Tracer Particle and Gas Releases for the Underground Transport Restoration ({UTR}) Project},
  author = {{Department of Homeland Security Science and Technology Directorate}},
  institution = {U.S. Department of Homeland Security},
  year = {2024},
  type = {Environmental Assessment},
  note = {Prepared for Department of Homeland Security Science and Technology Directorate}
}

@article{sai_dispersion_subsystem,
author        = {Segev, Yair and Levine, E. Y. and Bar-Yoseph, Yair and Amsallem, O. and Dagan, Y. and Laor, E. and Rahamim, S. and Luski, A. and Daniel, E. and Hettiarachchi, E. and Spector, A.},
title         = {Efficient dispersal of submicron solid particles for stratospheric aerosol injection},
journal       = {arXiv preprint arXiv:2605.27414},
year          = {2026},
month         = {May},
eprint        = {2605.27414},
archivePrefix = {arXiv}
}

@techreport{sai_manufacturing_processes,
author      = {Kuzurbardov, Tamir and Yaverboim, Avi and Kislev, Tzemah and Abramov, Eli},
title       = {Feasibility Study for Industrial Scale Submicronic Engineered Amorphous Silica Particle ({SEASP}) Manufacturing for Stratospheric Aerosol Injection ({SAI})},
institution = {Stardust Labs Ltd.},
address     = {Ness Tziona, Israel},
year        = {2026},
month       = {May},
note        = {Preprint Dated May 14, 2026},
url         = {https://www.stardust-initiative.com}
}

@article{sai_atmospheric_transport,
author  = {Lederer, Yoav and Wygoda, Nahliel and Halbertal, Dorri and Rose, Brian E. J.},
title   = {Solid-particle stratospheric aerosol injection: a 2-{D} modeling exploration of the design space},
journal = {EGUsphere},
year    = {2026},
month   = {May},
note    = {Preprint},
doi     = {10.5194/egusphere-2026-2772}
}

@misc{sai_monitoring_subsystem,
  title = {Monitoring Subsystem Design for {SAI} Operations},
  note  = {Companion paper in preparation}
}

@article{sai_technological_capabilities_for_governance,
author        = {Yahav, Roby and Spector, Amyad and Kushnir, Doron and Waxman, Matthew C.},
title         = {From Particles to Policy: {T}echnical {B}uilding {B}locks for {M}ulti-{S}tate {SAI} {C}oordination},
journal       = {arXiv preprint arXiv:2605.14947},
year          = {2026},
month         = {May},
eprint        = {2605.14947},
archivePrefix = {arXiv},
primaryClass  = {physics.ao-ph},
doi           = {10.48550/arXiv.2605.14947}
}

@misc{sai_climatic_response,
  author = {Stardust Labs},
  title  = {Climatic Response as a Function of Controlled Radiative Forcing in {SAI} Systems},
  year   = {2026},
  note   = {Companion paper in preparation}
}

@misc{sai_particle_architectures,
  title = {{GEN2} Particle Architectures for {SAI}},
  note  = {Companion paper in preparation}
}

@misc{sai_fabrication_heterogeneous_chemistry,
  title = {Particle Fabrication, Surface-Property Validation, and Heterogeneous Chemistry for {SAI} Systems},
  note  = {Companion paper in preparation}
}

@article{sai_uptake_coefficients,
author        = {Lostier, Anais and Segev, Yair and Kislev, Tzemah and Schwartz Roitman, Gal and Locoge, Nadine and Romanias, Manolis N.},
title         = {Uptake of stratospheric species on minerals proposed for stratospheric aerosol injection},
journal       = {arXiv preprint arXiv:2605.14826},
year          = {2026},
month         = {May},
eprint        = {2605.14826},
archivePrefix = {arXiv},
primaryClass  = {physics.chem-ph},
doi           = {10.48550/arXiv.2605.14826}
}

@misc{sai_tagging_encoding_technology,
  title = {Tagging and Encoding Technologies for Traceable {SAI} Particles},
  note  = {Companion paper in preparation}
}

@misc{sai_biosafety,
  title = {Biosafety Assessment of Candidate Particles for {SAI}},
  note  = {Companion paper in preparation}
}

@misc{sai_degradability_dissolution,
  title = {Degradability and Dissolution Optimization of Candidate Particles for {SAI}},
  note  = {Companion paper in preparation}
}

@techreport{OECD_AcuteToxicity,
  author       = {{OECD}},
  title        = {OECD Guidelines for the Testing of Chemicals, Section 4: Health Effects},
  institution  = {Organisation for Economic Co-operation and Development},
  year         = {2002--2022},
  address      = {Paris},
  note         = {Acute toxicity guidelines referenced:
                  TG 402 -- Acute Dermal Toxicity (2017);
                  TG 403 -- Acute Inhalation Toxicity (2009);
                  TG 423 -- Acute Oral Toxicity, Acute Toxic Class Method (2002);
                  TG 425 -- Acute Oral Toxicity, Up-and-Down Procedure (2022);
                  TG 436 -- Acute Inhalation Toxicity, Acute Toxic Class Method (2009)}
}

@techreport{OECD_RepeatedDoseChronic,
  author       = {{OECD}},
  title        = {OECD Guidelines for the Testing of Chemicals, Section 4: Health Effects},
  institution  = {Organisation for Economic Co-operation and Development},
  year         = {2009--2018},
  address      = {Paris},
  note         = {Repeated dose and chronic toxicity guidelines referenced:
                  TG 407 -- Repeated Dose 28-Day Oral Toxicity Study in Rodents;
                  TG 408 -- Repeated Dose 90-Day Oral Toxicity Study in Rodents;
                  TG 412 -- Subacute Inhalation Toxicity: 28-Day Study;
                  TG 413 -- Subchronic Inhalation Toxicity: 90-Day Study;
                  TG 451 -- Carcinogenicity Studies;
                  TG 452 -- Chronic Toxicity Studies;
                  TG 453 -- Combined Chronic Toxicity/Carcinogenicity Studies}
}

@techreport{OECD_SkinEyeSensitisation,
  author       = {{OECD}},
  title        = {OECD Guidelines for the Testing of Chemicals, Section 4: Health Effects},
  institution  = {Organisation for Economic Co-operation and Development},
  year         = {2004--2024},
  address      = {Paris},
  note         = {Skin, eye, and sensitisation guidelines referenced:
                  TG 430 -- In Vitro Skin Corrosion: Transcutaneous Electrical Resistance Test Method;
                  TG 431 -- In Vitro Skin Corrosion: Reconstructed Human Epidermis Test Method;
                  TG 435 -- In Vitro Membrane Barrier Test Method for Skin Corrosion;
                  TG 439 -- In Vitro Skin Irritation: Reconstructed Human Epidermis Test Method;
                  TG 442C -- In Chemico Skin Sensitisation: Direct Peptide Reactivity Assay;
                  TG 442D -- In Vitro Skin Sensitisation: ARE-Nrf2 Luciferase Test Method;
                  TG 442E -- In Vitro Skin Sensitisation: Human Cell Line Activation Test;
                  TG 467 -- Defined Approaches for Serious Eye Damage and Eye Irritation;
                  TG 492 -- Reconstructed Human Cornea-like Epithelium (RhCE) Test Method;
                  TG 496 -- In Vitro Macromolecular Test Method for Identifying Chemicals Inducing Serious Eye Damage and Not Requiring Classification for Eye Irritation or Serious Eye Damage}
}

@techreport{OECD_Mutagenicity,
  author       = {{OECD}},
  title        = {OECD Guidelines for the Testing of Chemicals, Section 4: Health Effects},
  institution  = {Organisation for Economic Co-operation and Development},
  year         = {1997--2023},
  address      = {Paris},
  note         = {Mutagenicity guidelines referenced:
                  TG 471 -- Bacterial Reverse Mutation Test (Ames Test);
                  TG 474 -- Mammalian Erythrocyte Micronucleus Test (in vivo, if in vitro positive);
                  TG 487 -- In Vitro Mammalian Cell Micronucleus Test;
                  TG 489 -- In Vivo Mammalian Alkaline Comet Assay (if in vitro positive)}
}

@techreport{OECD_ReproductiveDevelopmental,
  author       = {{OECD}},
  title        = {OECD Guidelines for the Testing of Chemicals, Section 4: Health Effects},
  institution  = {Organisation for Economic Co-operation and Development},
  year         = {2018--2022},
  address      = {Paris},
  note         = {Reproductive and developmental toxicity guidelines referenced:
                  TG 414 -- Prenatal Developmental Toxicity Study;
                  TG 443 -- Extended One-Generation Reproductive Toxicity Study}
}

@techreport{OECD_Neurotoxicity,
  author       = {{OECD}},
  title        = {OECD Guidelines for the Testing of Chemicals, Section 4: Health Effects},
  institution  = {Organisation for Economic Co-operation and Development},
  year         = {2004--2007},
  address      = {Paris},
  note         = {Neurotoxicity guidelines referenced:
                  TG 424 -- Neurotoxicity Study in Rodents;
                  TG 426 -- Developmental Neurotoxicity Study}
}

@techreport{OECD_AquaticEcotoxicity,
  author       = {{OECD}},
  title        = {OECD Guidelines for the Testing of Chemicals, Section 2: Effects on Biotic Systems},
  institution  = {Organisation for Economic Co-operation and Development},
  year         = {2004--2019},
  address      = {Paris},
  note         = {Aquatic ecotoxicity guidelines referenced:
                  TG 201 -- Freshwater Alga and Cyanobacteria, Growth Inhibition Test;
                  TG 202 -- Daphnia sp. Acute Immobilisation Test;
                  TG 203 -- Fish, Acute Toxicity Test;
                  TG 210 -- Fish, Early-Life Stage Toxicity Test;
                  TG 211 -- Daphnia magna Reproduction Test}
}

@techreport{OECD_TerrestrialEcotoxicity,
  author       = {{OECD}},
  title        = {OECD Guidelines for the Testing of Chemicals, Section 2: Effects on Biotic Systems},
  institution  = {Organisation for Economic Co-operation and Development},
  year         = {2000--2017},
  address      = {Paris},
  note         = {Terrestrial ecotoxicity guidelines referenced:
                  TG 207 -- Earthworm, Acute Toxicity Tests;
                  TG 208 -- Terrestrial Plant Test: Seedling Emergence and Seedling Growth Test;
                  TG 216 -- Soil Microorganisms: Nitrogen Transformation Test;
                  TG 217 -- Soil Microorganisms: Carbon Transformation Test;
                  TG 222 -- Earthworm Reproduction Test;
                  TG 226 -- Predatory Mite Reproduction Test in Soil;
                  TG 245 -- Honey Bee (Apis mellifera) Larval Toxicity Test, Single Exposure}
}

@techreport{OECD_PersistenceBioaccumulation,
  author       = {{OECD}},
  title        = {OECD Guidelines for the Testing of Chemicals, Section 3: Environmental Fate and Behaviour},
  institution  = {Organisation for Economic Co-operation and Development},
  year         = {1992--2012},
  address      = {Paris},
  note         = {Persistence and bioaccumulation guidelines referenced:
                  TG 301 -- Ready Biodegradability;
                  TG 305 -- Bioaccumulation in Fish: Aqueous and Dietary Exposure;
                  TG 310 -- Ready Biodegradability: CO2 in Sealed Vessels (Headspace Test)}
}

@misc{EU_CLP,
  author       = {{European Parliament and Council of the European Union}},
  title        = {Regulation (EC) No 1272/2008 on Classification, Labelling and Packaging of Substances and Mixtures (CLP Regulation)},
  year         = {2008},
  address      = {Brussels},
  note         = {Implements the UN GHS in the European Union.}
}

@misc{EU_REACH_AnnexXIII,
  author       = {{European Parliament and Council of the European Union}},
  title        = {Regulation (EC) No 1907/2006 (REACH), Annex XIII: Criteria for the Identification of Persistent, Bioaccumulative and Toxic Substances, and Very Persistent and Very Bioaccumulative Substances},
  year         = {2006},
  address      = {Brussels},
  note         = {PBT/vPvB assessment criteria as amended by Commission Regulation (EU) No 253/2011.}
}

@misc{EU2018_1881,
  title        = {Commission Regulation ({EU}) 2018/1881 of 3 December 2018 amending Regulation ({EC}) No 1907/2006 of the {European Parliament} and of the {Council} on the Registration, Evaluation, Authorisation and Restriction of Chemicals ({REACH}) as regards Annexes {I}, {III}, {VI}, {VII}, {VIII}, {IX}, {X}, {XI} and {XII} to address nanoforms of substances},
  author       = {{European Commission}},
  journal      = {Official Journal of the European Union},
  volume       = {L 308},
  pages        = {1--20},
  year         = {2018},
  date         = {2018-12-04},
  url          = {https://eur-lex.europa.eu/eli/reg/2018/1881/oj},
  note         = {Document 32018R1881}
}

@techreport{IARC_Monographs,
  author       = {{International Agency for Research on Cancer}},
  title        = {IARC Monographs on the Identification of Carcinogenic Hazards to Humans},
  institution  = {World Health Organization},
  address      = {Lyon},
  year         = {1972--2024},
  note         = {Classification system: Group 1 (carcinogenic), Group 2A (probably carcinogenic), Group 2B (possibly carcinogenic), Group 3 (not classifiable).}
}

@misc{EU_REACH_2006,
  title        = {Regulation ({EC}) {No} 1907/2006 of the {European Parliament} and of the {Council} of 18 {December} 2006 concerning the {Registration}, {Evaluation}, {Authorisation} and {Restriction} of {Chemicals} ({REACH})},
  author       = {{European Union}},
  year         = {2006},
  note         = {Official Journal of the European Union, L 396, 30.12.2006, p. 1--849},
  url          = {https://eur-lex.europa.eu/legal-content/EN/TXT/?uri=CELEX:02006R1907-20220301},
  number       = {2}
}

@misc{EPA_OPPTS_870,
  title        = {Health Effects Test Guidelines ({OCSPP}/{OPPTS} 870 Series)},
  author       = {{U.S. Environmental Protection Agency}},
  year         = {1998},
  note         = {Office of Chemical Safety and Pollution Prevention ({OCSPP}), formerly Office of Prevention, Pesticides and Toxic Substances ({OPPTS})},
  url          = {https://www.epa.gov/test-guidelines-pesticides-and-toxic-substances/series-870-health-effects-test-guidelines},
  number       = {4}
}

@article{EFSA_MOE_2012,
  title        = {Guidance on the use of the {Margin} of {Exposure} ({MOE}) approach in risk assessment},
  author       = {{European Food Safety Authority (EFSA)}},
  journal      = {EFSA Journal},
  year         = {2012},
  volume       = {10},
  number       = {6},
  pages        = {2571},
  doi          = {10.2903/j.efsa.2012.2571},
  url          = {https://efsa.onlinelibrary.wiley.com/doi/abs/10.2903/j.efsa.2012.2571},
  note         = {Reference [6]}
}

@article{plane2012cosmic_dust,

  author  = {Plane, John M. C.},

  title   = {Cosmic dust in the Earth's atmosphere},

  journal = {Chemical Society Reviews},

  year    = {2012},

  volume  = {41},

  number  = {19},

  pages   = {6507--6518},

  doi     = {10.1039/C2CS35132C}

}

@misc{echa_calcium_carbonate,
  author       = {{European Chemicals Agency}},
  title        = {{Calcium carbonate: Registration Dossier / Substance Information}},
  year         = {2026},
  howpublished = {\url{https://chem.echa.europa.eu/100.006.765}},
  note         = {EC No. 207-439-9; CAS No. 471-34-1. Accessed 5 May 2026}
}

@article{treguer2021biogeochemical,

  title={Reviews and syntheses: The biogeochemical cycle of silicon in the modern ocean},

  author={Tr{\'e}guer, Paul J. and Sutton, Jill N. and Brzezinski, Mark and Charette, Matthew A. and Devries, Timothy and Dutkiewicz, Stephanie and Ehlert, Claudia and Hawkings, Jon and Leynaert, Aude and Liu, Su Mei and Llopis Monferrer, Natalia and L{\'o}pez-Acosta, Mar{\'i}a and Maldonado, Manuel and Rahman, Shaily and Ran, Lihua and Rouxel, Olivier},

  journal={Biogeosciences},

  volume={18},

  pages={1269--1289},

  year={2021},

  doi={10.5194/bg-18-1269-2021}

}

@techreport{EFSA_2005,
  author       = {{EFSA Scientific Committee}},
  title        = {Opinion of the Scientific Committee on a request from
                  {EFSA} related to a harmonised approach for risk assessment
                  of substances which are both genotoxic and carcinogenic},
  institution  = {European Food Safety Authority},
  journal      = {EFSA Journal},
  volume       = {282},
  pages        = {1--31},
  year         = {2005},
  doi          = {10.2903/j.efsa.2005.282}
}

@techreport{ECHA_R8,
  author       = {{ECHA}},
  title        = {Guidance on Information Requirements and Chemical Safety
                  Assessment. Chapter {R.8}: Characterisation of Dose
                  [Concentration]--Response for Human Health},
  institution  = {European Chemicals Agency},
  address      = {Helsinki, Finland},
  year         = {2012},
  url          = {https://echa.europa.eu/guidance-documents/guidance-on-information-requirements-and-chemical-safety-assessment}
}

@misc{codex_gsfa_magnesium_carbonate_504i,

  author       = {{Codex Alimentarius Commission}},

  title        = {{GSFA Online Food Additive Details: Magnesium Carbonate (INS 504(i))}},

  howpublished = {\url{https://www.fao.org/gsfaonline/additives/details.html?id=76}},

  note         = {Codex General Standard for Food Additives (GSFA), updated up to the 48th Session of the Codex Alimentarius Commission, 2025. Accessed 2026-05-02}

}

@misc{codex_gsfa_magnesium_hydroxide_carbonate_504ii,

  author       = {{Codex Alimentarius Commission}},

  title        = {{GSFA Online Food Additive Details: Magnesium Hydroxide Carbonate (INS 504(ii))}},

  howpublished = {\url{https://www.fao.org/gsfaonline/additives/details.html?id=67}},

  note         = {Codex General Standard for Food Additives (GSFA), updated up to the 48th Session of the Codex Alimentarius Commission, 2025. Accessed 2026-05-02}

}

@article{blackstock2009climate,
  author        = {Blackstock, Jason J. and others},
  title         = {Climate engineering responses to climate emergencies},
  journal       = {arXiv preprint arXiv:0907.5140},
  year          = {2009},
  eprint        = {0907.5140},
  archivePrefix = {arXiv},
  url           = {https://arxiv.org/abs/0907.5140}
}

@techreport{2023OSTP.SRM,
  author      = {{Office of Science and Technology Policy}},
  title       = "{Congressionally Mandated Research Plan and an Initial Research Governance Framework Related to Solar Radiation Modification}",
  institution = {Office of Science and Technology Policy, Executive Office of the President},
  address     = {Washington, DC, USA},
  year        = 2023,
  month       = jun,
  url         = {https://bidenwhitehouse.archives.gov/wp-content/uploads/2023/06/Congressionally-Mandated-Report-on-Solar-Radiation-Modification.pdf},
  note        = {Accessed 15 September 2025}
}

@techreport{2023UNEP.SRMReport,
  author      = {{United Nations Environment Programme}},
  title       = "{One Atmosphere: An Independent Expert Review on Solar Radiation Modification Research and Deployment}",
  institution = {United Nations Environment Programme},
  address     = {Nairobi, Kenya},
  year        = 2023,
  month       = feb,
  day         = 28,
  url         = {https://www.unep.org/resources/report/Solar-Radiation-Modification-research-deployment},
  note        = {Accessed 15 September 2025}
}

@techreport{2024EC.SRM,
  author      = {{European Commission} and {Directorate-General for Research and Innovation} and {Group of Chief Scientific Advisors}},
  title       = "{Solar Radiation Modification}",
  institution = {European Commission},
  address     = {Brussels, Belgium},
  year        = 2024,
  month       = dec,
  day         = 9,
  doi         = {10.2777/391614},
  url         = {https://data.europa.eu/doi/10.2777/391614},
  note        = {Accessed 15 September 2025}
}

@techreport{2025RoyalSoc.SRM,
  author      = {{The Royal Society}},
  title       = "{Solar Radiation Modification: Policy Briefing}",
  institution = {The Royal Society},
  address     = {London, UK},
  year        = 2025,
  url         = {https://royalsociety.org/news-resources/projects/solar-radiation-modification/},
  note        = {Accessed 15 September 2025; independent scientific policy briefing on SRM techniques and implications}
}

@article{dykema2016improved,
  title={Improved aerosol radiative properties as a foundation for solar geoengineering risk assessment},
  author={Dykema, John A and Keith, David W and Keutsch, Frank N},
  journal={Geophysical Research Letters},
  volume={43},
  number={14},
  pages={7758--7766},
  year={2016},
  doi={10.1002/2016GL069258},
  publisher={Wiley Online Library}
}

@article{crutzen2006albedo,
  author    = {Crutzen, Paul J.},
  title     = {Albedo Enhancement by Stratospheric Sulfur Injections: A Contribution to Resolve a Policy Dilemma?},
  journal   = {Climatic Change},
  volume    = {77},
  number    = {3--4},
  pages     = {211--220},
  year      = {2006},
  doi       = {10.1007/s10584-006-9101-y}
}

@article{pope2012stratospheric,
  title={Stratospheric aerosol particles and solar-radiation management},
  author={Pope, Francis D and Braesicke, Peter and Grainger, Roy G and Kalberer, Markus and Watson, I Matthew and Davidson, Peter J and Cox, Richard A},
  journal={Nature Climate Change},
  volume={2},
  number={10},
  pages={713--719},
  year={2012},
  doi = {10.1038/nclimate1528},
  publisher={Nature Publishing Group}
}

@article{rasch2008overview,
  title={An overview of geoengineering of climate using stratospheric sulphate aerosols},
  author={Rasch, Philip J and Tilmes, Simone and Turco, Richard P and Robock, Alan and Oman, Luke and Chen, Chih-Chieh and Stenchikov, Georgiy L and Garcia, Rolando R},
  journal={Philosophical Transactions of the Royal Society A: Mathematical, Physical and Engineering Sciences},
  volume={366},
  number={1882},
  pages={4007--4037},
  doi = {10.1098/rsta.2008.0131},
  year={2008},
  publisher={The Royal Society London}
}

@inproceedings{teller1997global,
  title={Global Warming and Ice Ages: I. Prospects for Physics-Based Modulation of Global Change},
  author={Teller, Edward and Wood, Lowell and Hyde, Roderick},
  booktitle={International Symposium on Planetary Emergencies},
  address={Erice, Italy},
  year={1997}
}

@article{Vattioni2024,
  author    = {Vattioni, Sandro and Weber, Rahel and Feinberg, Aryeh and Stenke, Andrea and Dykema, John A. and Luo, Beiping and Kelesidis, Georgios A. and Bruun, Christian A. and Sukhodolov, Timofei and Keutsch, Frank N. and Peter, Thomas and Chiodo, Gabriel},
  title     = {A fully coupled solid-particle microphysics scheme for stratospheric aerosol injections within the aerosol--chemistry--climate model {SOCOL-AERv2}},
  journal   = {Geoscientific Model Development},
  volume    = {17},
  pages     = {7767--7793},
  year      = {2024},
  doi       = {10.5194/gmd-17-7767-2024},
  url       = {https://gmd.copernicus.org/articles/17/7767/2024/}
}

@article{Vattioni2023,
  author    = {Vattioni, Sandro and Luo, Beiping and Feinberg, Aryeh and Stenke, Andrea and Vockenhuber, Christof and Weber, Rahel and Dykema, John A. and Krieger, Ulrich K. and Ammann, Markus and Keutsch, Frank N. and Peter, Thomas and Chiodo, Gabriel},
  title     = {Chemical Impact of Stratospheric Alumina Particle Injection for Solar Radiation Modification and Related Uncertainties},
  journal   = {Geophysical Research Letters},
  volume    = {50},
  number    = {24},
  pages     = {e2023GL105889},
  year      = {2023},
  doi       = {10.1029/2023GL105889}
}

@article{weisenstein2015solar,
  title={Solar geoengineering using solid aerosol in the stratosphere},
  author={Weisenstein, Debra K and Keith, David W and Dykema, John A},
  journal={Atmospheric Chemistry and Physics},
  volume={15},
  number={20},
  pages={11835--11859},
  year={2015},
  publisher={Copernicus GmbH}
}

@article{dai2020experimental,
  title={Experimental reaction rates constrain estimates of ozone response to calcium carbonate geoengineering},
  author={Dai, Zhen and Weisenstein, Debra K and Keutsch, Frank N and Keith, David W},
  journal={Communications Earth \& Environment},
  volume={1},
  number={1},
  pages={1--9},
  year={2020},
  doi={10.1038/s43247-020-00058-7},
  publisher={Nature Publishing Group}
}

@article{Liu2012,
  author  = {Liu, Y. and Zhao, X. and Li, W. and Zhou, X.},
  title   = {Background stratospheric aerosol variations deduced from satellite observations},
  journal = {Journal of Applied Meteorology and Climatology},
  volume  = {51},
  number  = {4},
  pages   = {799--812},
  year    = {2012},
  doi     = {10.1175/JAMC-D-11-0120.1}
}

@techreport{WMO2025StateClimate,
    author      = {{World Meteorological Organization}},
    title       = {State of the Global Climate 2024},
    institution = {World Meteorological Organization},
    year        = {2025},
    number      = {WMO-No. 1368},
    address     = {Geneva},
    isbn        = {978-92-63-11368-5},
    url         = {https://wmo.int/publication-series/state-of-global-climate/state-of-global-climate-2024}
}

@techreport{unep2025emissions,
    author      = {{United Nations Environment Programme}},
    title       = {Emissions Gap Report 2025: Off Target -- Continued Collective Inaction Puts Global Temperature Goal at Risk},
    institution = {United Nations Environment Programme (UNEP)},
    year        = {2025},
    address     = {Nairobi}
}

@techreport{lee2023ipcc,
    author  = {Lee, Hoesung and Romero, Jos{\'e}},
    title   = {{IPCC, 2023: Climate Change 2023: Synthesis Report. Summary for Policymakers. Contribution of Working Groups I, II and III to the Sixth Assessment Report of the Intergovernmental Panel on Climate Change}},
    institution = {Intergovernmental Panel on Climate Change (IPCC)},
    address = {Geneva, Switzerland},
    year    = {2023},
    note    = {Core Writing Team, H. Lee and J. Romero (eds.)}
}

@techreport{IPCC_WGIII,
  author       = {{Intergovernmental Panel on Climate Change}},
  title        = {Climate Change 2022: Mitigation of Climate Change. Contribution of Working Group III to the Sixth Assessment Report of the Intergovernmental Panel on Climate Change},
  institution  = {Cambridge University Press},
  year         = {2022},
  address      = {Cambridge, UK and New York, NY, USA},
  note         = {Edited by P.R. Shukla, J. Skea, R. Slade, A. Al Khourdajie, R. van Diemen, D. McCollum, M. Pathak, S. Some, P. Vyas, R. Fradera, M. Belkacemi, A. Hasija, G. Lisboa, S. Luz, J. Malley.}
}

@techreport{sparc2006assessment,
  title={SPARC Assessment of Stratospheric Aerosol Properties (ASAP)},
  author={{SPARC}},
  editor={Thomason, L and Peter, Th},
  institution={SPARC},
  number={Report No. 4, WCRP-124, WMO/TD-No. 1295},
  year={2006}
}

@techreport{sparc2017data,
  title={The SPARC Data Initiative: Assessment of stratospheric trace gas and aerosol climatologies from satellite limb sounders},
  author={{SPARC}},
  editor={Hegglin, Michaela I and Tegtmeier, Susann},
  institution={SPARC},
  number={Report No. 8, WCRP-5/2017},
  year={2017}
}

@article{Kovilakam2020GloSSAC,
  author  = {Kovilakam, Mahesh and Thomason, Larry W. and Ernest, Natalie and Rieger, Landon and Bourassa, Adam and Mill{\'a}n, Luis},
  title   = {The {G}lobal {S}pace-based {S}tratospheric {A}erosol {C}limatology (version 2.0): 1979--2018},
  journal = {Earth System Science Data},
  year    = {2020},
  volume  = {12},
  pages   = {2607--2634},
  doi     = {10.5194/essd-12-2607-2020}
}

\appendix

\section{Particle-Related Requirements}
\label{app:requirements}

This appendix provides a concise summary of the particle properties requirements derived from the safety requirements of \cite{2026SafetyWP} (tables \ref{tab:reqA}--\ref{tab:reqC}) and the functionality requirements derived in this paper (tables \ref{tab:req_eff_or}--\ref{tab:req_eff_df}). 
\begin{itemize}
    \item {\it Safety and controllability derived requirements}.Table~\ref{tab:reqA} summarizes the requirements related to direct compatibility with human health, biota, and the broader environment. These requirements act as admissibility and life-cycle constraints on particle composition, impurity content, size distribution, exposure, and environmental fate. Table~\ref{tab:reqB} summarizes the requirements related to atmospheric chemistry and composition. These constraints limit unintended perturbations to ozone chemistry, halogen cycling, sulfate surface area, PSC behavior, and cloud radiative forcing. Table~\ref{tab:reqC} summarizes the requirements related to controllability, predictability, and monitoring of the radiative forcing (RF) that impact the climate. Although the latter are system-level requirements, they imply constraints on particle properties. 
    \item {\it Functionality derived requirements}. 
    Table~\ref{tab:req_eff_or} groups optical performance and stratospheric residence requirements,
    which define the core trade space between forcing efficiency, absorption, and lifetime. Table~\ref{tab:req_eff_m} summarizes manufacturing ability requirements that should be met to ensure reproducible production at scale with practical supply-chain and logistics support. Table~\ref{tab:req_eff_df} summarizes dispersion compatibility requirements together with aircraft fleet and operational requirements, which capture the interface between particle properties and a practical deployment capability.
\end{itemize}
Each distinct quantitative requirement, capability constraint, or regulatory standard is listed in a separate row for direct reference. Throughout, ``local average'' denotes an average over one month and on a horizontal scale of approximately 1000~km.

\begin{table*}[ht]
\caption{Safety requirement: A. Safety for humans, biota, and environment (source: \S\,III.A of the companion paper~\cite{2026SafetyWP}).}
\label{tab:reqA}
\begin{ruledtabular}
\begin{tabular}{p{0.035\textwidth} p{0.095\textwidth} p{0.36\textwidth} p{0.29\textwidth} p{0.05\textwidth}}
\textbf{ID} & \textbf{Domain} & \textbf{Requirement summary} & \textbf{Quantitative target} & \textbf{Source} \\
\hline
\reqid{A1} & Chemical compound safety &
Particles must be constructed of compounds demonstrated to be compatible with safety requirements for humans, biota, and environment. &
Qualitative material screening; compliance per OECD test guidelines, EU-REACH, and EPA OCSPP frameworks. &
\cite{2026SafetyWP,OECD_AcuteToxicity,OECD_RepeatedDoseChronic,OECD_SkinEyeSensitisation,OECD_Mutagenicity,OECD_ReproductiveDevelopmental,OECD_Neurotoxicity,OECD_AquaticEcotoxicity,OECD_TerrestrialEcotoxicity,OECD_PersistenceBioaccumulation,EU_REACH_AnnexXIII,EU_REACH_2006,EPA_OPPTS_870} III.A \\[4pt]

\reqid{A1a} & Impurity limits &
Chemical compounds used must meet defined impurity limits. &
Per EU-CLP Reg.\ 1272/2008 thresholds and EU-REACH substance identity requirements. &
\cite{2026SafetyWP,EU_CLP,EU_REACH_2006} III.A \\[4pt]

\reqid{A1b} & Hazardous material exclusion &
Materials classified as hazardous by authoritative bodies should be avoided. &
Exclude per EU-REACH CLP (Reg.\ 1272/2008) and IARC (Groups~1, 2A, 2B). &
\cite{2026SafetyWP,EU_CLP,IARC_Monographs,EU_REACH_2006} III.A \\[4pt]

\reqid{A1c} & Nanoform avoidance &
Materials with non-negligible sub-0.1~$\mu$m content should be avoided. &
Per EU Reg.\ 2018/1881 amending REACH Annexes for nanoforms. &
\cite{2026SafetyWP,EU2018_1881} III.A \\[4pt]

\reqid{A2} & Human NOAEC/NOAEL and MOE &
NOAEC/NOAEL must be established for all relevant dose--response endpoints. Ground-level concentrations must maintain a benchmark MOE. &
Per OECD TG~407--453, 430--496, 414--443, 424--426; EPA OCSPP 870. MOE~$\geq$~100$\times$. &
\cite{2026SafetyWP,OECD_RepeatedDoseChronic,OECD_SkinEyeSensitisation,OECD_ReproductiveDevelopmental,OECD_Neurotoxicity,EPA_OPPTS_870,EFSA_MOE_2012,ECHA_R8,EFSA_2005} III.A \\[4pt]

\reqid{A3} & PM$_{2.5}$ compliance &
Ground-level particle density must remain well below PM$_{2.5}$ limits, with a separate population-level standard applied. &
Much lower than PM$_{2.5}$ ambient limits. &
\cite{2026SafetyWP} III.A \\[4pt]

\reqid{A4} & NOEC for biota &
Environmental concentrations must comply with NOECs for aquatic and terrestrial biota. &
Below NOEC/COC per OECD TG~201--211, 207--245. &
\cite{2026SafetyWP,OECD_AcuteToxicity,OECD_TerrestrialEcotoxicity} III.A \\[4pt]

\reqid{A5} & Mutagenicity and carcinogenicity &
Categorical assessment. Mutagenicity is binary: a positive result disqualifies. Carcinogenicity follows IARC categorical classification, with Groups~1 and 2A excluded. &
Mutagenicity: TG~471, 474, 487, 489 (pass/fail). Carcinogenicity: IARC; TG~451. &
\cite{2026SafetyWP,OECD_Mutagenicity,OECD_RepeatedDoseChronic,IARC_Monographs,EFSA_2005} III.A \\[4pt]

\reqid{A6} & Bioaccumulation and persistence &
Assess bioaccumulation; demonstrate viable elimination pathways. &
Per EU-REACH PBT/vPvB; OECD TG~305, 301, 310. &
\cite{2026SafetyWP,OECD_PersistenceBioaccumulation,EU_REACH_AnnexXIII} III.A \\[4pt]

\reqid{A7} & Aging and byproducts &
Particle aging and aging byproducts must not undermine compliance with A1--A6. &
Experimental demonstration of compliance across the particles' life-cycle. &
\cite{2026SafetyWP} III.A \\
\end{tabular}
\end{ruledtabular}
\end{table*}

\begin{table*}[ht]
\caption{Safety requirements: B. Atmospheric chemistry and composition (source: \S\,III.B of the companion paper~\cite{2026SafetyWP}).}
\label{tab:reqB}
\begin{ruledtabular}
\begin{tabular}{p{0.035\textwidth} p{0.095\textwidth} p{0.36\textwidth} p{0.29\textwidth} p{0.05\textwidth}}
\textbf{ID} & \textbf{Domain} & \textbf{Requirement summary} & \textbf{Quantitative target} & \textbf{Source} \\
\hline
\reqid{B1a} & Trace gas column &
The impact of direct interactions with stratospheric gases should be negligible. &
$\leq$~0.01~DU. &
\cite{2026SafetyWP} III.B \\[4pt]

\reqid{B1b} & Halogen species &
Fractional change in halogen (Cl, Br) column density must be limited. &
$\leq$~1\% fractional change. &
\cite{2026SafetyWP} III.B \\[4pt]

\reqid{B2} & Ozone: near-global TCO &
Near-global TCO (60$^\circ$S--60$^\circ$N) modification must remain below natural variability. &
$\leq\!\sim$1~DU. &
\cite{2026SafetyWP} III.B \\[4pt]

\reqid{B2'} & Ozone: high-latitude spring &
Modification must remain below natural variability&
$\leq\!\sim$3~DU (well below CFC-era depletion). &
\cite{2026SafetyWP} III.B \\[4pt]

\reqid{B2a} & Sulfate surface area &
Increased stratospheric sulfate surface area must be limited. &
Comply with B2/B2$'$; The natural variability is $\sim$10\%. &
\cite{2026SafetyWP} III.B \\[4pt]

\reqid{B2b} & PSC amplification &
Particles must not extend PSC formation in a way that amplifies ozone depletion. &
Comply with B2/B2$'$. &
\cite{2026SafetyWP} III.B \\[4pt]

\reqid{B2c} & Climatic feedback &
SAI-driven climatic changes must not cause excess ozone modification. &
Comply with B2/B2$'$. &
\cite{2026SafetyWP} III.B \\[4pt]

\reqid{B3} & Cloud radiative forcing &
Cloud RF modification due to enhanced INP activity must be limited. &
$\leq$~10\% of the IRF. &
\cite{2026SafetyWP} III.B \\[4pt]

\reqid{B4} & Aging compliance &
Particle aging must not affect compliance with B1--B3. &
Experimental demonstration that B1--B3 are retained over the particles' lifetime. &
\cite{2026SafetyWP} III.B \\
\end{tabular}
\end{ruledtabular}
\end{table*}

\begin{table*}[ht]
\caption{Safety requirements: C. Climatic effects (source: \S\,III.C of companion white paper~\cite{2026SafetyWP}).}
\label{tab:reqC}
\begin{ruledtabular}
\begin{tabular}{p{0.035\textwidth} p{0.095\textwidth} p{0.36\textwidth} p{0.29\textwidth} p{0.05\textwidth}}
\textbf{ID} & \textbf{Domain} & \textbf{Requirement summary} & \textbf{Quantitative target} & \textbf{Source} \\
\hline
\reqid{C1} & Ramp-up RF monitoring &
Gradual ramp-up is required. At small-scale dispersion, corresponding to IRF $\ll0.3{\rm W/m^2}$, IRF is inferred from particle density and size distribution; at intermediate scale, corresponding to IRF $\sim0.3{\rm W/m^2}$, RF is measured directly. &
Small scale: $\sim$10\% IRF determination accuracy; Intermediate scale: $\sim$0.1\% IRF determination accuracy. &
\cite{2026SafetyWP} III.C \\[4pt]

\reqid{C2} & Operation RF monitoring &
Measure global net-flux change and hemispheric asymmetry at TOA. &
Measurement capabilities: (i)~0.3~W/m$^2$ global; (ii)~0.2~W/m$^2$ hemispheric. &
\cite{2026SafetyWP} III.C \\[4pt]

\reqid{C3a} & Natural decay &
Particle burden must naturally diminish quickly after the injection halt. &
Decay by 1 e-fold in $\sim$1~yr, and to a negligible residual level by $\sim$5~yr. &
\cite{2026SafetyWP} III.C \\[4pt]

\reqid{C3b} & Phase-down &
The system must support a gradual, controlled phase-down. &
Controlled phase down over a years-to-decade time scale. &
\cite{2026SafetyWP} III.C \\[4pt]

\reqid{C4a} & Spatial and temporal control &
The system must allow for flexible latitude and temporal dependence in the IRF. &
Allowing dispersion at a similar rate per surface area in 20$^\circ$ latitude ($\sim$2000~km) bins, and monthly cadence.
&
\cite{2026SafetyWP} III.C \\[4pt]

\reqid{C4b} & RF predictability &
Locally averaged IRF/SARF must be predictable; hemispheric IRF/SARF must be predictable and controllable. &
Locally (1~month, 1000~km) averaged IRF should be predictable to $<$10\%; Hemispherically averaged IRF should be controlled to better than 0.2~W/m$^2$. &
\cite{2026SafetyWP} III.C \\[4pt]

\reqid{C4c} & Stratospheric heating &
The heating of the tropical lower stratosphere should be limited, imposing constraints on particles' IR absorption. &
Heating at 25$^\circ$N--25$^\circ$S, 50--100~hPa limited to $\leq$~1.5~K; Monitoring should enable $<$~1~K sensitivity. &
\cite{2026SafetyWP} III.C \\[4pt]

\reqid{C4d} & Tagging and attribution &
Encode batch, injection site, and date to enable full attribution. &
Tagging should enable attribution via in-situ or ground measurements of collected particle samples. &
\cite{2026SafetyWP} III.C \\
\end{tabular}
\end{ruledtabular}

\vspace{3pt}
\begin{minipage}{0.96\textwidth}
\footnotesize\itshape
Note. DU = Dobson Unit; TCO = Total Column Ozone; IRF = Instantaneous Radiative Forcing; SARF = Stratospherically Adjusted Radiative Forcing; TOA = Top of Atmosphere; NOAEC = No Observed Adverse Effect Concentration; NOAEL = No Observed Adverse Effect Level; NOEC = No Observed Effect Concentration; PSC = Polar Stratospheric Cloud; INP = Ice-Nucleating Particle; MOE = Margin of Exposure; COC = Concentration of Concern.
\end{minipage}
\end{table*}

\begin{table*}[t]
\caption{Functionality requirements: I. optical properties (O) and stratospheric residence (R) (source: Sec.~\ref{sub-sec:functionality} of the present paper).}
\label{tab:req_eff_or}
\small
\renewcommand{\arraystretch}{1.05}
\begin{ruledtabular}
\begin{tabular}{p{0.045\textwidth} p{0.11\textwidth} p{0.34\textwidth} p{0.30\textwidth} p{0.055\textwidth}}
\textbf{ID} & \textbf{Domain} & \textbf{Requirement summary} & \textbf{Quantitative target} & \textbf{Source} \\
\hline

\reqid{O} & Optical properties &
Particles must provide efficient shortwave scattering while limiting longwave absorption and maintaining stable optical performance throughout their stratospheric lifetime. &
Enable approximately 1\% reflection of incoming solar radiation, corresponding to roughly 2.5~W/m$^2$ of radiative offset, at operationally plausible dispersion rates of order $10$ million tons per year. &
\S\ref{sub-sec:functionality} \\[4pt]

\reqid{Oa} & Shortwave scattering &
Efficient scattering of incoming solar radiation requires particle sizes comparable to visible wavelengths together with a dielectric constant significantly larger than unity at optical wavelengths. &
Particle diameter of roughly 0.2--0.8~$\mu$m; refractive index well above 1 at $\lambda\sim 550$~nm; radiative forcing efficiency of order several $10^{-1}$~W\,m$^{-2}$\,Tg$^{-1}$. &
\S\ref{sub-sec:functionality} \\[4pt]

\reqid{Ob} & Longwave absorption and stratospheric heating &
Absorption of Earth-emitted infrared radiation must be minimized to limit stratospheric heating and positive RF contribution. 
&
Absorption minimized within the 8--13~$\mu$m atmospheric window; supports compliance with C4c ($\Delta T\leq 1.5$~K in the tropical lower stratosphere). &
\S\ref{sub-sec:functionality} \\[4pt]

\reqid{Oc} & Optical stability over lifetime &
Interactions with the stratospheric environment must not alter optical properties beyond specification over the particle lifetime. &
No change in refractive index or absorption coefficient sufficient to violate Oa or Ob over an approximately one-year residence time. &
\S\ref{sub-sec:functionality} \\[4pt]

\reqid{R} & Stratospheric residence &
Particles must remain aloft long enough to provide useful radiative forcing while preserving controllability and phase-down capability. &
Residence time of order 1~year. &
\S\ref{sub-sec:functionality} \\[4pt]

\reqid{Ra} & Sedimentation constraint &
Particle sedimentation must remain sufficiently slow relative to Brewer--Dobson upwelling. &
Sedimentation velocity $\le0.1~mm/s$.
&
\S\ref{sub-sec:functionality} \\[4pt]

\reqid{Rb} & Joint optical--lifetime optimization &
Particle size must be chosen by jointly optimizing scattering efficiency and residence time rather than optical performance alone. &
Optimal particle diameter somewhat smaller than an optimum based purely on optical performance. &
\S\ref{sub-sec:functionality} \\[4pt]

\reqid{Rc} & Stability of size and density distributions&
Interactions with the stratospheric environment must not change particle size or density enough to violate the sedimentation constraint over the particles' lifetime. &
No change sufficient to violate Ra over the particle lifetime. &
\S\ref{sub-sec:functionality} \\
\end{tabular}
\end{ruledtabular}
\normalsize
\end{table*}

\begin{table*}[t]
\caption{Functionality requirements: II. Manufacturability (M) (source: Sec.~\ref{sub-sec:functionality} of the present paper).}
\label{tab:req_eff_m}
\small
\renewcommand{\arraystretch}{1.05}
\begin{ruledtabular}
\begin{tabular}{p{0.045\textwidth} p{0.11\textwidth} p{0.34\textwidth} p{0.30\textwidth} p{0.055\textwidth}}
\textbf{ID} & \textbf{Domain} & \textbf{Requirement summary} & \textbf{Quantitative target} & \textbf{Source} \\
\hline

\reqid{M} & Manufacturability &
Production must be feasible at global deployment scale, at sufficiently low cost, with reproducible particle properties and practical logistics. &
Production on the order of several million tons per year at a fabrication cost of roughly a few dollars per kilogram. &
\S\ref{sub-sec:functionality} \\[4pt]

\reqid{Ma} & Supply-chain feasibility &
Feedstocks and manufacturing equipment must be available at the necessary scale without impractical market bottlenecks. &
Access to feedstocks and manufacturing equipment at the necessary scale. &
\S\ref{sub-sec:functionality} \\[4pt]

\reqid{Mb} & QA/QC and batch reproducibility &
Critical particle properties must remain reproducible within and across production batches under mass manufacturing. &
Reproducibility of all critical particle parameters across production batches. &
\S\ref{sub-sec:functionality} \\[4pt]

\reqid{Mc} & Ramp-up and capital feasibility &
Factories must be capable of reaching required production volumes within a realistic timeline and capital budget. &
Production ramp-up and associated capital expenditure compatible with realistic deployment timelines and budgets. &
\S\ref{sub-sec:functionality} \\[4pt]

\reqid{Md} & Storage and transport logistics &
Storage, handling, and transport from production sites to airfields must remain practical and economical at the required throughput. &
Practical storage and transport logistics from factory to airfield at deployment-scale throughput. &
\S\ref{sub-sec:functionality} \\[4pt]

\reqid{Me} & Process maturity and scale-up path &
Fabrication should rely on well-established chemistry together with a credible scale-up strategy. &
Use of well-established chemical processes supported by a credible scale-up plan. &
\S\ref{sub-sec:functionality} \\
\end{tabular}
\end{ruledtabular}
\normalsize
\end{table*}

\begin{table*}[t]
\caption{Functionality requirements: III. Dispersion compatibility (D) and fleet and operations (F) (source: Sec.~\ref{sub-sec:functionality} of the present paper).}
\label{tab:req_eff_df}
\small
\renewcommand{\arraystretch}{1.05}
\begin{ruledtabular}
\begin{tabular}{p{0.045\textwidth} p{0.11\textwidth} p{0.34\textwidth} p{0.30\textwidth} p{0.055\textwidth}}
\textbf{ID} & \textbf{Domain} & \textbf{Requirement summary} & \textbf{Quantitative target} & \textbf{Source} \\
\hline

\reqid{D} & Dispersion compatibility &
Particle properties must enable onboard conveyance, deagglomeration, and airborne dispersion while preserving the target particle state. &
Release of deagglomerated particles while preserving the target PSD at the nozzle exit and after wake expansion, at mass-release rates of order a few tons per hour. &
\S\ref{sub-sec:functionality} \\[4pt]

\reqid{Da} & Deagglomeration and PSD fidelity &
The post-dispersion particle size distribution must match the intended target both at the nozzle exit and after wake evolution. &
Nozzle-exit and post-wake PSD must match the target PSD at deployment-relevant mass-release rates of order a few tons per hour. &
\S\ref{sub-sec:functionality} \\[4pt]

\reqid{Db} & Fracture avoidance &
Particle's fracture and ultra-fine fragments production during dispersion should be avoided. &
No significant generation of sub-0.1~$\mu$m fragments; supports compliance with A1c. &
\S\ref{sub-sec:functionality} \\[4pt]

\reqid{Dc} & Reagglomeration control &
Electrostatic charging and other mechanisms of reagglomeration may need to be controlled in order to preserve the target PSD. &
Charge control sufficient to prevent significant reagglomeration, where needed. &
\S\ref{sub-sec:functionality} \\[4pt]

\reqid{Dd} & Conveyance and aircraft compatibility &
The particle powder must be conveyable onboard with acceptable energy expenditure, and the dispersion system must remain compatible with aircraft payload constraints. &
Acceptable conveyance energy; system mass and volume compatible with aircraft payload constraints. &
\S\ref{sub-sec:functionality} \\[4pt]

\reqid{De} & Post-dispersion physicochemical stability &
Engine exhaust, nozzle conditions, and plume evolution must not degrade coatings or alter critical particle properties. &
Optical, agglomeration, sedimentation, and surface-chemistry properties remain within specification after dispersion. &
\S\ref{sub-sec:functionality} \\[4pt]

\reqid{Df} & Mission completion &
The combined particle--disperser system must support mass-release rates high enough that the required payload can be dispersed within the operational envelope of a sortie. &
Required payload dispersed within a practical flight duration at the required altitude, consistent with feasible sortie rates and fleet size. &
\S\ref{sub-sec:functionality} \\[4pt]

\reqid{F} & Fleet and operations &
The operational system must be capable of delivering the required annual mass budget at the required altitude, latitude bands, and cost. &
Aircraft fleet capable of reaching the required dispersion altitude with sufficient payload and sortie rate for sustained multi-year operation. &
\S\ref{sub-sec:functionality} \\[4pt]

\reqid{Fa} & Throughput and altitude access &
Aircraft must reach the required altitude with sufficient payload per sortie. &
Payload and altitude capability sufficient for deployment operations above the tropopause, including roughly $18.5$~km ($60$~kft) for reliable tropical injection. &
\S\ref{sub-sec:functionality} \\[4pt]

\reqid{Fb} & Annual mass delivery &
The flight rate must be high enough to deliver the annual mass budget across the intended latitude bands. &
Sortie rate sufficient to meet annual mass targets across intended latitude bands. &
\S\ref{sub-sec:functionality} \\[4pt]

\reqid{Fc} & Geographic coverage &
Fleet size, operating bases, and sortie frequency must provide adequate latitude coverage for the desired forcing distribution. &
Adequate latitude coverage through the number of operational bases and sortie frequency. &
\S\ref{sub-sec:functionality} \\[4pt]

\reqid{Fd} & Multi-year cost viability &
The per-kilogram lofting cost must remain economically feasible for sustained operation. &
Per-kilogram lofting cost compatible with sustained multi-year deployment. &
\S\ref{sub-sec:functionality} \\[4pt]

\reqid{Fe} & Tagging and measurement operations &
If tagging is used, aerial platforms must be able to collect or measure tagged particles routinely; these platforms may differ from the dispersion fleet. &
Routine in situ and/or collected-particle measurement capability across relevant latitudes and timescales; measurement platforms may differ from the dispersion fleet. &
\S\ref{sub-sec:functionality} \\
\end{tabular}
\end{ruledtabular}

\vspace{3pt}
\begin{minipage}{0.96\textwidth}
\footnotesize\itshape
Note. PSD = particle size distribution; RF = radiative forcing; GHG = greenhouse gas. The functionality requirements use a thematic identifier convention: O (optical), R (residence), M (manufacturability), D (dispersion), and F (fleet and operations).
\end{minipage}

\normalsize
\end{table*}
\clearpage
\section{Estimate of Ground-Level Tropospheric Concentration}
\label{app:groundconc}

Ground-level airborne concentrations of descended SAI particles are expected to be far below those of existing natural and anthropogenic aerosols, owing to the much shorter aerosol lifetime in the troposphere than in the stratosphere~\citep{colarco2010online,toohey2025stratospheric_lifetime}.

An approximate estimate of ground level concentration is obtained by treating the tropospheric aerosol column as a steady reservoir fed by particles descending from the stratosphere at a global-mean flux $\dot{M}/(4\pi R_\oplus^2)$, with effective \emph{tropospheric} residence time $\tau_{\mathrm{eff}}$ (
which is much shorter than the stratospheric lifetime), and mapping the resulting column burden into a near-surface concentration assuming a mixing depth $H_g(\mathbf{x})$. This gives
\begin{equation}
C_g(\mathbf{x}) = \eta_s(\mathbf{x})
\frac{\,\dot{M}\,\tau_{\mathrm{eff}}}{4\pi R_\oplus^2\,H_g(\mathbf{x})},
\end{equation}
where $\dot{M}$ is the global injection rate and $\eta_s(\mathbf{x})$ is a dimensionless order unity parameter 
that absorbs the effects of descent pathways, preferential scavenging aloft, incomplete mixing into the boundary layer, and regional inhomogeneity (with $\bar{\eta}_s = 1$ 
for the global mean).
Natural analogs such as the 1961--62 fallout record and the weak meteoric background provide qualitative support to this scaling estimate, and also demonstrate that real atmospheric transport and scavenging generate substantial variability around the average $\eta_s(\mathbf{x})$ \citep{feely1966fallout,plane2012cosmic_dust}.

Using tropospheric aerosol lifetimes reported for fine aerosol species \citep{colarco2010online}(4.4~d for sulfate, 6.9~d for particulate organic matter, and 8.8~d for black carbon), so that $\tau_{\mathrm{eff}} \approx 4$--$9$~d with a fiducial value near 6--7~d, the global-mean ground-level concentration evaluates to
\begin{equation}
\bar{C}_g \approx (0.02\text{--}0.05)\;\mu\mathrm{g\,m^{-3}}\;
\left(\frac{\bar{\eta}_s}{1}\right)
\left(\frac{\dot{M}}{1\;\mathrm{Mt\,yr^{-1}}}\right)
\left(\frac{H_g}{1\;\mathrm{km}}\right)^{-1}.
\end{equation}
To place this in context, the global population-weighted annual-mean PM$_{2.5}$ concentration is $\sim 28\;\mu\mathrm{g\,m^{-3}}$ \citep{ihme2024gbd2021_air_pollution}, and the WHO annual guideline is $5\;\mu\mathrm{g\,m^{-3}}$~\citep{ihme2024gbd2021_air_pollution,who2021air_quality_guidelines}; the estimated SAI ground-level contribution is therefore $\sim 600$--$1200\times$ smaller than the existing PM$_{2.5}$ background. The margin relative to toxicological thresholds is even wider: the sub-chronic inhalation NOAEC for synthetic amorphous silica (the material class closest to the envisaged SAI aerosol \citep{atsdr2019silica,ecetoc2014silica,oecd2004silica}) is $\sim 1\;\mathrm{mg\,m^{-3}}$ ($= 1000\;\mu\mathrm{g\,m^{-3}}$) for sub-micron particles, roughly $2\times 10^4$--$4\times 10^4$ times higher than $\bar{C}_g$.

\section{Estimate of Dissolution Timescale of Deposited Silica Spheres}
\label{app:dissolution}

Amorphous silica dissolves in seawater to form silicic acid, which is continuously cycled through biological uptake: diatoms, radiolaria, and siliceous sponges extract nearly all dissolved silica from surface waters to biomineralize their skeletal structures, most of which redissolves upon sinking~\cite{demaster1981supply,mann2001biomineralization}. The total natural input from rivers and hydrothermal emanations is $\sim 6\times 10^{14}\;\mathrm{g\,SiO_2\,yr^{-1}}$, with a mean oceanic residence time of $\sim 20$~kyr. A dissolution rate of $1\;\mathrm{Mt\,yr^{-1}}$ ($10^{12}\;\mathrm{g\,SiO_2\,yr^{-1}}$) would constitute $<0.2\%$ of this flux, a negligible perturbation to the marine silica cycle.

Well below saturation, the dissolution timescale of an amorphous silica sphere scales linearly with its diameter, $\tau_{\mathrm{diss}}\propto D$, since the mass-normalized dissolution rate scales inversely with $D$. For untreated 500~nm spheres in near-neutral water at room temperature, $\tau_{\mathrm{diss}}$ is of order several weeks . warm or alkaline conditions, typical of surface seawater, shorten this to several days. Optimization of the surface treatment and calcination process may be controlled~\citep{spitzmuller2023dissolution}. quantitative measurements will be presented in a companion study \citep{sai_biosafety}.

\end{document}